\documentclass[%
 reprint,
superscriptaddress,
nofootinbib,
 amsmath,amssymb,
 aps,
prb,
]{revtex4-2}

\usepackage{graphicx}
\usepackage{dcolumn}
\usepackage{bm}
\usepackage{xcolor}
\usepackage{stmaryrd}
\usepackage{natbib}
\usepackage{lipsum}  
\usepackage{dsfont}
\usepackage{amssymb}
\usepackage{times}
\usepackage{BOONDOX-calo}

\usepackage{url}

\begin{document}

\preprint{APS/123-QED}

\title{Theory of the $\beta$-Relaxation Beyond Mode-Coupling Theory: A Microscopic Treatment}

\author{Corentin C. L. Laudicina}
\affiliation{Soft Matter and Biological Physics, Department of Applied Physics, Eindhoven University of Technology,
P.O. Box 513, 5600 MB Eindhoven, Netherlands
}
\author{Liesbeth M. C. Janssen}
\affiliation{Soft Matter and Biological Physics, Department of Applied Physics, Eindhoven University of Technology,
P.O. Box 513, 5600 MB Eindhoven, Netherlands
}
\affiliation{Institute for Complex Molecular Systems, Eindhoven University of Technology, P.O. Box 513, 5600MB Eindhoven, The Netherlands}

\author{Grzegorz Szamel}
\affiliation{Department of Chemistry, Colorado State University, Fort Collins, Colorado 80523, USA}

\date{\today}

\begin{abstract}
We develop a systematic extension of mode-coupling theory (MCT) that incorporates critical dynamical fluctuations. Starting from a microscopic diagrammatic theory, we identify dominant classes of divergent diagrams near the mode-coupling transition and show that the corresponding asymptotic series dominates the mean-field prediction below an upper critical dimension $d_c=8$. To resum these divergences, we construct a mapping to a stochastic dynamical process in which the order parameter evolves under random spatio-temporal fields. This reformulation provides a controlled, fully dynamical derivation of an effective theory for $\beta$-relaxation that remarkably coincides with stochastic beta-relaxation theory [T. Rizzo, EPL \textbf{106}, 56003 (2014)]. All coupling constants of the latter theory are expressed microscopically in terms of the liquid static structure factor and are computed for the paradigmatic hard-sphere system in the Percus-Yevick approximation. The analysis demonstrates that critical fluctuations alone restore ergodicity and replace the putative mean-field transition with a smooth crossover. Our results establish a predictive framework for structural relaxation beyond mean-field. 
\end{abstract}

\maketitle

\section{Introduction}

The theoretical description of the dynamical glass transition has long relied on mean-field approaches. In recent decades, two complementary frameworks have emerged: a dynamical one known as mode-coupling theory (MCT) \cite{leutheusser1984dynamical, bengtzelius1984dynamics, gotze2009complex} and a static one based on the replica technique \cite{parisi2010mean,mezard2012glasses}. MCT identifies the transition dynamically, through 
time-dependent correlation functions that fail to fully decay, producing low-frequency singularities in the relaxation spectra. In contrast, in the replica approach, the transition is signaled by the appearance of a secondary minimum in the Franz-Parisi potential (\textit{i.e.}\ the Landau free-energy functional associated with an appropriate order parameter), reflecting the proliferation of exponentially numerous metastable amorphous states separated by diverging barriers and the concomitant loss of ergodicity. More recently, an exact theory for the dynamics of simple liquids in the infinite-dimensional limit (DMFT) has also been developed \cite{maimbourg2016solution,liu2021dynamics}.

Despite their different formulations, both MCT and replica-based theories predict a mean-field dynamical arrest at a critical temperature $T_c$, and their formal equivalence has been established through mappings that reproduce the asymptotic mode-coupling dynamic scaling near the transition from replica calculations \cite{parisi2013critical, rizzo2013supercooled}. While MCT is known to be incorrect in the limit of infinite dimensions \cite{schmid2010glass, ikeda2010mode, laudicina2024simple}, it is nevertheless conceptually related to the DMFT of the glass transition. In fact, all three theories predict qualitatively similar scaling behavior near the dynamic transition (which we know to exist in the limit of infinite dimensions). In contrast to MCT, the replica theory does becomes exact in the limit of infinite dimensions \cite{kurchan2012exact, kurchan2013exact, charbonneau2014exact}. While the precise location of the critical point depends on the approach, by concentrating on the similarity of the critical scaling we can nevertheless say that these three theories belong to the same universality class. The discrepancies in the location of the transition are purely a technical matter reflecting different approximations involved in each scheme \cite{szamel2010dynamic}.

In finite dimensions, however, the dynamical phase transition is never observed as a true singularity. Simulations \cite{berthier2020finite, scalliet2022thirty, charbonneau2022dimensional} and experiments \cite{brambilla2009probing, mallamace2010transport,schmidtke2012boiling} on supercooled liquids consistently show that although structural relaxation times grow rapidly as temperature decreases (or density increases), they remain finite at all accessible conditions. Instead of a genuine ergodicity-breaking transition at $T_c$, one typically observes a crossover: transport coefficients such as the viscosity evolve smoothly from an apparent super-Arrhenius behavior to a weaker Arrhenius-like one \cite{mallamace2010transport}. Concomitantly, supercooled liquids in this parameter regime tend to display strongly heterogeneous dynamics \cite{berthier2011overview}. Yet, dynamic susceptibilities and associated correlation lengths that measure the magnitude and extent of such dynamic heterogeneities grow only modestly without any obvious singularities \cite{berthier2011overview, dauchot2023glass}, contrary to mean-field predictions \cite{franz2000onnon, donati2002theory,biroli2006inhomogeneous}. This rounding of the mean-field singularity into a continuous crossover is a widely discussed \cite{kirkpatrick1989scaling, gotze1998essentials}, but by no means universally accepted, scenario \cite{biroli2013perspective, tanaka2025structural}.

Understanding why the transition is avoided requires going beyond mean-field theory. For any continuous phase transition (such as the dynamical arrest predicted by the aforementioned mean-field theories), this means assessing the role of critical fluctuations of the order parameter. It is well understood that in physical spatial dimensions, mean-field critical behavior is generally unstable to the inclusion of critical fluctuations. A paradigmatic example is provided by the Ising universality class, where the Wilson–Fisher fixed point supplants the Gaussian (mean-field) one below four dimensions \cite{wilson1972critical}. This program has been pursued for replicated field theories of structural glasses, where it was shown that the mean-field critical point is unstable below an upper critical dimension $d_c=8$ \cite{franz2011field}. More precisely, the non-ergodic mean-field solution becomes unstable, suggesting the absence of a true dynamical transition in physical dimensions. Yet this analysis is entirely static: it constrains the structure of metastable states and the barriers between them, and does not directly address how the dynamical features of glass formation, such as time-dependent correlations, are modified by fluctuations. These aspects can only be accessed indirectly, through the mapping of replicated field theories onto dynamical supersymmetric field theories \cite{rizzo2016dynamical}, leading to an effective theory known as stochastic beta-relaxation (SBR) \cite{rizzo2014long, rizzo2015qualitative, rizzo2016dynamical, rizzo2020solvable}; a beautiful construction, but one that does not fully incorporate real microscopic dynamics.

Although the replica framework provides a powerful conceptual picture of metastable states, its focus on static quantities makes direct comparison with experiments and simulations somewhat difficult. In contrast, MCT has achieved broad success precisely because it yields detailed, falsifiable predictions for dynamical observables in finite dimensions \cite{kob1995testing, kob1995testing2, gotze1999recent, weysser2010structural}. Thus, even though it predicts the incorrect large dimensional physics, MCT remains an excellent starting point for improvements and refinements. A majority of these improvements attempt to remedy the shortcomings of MCT by invoking additional relaxation channels or processes that are generically called `activated' or `hopping' mechanisms \cite{Schweizer2003NLE, mayer2006cooperativity, MedinaNoyola2007PRE,  bhattacharyya2008facilitation, janssen2015microscopic, Schweizer2013EC}. The precise physical nature of these processes and how to appropriately incorporate them in a theoretical framework is a subject of intense debate. What is still lacking, however, is a systematic framework to incorporate critical dynamical fluctuations beyond the mean-field level. Developing such an extension is the central aim of this work. To this end, we build on a diagrammatic formulation of MCT previously derived by one of us \cite{szamel2007dynamics}. 
In essence, we 
go beyond the mean-field scenario of MCT by including certain classes of spatio-temporal density fluctuations, which changes the singular transition into a smooth crossover. 

Our main result is that the mean-field dynamical transition is destabilized by critical fluctuations solely, in all physical dimensions. No new relaxation processes are necessary to destroy the transition. Beyond this qualitative result, we demonstrate that our analysis leads to an equation identical to SBR. Our results therefore provide, for the first time, a fully microscopic and dynamical derivation of this effective theory for a system of interacting particles. Crucially, our treatment yields a complete microscopic characterization of all coupling constants of SBR, which can be computed directly from the liquid's static structure factor alone.

The rest of the paper is structured as follows. In Sec.~\ref{sec:angle_of_attack}, we recall the general theoretical description of dynamic heterogeneities and outline the general strategy of our calculation. In Sec.~\ref{sec:diagrammatic_rules} we briefly review the diagrammatic kinetic theory of mode–coupling, which provides the microscopic foundation for our analysis. Section~\ref{sec:brief_mean_field_scenario} summarizes the mean-field transition scenario, from the behavior of the order parameter to the critical dynamical susceptibilities. In Sec.~\ref{sec:leading_divergences_diagrammatic} we identify and study the leading divergent contributions that arise in the diagrammatic expansion near criticality. This allows us to compute the upper critical dimension. Sec.~\ref{sec:mapping_stochastic_process} introduces the effective stochastic process that resums these divergences. In Sec.~\ref{sec:asymptotic_dynamics_beyond_mean-field} we analyze the asymptotic dynamics beyond mean-field, and demonstrate that we are able to recover SBR. We continue with a brief discussion of the theory and its properties in Sec.~\ref{sec:properties_SBR}. We conclude in Sec.~\ref{sec:conclusions} with a discussion of the implications of these results and future perspectives. This manuscript is also accompanied by an extended set of appendices which provide detailed derivations of the key expressions obtained in this work.

A concise overview of the central ideas underlying the calculations shown here are presented in a companion work \cite{companionpaper}. Here we provide the details of the diagrammatic structure, the derivations, and the systematic dynamical treatment of the perturbative expansion.

\section{Angle of Attack}\label{sec:angle_of_attack}

In this section we motivate our approach and present it in very general terms. The discussion is deliberately qualitative, prioritizing physical intuition over technical precision.

\subsection{Dynamic Heterogeneity \& Critical Fluctuations}\label{sec:dyn_het_crit_fluct}

To address the problem of fluctuation corrections to the dynamical transition, we first recall the observables through which both the structural relaxation and its spatio-temporal fluctuations are typically characterized in supercooled liquids.

The overall picture of structural relaxation is captured by the intermediate scattering function $F(k;t)$, where isotropy dictates that $F$ depends only on the modulus $k$ of the probed wavevector $\boldsymbol{k}$. Typically, this observable is evaluated at the wavenumber corresponding to the first peak of the static structure factor $S(k)$, to probe relaxation on the scale of the typical cage size. Importantly, the long-time limit of $F(k;t)$ constitutes the order parameter of the dynamical transition: in the mean-field picture it develops a non-vanishing long-time limit $F(k) = \lim_{t\to\infty}F(k;t)>0$ upon entering the non-ergodic glass phase \cite{gotze2009complex}. 

Dynamic heterogeneities, which correspond to spatio-temporal correlations of local fluctuations of relaxation processes, are precisely the type of fluctuations we aim to treat. To quantify them, one may consider correlations between local mobility fluctuations at positions $\boldsymbol{r}$ and $\boldsymbol{r}'$ in the fluid over a time interval $t$. Since mobility indicators are already two-point functions, correlations between their local fluctuations at two points in space naturally take the form of four-point functions. If one focuses on fluctuations of local relaxations at wavevector $\boldsymbol{k}$, one arrives at a four-point function denoted $G_4(\boldsymbol{r} ; \boldsymbol{k}; t)$, which measures the correlation between fluctuations separated by distance $\boldsymbol{r}=\boldsymbol{r}_1-\boldsymbol{r}_2$. More generally, disparate microscopic definitions of $G_4$ have been proposed in the literature \cite{lavcevic2002growing, berthier2011overview}, but they typically probe the same underlying physics. 

At fixed $\boldsymbol{k}$ and $t$, $G_4(\boldsymbol{r} ; \boldsymbol{k}; t)$ is expected to decay to zero as $r=|\boldsymbol{r}|\to\infty$, reflecting the fact that sufficiently well-separated regions relax independently. The characteristic lengthscale over which this decay occurs defines the dynamic correlation length $\xi_{\mathrm{d}}(t)$, which quantifies the typical linear size of regions that relax cooperatively on time scale $t$. A growing $\xi_\mathrm{d}(t)$ signals that growing regions of the system must move in concert for structural relaxation to occur. The Fourier transform of $G_4(\boldsymbol{r} ; \boldsymbol{k}; t)$ with respect to $\boldsymbol{r}$ defines the four-point dynamic structure factor $S_4(\boldsymbol{q} ; \boldsymbol{k}; t)$, thereby resolving correlations in local relaxation fluctuations at wavevector $\boldsymbol{q}$. Its zero-wavevector limit yields the well studied dynamic susceptibility $\chi_4(\boldsymbol{k}; t) = \lim_{\boldsymbol{q}\to0}S_4(\boldsymbol{q} ; \boldsymbol{k}; t)$. As discussed later, both $\chi_4(\boldsymbol{k}; t)$ and $\xi_{\mathrm{d}}(t)$ diverge at the mean-field transition.

An important advance made by Berthier \emph{et al.}~\cite{berthier2007spontaneousI, berthier2007spontaneousII} was to establish that the microscopic mechanisms driving the critical behavior of four-point functions are encoded in the so-called three-point dynamical susceptibilities $\chi_{\boldsymbol{q}}(\boldsymbol{k}; t)$ \cite{biroli2006inhomogeneous}, which measure the response of the correlator $F(k;t)$ to a spatially modulated perturbation at wavevector $\boldsymbol{q}$. Schematically, $S_4(\boldsymbol{q} ; \boldsymbol{k}; t) \sim \chi_{\boldsymbol{q}}(\boldsymbol{k}; t)^2$, so that divergences of $\chi_4(\boldsymbol{k}; t)$ and $\xi_{\mathrm{d}}(t)$ upon approaching the transition are traceable to critical behavior in $\chi_{\boldsymbol{q}}(\boldsymbol{k}; t)$. Incorporating fluctuation corrections to the dynamical glass transition therefore reduces to computing $\chi_{\boldsymbol{q}}(\boldsymbol{k}; t)$ within a theoretical framework for $F(k;t)$, and self-consistently accounting for $\chi_{\boldsymbol{q}}(\boldsymbol{k}; t)$'s feedback on the order parameter. The strategy by which this is achieved is outlined in the remainder of this section. 

\subsection{Outline of the Computation}
Having recalled the pivotal role played by the dynamical susceptibilities $\chi_{\boldsymbol{q}}(\boldsymbol{k},t)$ in the theoretical description of dynamic heterogeneity, we now outline the spirit of our calculation. Our overall strategy can be divided into two successive stages.

\textbf{Stage I: Perturbative analysis of fluctuation-dominated corrections}. Since the critical behaviour is governed by the $\beta$-regime of relaxation\footnote{Glassy relaxation generically exhibits two successive dynamical regimes: the $\beta$-regime, corresponding to dynamics in the vicinity of the plateau of $F(k;t)$, and the $\alpha$-regime, corresponding to complete structural relaxation. The critical properties of the mean-field transition are entirely encoded in the $\beta$-regime.}, we work asymptotically close to the mean-field transition. While the analysis could be carried out on the non-ergodic side by working directly with the long-time limit of the correlator $F(k) = \lim_{t\to\infty}F(k;t)$, we instead retain the full temporal dependence of the diagrammatic series. This allows us to treat both sides of the transition on an equal footing and will prove essential for the calculations carried out in Stage II below.

We decompose the full dynamical order parameter as $F(k;t) = F_{\mathrm{MCT}}(k; t) + \Delta F(k;t)$, where $F_{\mathrm{MCT}}(k; t)$ is the standard MCT solution. We study contributions to the correction term $\Delta F(k;t)$ arising from specific classes of diagrams that are neglected in MCT. Specifically, we focus on diagrams that upon resummation can be expressed in terms of the three-point susceptibilities $\chi_{\boldsymbol{q}}(\boldsymbol{k}; t)$ discussed above. In this way we incorporate the feedback of spatio-temporal fluctuations on the order parameter in a controlled, diagrammatic fashion.

An explicit perturbative analysis reveals that the terms in the resulting series for $\Delta F(k;t)$ at each successive order grow increasingly singular near the transition, signaling the complete breakdown of the mean-field description below an upper critical dimension $d_c=8$. We show that the same value of the upper critical dimension follows from a Ginzburg criterion. Notably, this value is consistent with the result obtained from a replica field-theoretic approach \cite{franz2011field}.

\textbf{Stage II: Resummation through a stochastic process.} To overcome the breakdown of perturbation theory and formally resum the divergent series, we take inspiration from techniques developed in the theory of critical phenomena in disordered systems \cite{parisi1979random, parisi1981critical}. We establish a mapping between $F(k;t)$, with $\Delta F(k;t)$ given by the divergent diagrammatic series, and the solution to an auxiliary stochastic problem. To this end, we introduce random Gaussian fields and a dynamical correlation function  $F_u(\boldsymbol{k}_+, \boldsymbol{k}_- ; t, 0)$ that obeys a mode-coupling-like equation of motion driven by these fields. We show that the disorder averaged solution of $F_u$ gives the full correlator $F(k;t)$ and thus provides a resummation of the divergent corrections identified above. The key insight is that the diagrammatic structure of the contributions to $\Delta F(k;t)$ identified in Stage I maps precisely onto the perturbative expansion of $F_u$ in powers of the random fields, so that the disorder average systematically resums the dominant contributions to all orders.

\textbf{Outcome.} Analyzing the time evolution of $F_u$ asymptotically in the $\beta$-relaxation regime, we demonstrate that ergodicity is restored solely through the inclusion of critical fluctuations. Moreover, the effective theory governing $F_u$ coincides with the basic equation of SBR \cite{rizzo2014long, rizzo2015qualitative, rizzo2016dynamical, rizzo2020solvable}. Crucially, all coupling constants of the SBR are determined directly from the static structure factor, yielding parameter-free predictions beyond mean field. 

\section{Diagrammatic Theory for Interacting Brownian Particles}\label{sec:diagrammatic_rules}

We consider a monodisperse system of strongly interacting Brownian particles moving in a $d$-dimensional space with average number density $n$. The bare diffusion constant of an isolated particle is denoted $D_0$. The system is assumed to be in thermal equilibrium such that its configurations are sampled from the Gibbs distribution. In this work, we treat $n$ as the sole control parameter. However, our framework readily generalizes to include additional parameters such as temperature. Moreover, we expect the results to be largely insensitive to the specific form of the equilibrium microscopic dynamics, whether overdamped or underdamped, dissipative or conservative. This expectation is grounded in the fact that, in the supercooled regime, the structural relaxation of the fluid becomes largely independent of microscopic details \cite{gleim1998does,szamel2004independence}.

The starting point of our analysis is a diagrammatic theory developed previously by one of us \cite{szamel2007dynamics}. In the present work, we employ a slightly reformulated version of the diagrammatic rules of Ref. \cite{szamel2007dynamics} in order to make the subsequent analysis easier to present. Specifically, here we use a diagrammatic expansion for the intermediate scattering function $F(k;t)$ in terms of diagrams involving the bare propagator $F_0(k;t) = S(k)\exp[-D_0k^2t/S(k)]$, with $S(k)$ being the static structure factor. 
In contrast, in Ref. \cite{szamel2007dynamics} a diagrammatic theory for $G(k;t)=F(k;t)/S(k)$ was formulated in terms of diagrams involving $G_0(k;t)=\exp[-D_0k^2t/S(k)]$. This section summarizes the essential elements of the diagrammatic formalism. For a full account of the details of the derivation, we refer the reader to the original work. 

The central object of interest in the dynamical description of supercooled liquids presented here is the intermediate scattering function $F(k;t)$. This function satisfies a Dyson-like equation
\begin{widetext}
    \begin{equation}
    \begin{split}
        F(k;t) =&\ F_0(k;t)+ \int_0^{t}\mathrm{d}\tau_1\int_0^{\tau_1}\mathrm{d}\tau_2\int \frac{\mathrm{d}\boldsymbol{p}}{(2\pi)^d} F_0(k,t-\tau_1)\Sigma(\boldsymbol{k}, \boldsymbol{p}; \tau_1-\tau_2) F(p; \tau_2) 
    \end{split}
    \label{eq:dyson}
    \end{equation}
\end{widetext}
where $\Sigma(\boldsymbol{k}, \boldsymbol{k}'; t)$ denotes the self-energy. Translational invariance implies that the self-energy is diagonal in the space of wavevectors, $\Sigma(\boldsymbol{k}, \boldsymbol{k}'; t) \propto (2\pi)^d\delta(\mathbf{k}-\mathbf{k}')$. As is standard, the self-energy admits an infinite series representation which can be systematically organized and evaluated using diagrammatic techniques which we outline further below. It is important to note that the self-energy can be expressed in terms of a memory matrix $\boldsymbol{M}(\boldsymbol{k}, \boldsymbol{k}';t)$ as follows 
    \begin{equation}
        \Sigma(\boldsymbol{k}, \boldsymbol{k}'; t) = \frac{D_0}{S(k)} \boldsymbol{k}\cdot\boldsymbol{M}(\boldsymbol{k}, \boldsymbol{k}'; t) \cdot\boldsymbol{k}' \frac{1}{S(k')}.   \label{eq:decomposition_selfenergy_reducible_memory}
    \end{equation}
Naturally, the latter can be geometrically decomposed as 
    \begin{equation}
    \begin{split}
        \boldsymbol{M}(\boldsymbol{k}, \boldsymbol{k}'; t) \equiv&\ \left(\hat{\boldsymbol{k}}\otimes \hat{\boldsymbol{k}}'\right) M(k;t)(2\pi)^d\delta(\boldsymbol{k}-\boldsymbol{k}') \\
        &\ + (\mathcal{I} - \hat{\boldsymbol{k}}\otimes\hat{\boldsymbol{k}}') M_\perp(k;t)(2\pi)^d\delta(\boldsymbol{k}-\boldsymbol{k}')    
    \end{split}
    \label{eq:memory_matrix}
    \end{equation}
where $\mathcal{I}$ denotes the $d$-dimensional unit tensor. 
Eq. \eqref{eq:memory_matrix} defines a longitudinal memory function $M(k;t)$ and a transverse memory function $M_\perp(k;t)$. We note that while only the longitudinal function enters the Dyson equation \eqref{eq:dyson}, both functions have clear physical interpretation: the longitudinal memory kernel $M(k;t)$ is related to the bulk viscosity \cite{cichocki1987memory}, whereas the transverse contribution $M_\perp(k;t)$ pertains to the shear stresses \cite{vogel2020stress}.

\subsection{Diagrammatic Rules}
\begin{figure}
    \centering
    \includegraphics[width=\linewidth]{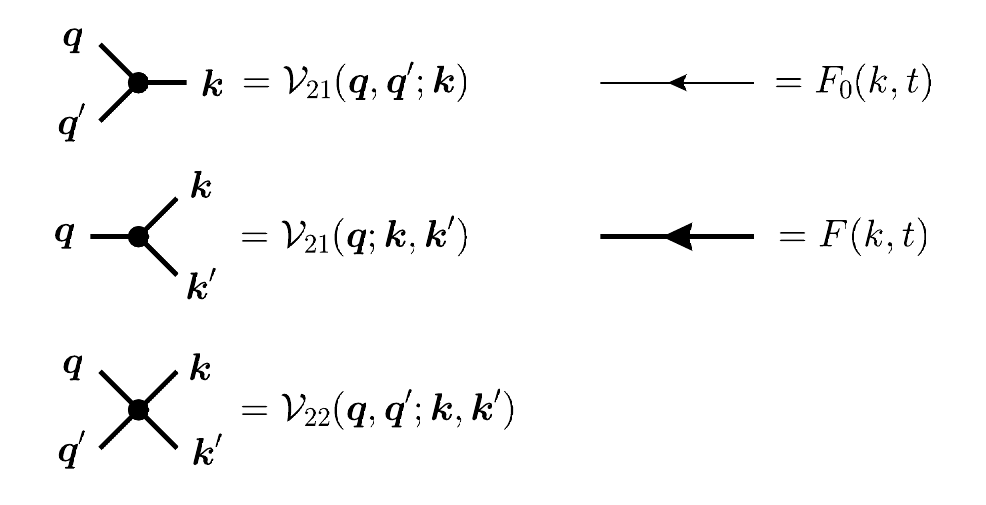}
    \vspace{-20pt}
    \caption{Diagrammatic rules of the kinetic theory to lowest order in a cluster expansion.}
    \label{fig:diagrammatic_rules}
\end{figure}

\begin{figure*}
    \centering
    \includegraphics[width=0.7\linewidth]{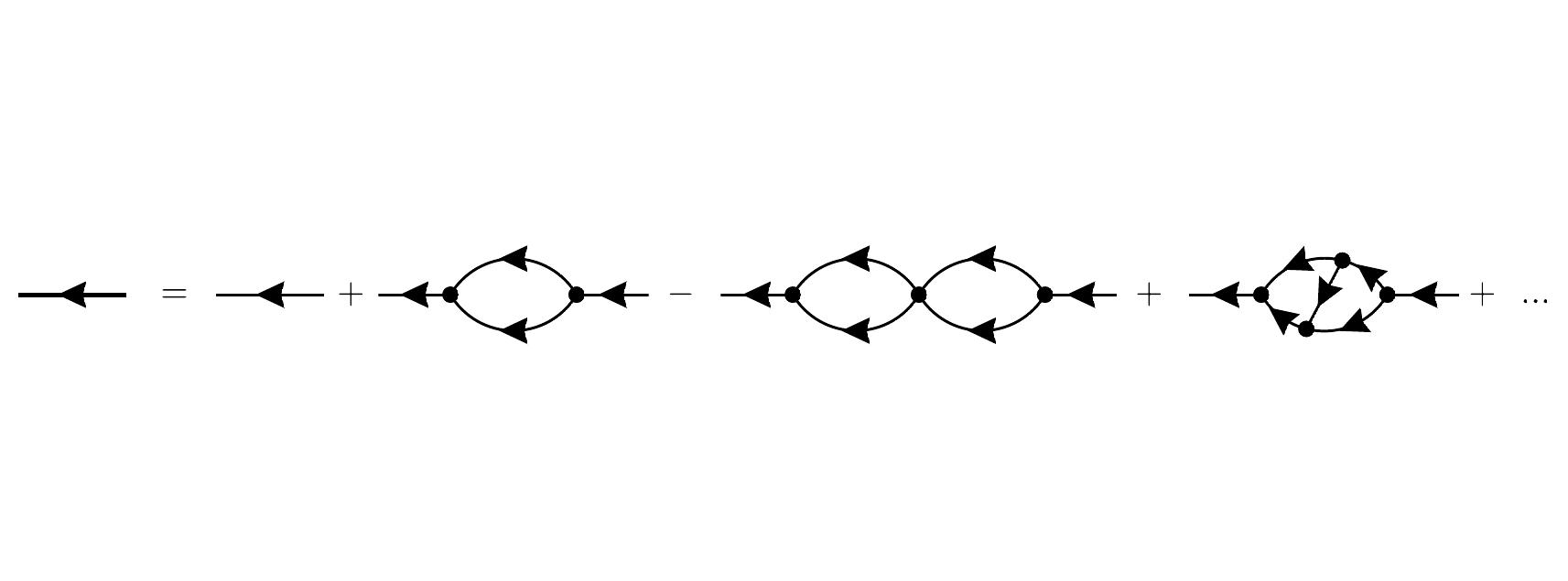}
    \vspace{-40pt}
    \caption{Diagrammatic expansion of the normalized correlation function $F(k;t)$ in terms of the bare propagator $F_0(k;t)$.}
    \label{fig:bare_PT}
\end{figure*}
The formally exact diagrammatic theory contains an infinite number of interaction vertices, but here we follow Ref. \cite{szamel2007dynamics} and retain only the lowest non-trivial ones in a cluster expansion. In this approximation three types of vertices emerge: two distinct cubic vertices referred to as left- and right-handed,  $\mathcal{V}_{21}$ and $\mathcal{V}_{12}$ (due to the causal nature of the diagrams) and one quartic vertex, $\mathcal{V}_{22}$, which can in fact be decomposed into a product of the cubic ones in a leading approximation. The corresponding diagrammatic rules are summarized in Fig.~\ref{fig:diagrammatic_rules} and all the vertices are formally defined below.

In the diagrammatic expansion, $F(k;t)$ is obtained as the sum over all topologically distinct diagrams that start and end with $F_0$ and include vertices $\mathcal{V}_{12}$, $\mathcal{V}_{21}$ and $\mathcal{V}_{22}$ connected by internal bonds $F_0$. At each vertex, one integrates over internal times and momenta, with momentum conservation and causality imposed. Diagrams with an odd number of $\mathcal{V}_{22}$ vertices contribute with an overall minus sign, while those with an even number contribute with a plus sign.  When writing the diagrams, time is taken to be flowing from the right to left as indicated by the arrow on the propagator. Propagators flowing backwards in time are naturally forbidden by causality of the theory. 

We show in Fig.~\ref{fig:bare_PT} the first few terms in the series expansion of the propagator $F(k;t)$. It is important to note that the series for $F(k;t)$ does not emerge from a perturbative expansion in any small parameter. Rather, it is more appropriate to think of it as a formal series representation whose physical content must be extracted through non-trivial resummations.

In turn, the self-energy $\Sigma(\boldsymbol{k}, \boldsymbol{k}'; t)$ also admits a diagrammatic representation: it is given by the sum over all topologically distinct, one-particle irreducible amputated diagrams (\textit{i.e.} those whose extremal legs have been removed). By virtue of the decomposition described by Eqs.~\eqref{eq:decomposition_selfenergy_reducible_memory}-\eqref{eq:memory_matrix}, diagrams composing the memory matrix $\boldsymbol{M}(\boldsymbol{k},\boldsymbol{k}';t)$ are the same as those composing the self-energy $\Sigma(\boldsymbol{k}, \boldsymbol{k}'; t)$ with the difference that the extremal vertices are given by ``roots" denoted $\mathbf{V}_{12}^{\mathrm{c}}$ (left-most) and $\mathbf{V}_{21}^{\mathrm{c}}$ (right-most). 

The vertex functions used in the diagrammatic expansion are defined explicitly as 
    \begin{equation}    
        \mathcal{V}_{21}(\boldsymbol{q}, \boldsymbol{q}' ; \boldsymbol{k})  = \frac{nD_0k^2}{S(k)} \tilde{v}_{\boldsymbol{k}}(\boldsymbol{q}, \boldsymbol{q}') (2\pi)^d\delta(\boldsymbol{q}+\boldsymbol{q}'-\boldsymbol{k})
    \label{eq:bare_left_handed_vertex}
    \end{equation}
and 
    \begin{equation}
        \mathcal{V}_{12}(\boldsymbol{q}; \boldsymbol{k}, \boldsymbol{k}') = \frac{D_0 q^2}{S(q)} \tilde{v}_{\boldsymbol{q}}(\boldsymbol{k}, \boldsymbol{k}') (2\pi)^d\delta(\boldsymbol{k}+\boldsymbol{k}'-\boldsymbol{q})
    \label{eq:bare_right_handed_vertex}
    \end{equation}
for the left-handed the right-handed cubic vertex, respectively. In Eqs.~\eqref{eq:bare_left_handed_vertex}-\eqref{eq:bare_right_handed_vertex} above, $\tilde{v}_{\boldsymbol{k}}(\boldsymbol{q}, \boldsymbol{q}') = k^{-1} v_{\boldsymbol{k}}(\boldsymbol{q}, \boldsymbol{q}')$, where $v_{\boldsymbol{k}}(\boldsymbol{q}, \boldsymbol{q}') = \hat{\boldsymbol{k}}\cdot \boldsymbol{v}(\boldsymbol{q}, \boldsymbol{q}')$ with $\boldsymbol{v}(\boldsymbol{q}, \boldsymbol{q}') = \boldsymbol{q}c(q) + \boldsymbol{q}'c(q')$. 
The quartic vertex can be written in terms of a left-handed and right-handed `roots'
    \begin{equation}
        \mathcal{V}_{22}(\boldsymbol{k}, \boldsymbol{k}'; \boldsymbol{q}, \boldsymbol{q}') = \int \frac{\mathrm{d}\boldsymbol{p}}{(2\pi)^d} \mathbf{V}^c_{21}(\boldsymbol{k}, \boldsymbol{k}' ; \boldsymbol{p}) \cdot \mathbf{V}^c_{12}(\boldsymbol{p} ; \boldsymbol{q}, \boldsymbol{q}')
    \label{eq:bare_quartic_vertex}
    \end{equation}
defined as
    \begin{equation}
        \mathbf{V}_{21}^{\mathrm{c}}(\boldsymbol{q}, \boldsymbol{q}' ; \boldsymbol{k}) = n (2\pi)^d\delta(\boldsymbol{q}+\boldsymbol{q}'-\boldsymbol{k}) \boldsymbol{v}(\boldsymbol{q}, \boldsymbol{q}') 
    \end{equation}
and
    \begin{equation}
        \mathbf{V}_{12}^{\mathrm{c}}(\boldsymbol{q}; \boldsymbol{k}, \boldsymbol{k}') = D_0 (2\pi)^d\delta(\boldsymbol{q} - \boldsymbol{k}-\boldsymbol{k}') \boldsymbol{v}(\boldsymbol{k}, \boldsymbol{k}'),
    \end{equation}
for the left-handed and right-handed roots, respectively. 

\subsection{Regularization of Spurious Divergences}\label{sec:regularization}

A key difference between the present diagrammatic theory and standard field-theoretic approaches \cite{andreanov2006dynamical, kim2008fluctuation} is the absence of explicit response functions, either bare $\partial_t F_0(k;t)$ or renormalized $\partial_t F(k;t)$, in the diagrammatic rules. This apparent absence complicates the analysis of long-time dynamics in certain instances. Suppose, for example, that one has identified a particular resummation of bare diagrams for the intermediate scattering functions that results in an approximate $F(k;t)$ that exhibits a long-lived plateau for some values of control parameters. It is then natural to examine corrections from additional classes of diagrams of diagrams perturbatively, by expressing them in terms of the approximate scattering function. However, when $F(k;t)$ remains approximately constant over an extended time window, the time integrals associated with these perturbative corrections typically give rise to spurious temporal divergences. Consider, for instance, the third diagram in Fig.~\ref{fig:bare_PT}, which, for the sake of argument, may be regarded as belonging to an additional class of diagrams. If there exists a long time interval over which $F(k;t)$ is approximately constant, then the contribution of this diagram will scale as $t^{n}$, where $n$ is an integer determined by the number of vertices. Consequently, evaluation at large times leads to unphysical divergences in the corresponding time integrals.

\begin{figure}
    \centering
    \includegraphics[width=\linewidth]{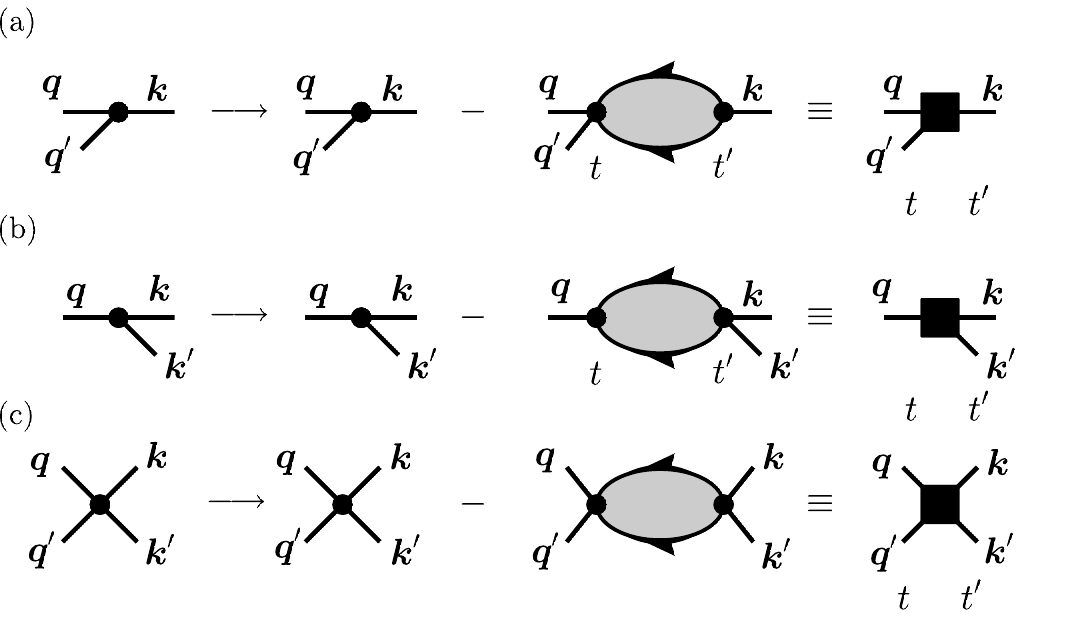}
    \vspace{-10pt}
    \caption{Illustration of the diagrammatic regularization procedure for the vertices of the diagrammatic theory. Regularization of the (a) $\mathcal{V}_{21}$, (b) $\mathcal{V}_{12}$ and (c) $\mathcal{V}_{22}$ vertices. In each, the grey bubble corresponds to the memory matrix defined in Eq.~\eqref{eq:memory_matrix}. The time labels refer to the time-slice of the renormalized vertices.}
    \label{fig:regularization_procedure}
\end{figure}

The resolution to these issues lies in reintroducing the response functions so as to precisely cancel the divergences outlined above. This can be achieved through a systematic targeted resummation of specific classes of diagrams. To this end, we introduce the longitudinal
    \begin{equation}
    R(k;t) = \left[\delta(t) - M(k;t)\right],
    \label{eq:resolvent}
    \end{equation}
and transverse resolvents
    \begin{equation}
        R_\perp(k;t) = \left[\delta(t) - M_\perp(k;t)\right],
    \end{equation}
that take care of the targeted resummation mentioned above and will enter explicitly in the definition of the regularized vertices. The regularization procedure for both three-legged vertices is outlined  diagrammatically in Fig.~\ref{fig:regularization_procedure}(a)-(b). In the perturbative treatment of the additional classes of diagrams, vertices\footnote{Note that one must be careful with over-counting certain diagrams when regularizing vertices. In such instances only a subset of the vertices must be replaced.} must be replaced by renormalized counterparts expressed in terms of the first, approximate resummation resummation. Specifically, the right- and left-handed vertices must be replaced by regularized counterparts given by 
\begin{widetext}
    \begin{equation}
    \begin{split}
        \mathcal{V}_{21}(\boldsymbol{q}, \boldsymbol{q}' ; \boldsymbol{k}) \longrightarrow\  \mathcal{V}_{21}(\boldsymbol{q}, \boldsymbol{q}' ; \boldsymbol{k})R(k; t-t') \equiv 
        \mathcal{v}_{21}^{\mathrm{reg}}(\boldsymbol{q}, \boldsymbol{q}' ; \boldsymbol{k} ; t-t')(2\pi)^d\delta(\boldsymbol{q}+\boldsymbol{q}'-\boldsymbol{k})
    \end{split}    
\label{eq:regularization_left_handed_vertex}
    \end{equation}
    and
    \begin{equation}
    \begin{split}
        \mathcal{V}_{12}(\boldsymbol{q}; \boldsymbol{k}, \boldsymbol{k}') \longrightarrow\  R(q; t-t')\mathcal{V}_{12}(\boldsymbol{q} ; \boldsymbol{k}, \boldsymbol{k}') \equiv \mathcal{v}_{12}^{\mathrm{reg}}(\boldsymbol{q}; \boldsymbol{k}, \boldsymbol{k}' ; t-t') (2\pi)^d\delta(\boldsymbol{k}+\boldsymbol{k}' - \boldsymbol{q}).
    \end{split}
\label{eq:regularization_right_handed_vertex}
    \end{equation}

Akin to the three-legged vertices, the four-vertex must also be regularized. To preserve the structure of the four-vertex $\mathcal{V}_{22}$, we perform a restricted resummation that retains the topological structure of the four-vertex. Specifically, we resum diagrams in which the bare four-vertex is renormalized by contributions from the full memory matrix $\boldsymbol{M}$, as illustrated diagrammatically in Fig.~\ref{fig:regularization_procedure}(c). The resummation is carried as follows 
    \begin{equation}
        \mathcal{V}_{22}(\boldsymbol{q}, \boldsymbol{q}' ; \boldsymbol{k}, \boldsymbol{k}') \longrightarrow \mathcal{v}_{22}^{\mathrm{reg}}(\boldsymbol{q}, \boldsymbol{q}' ; \boldsymbol{k}, \boldsymbol{k}' ; t-t') (2\pi)^d\delta(\boldsymbol{q}+\boldsymbol{q}'- \boldsymbol{k}-\boldsymbol{k}')
    \label{eq:regularization_four_vertex}
    \end{equation}    
where
    \begin{equation}
    \begin{split}
        \mathcal{v}_{22}^{\mathrm{reg}}(\boldsymbol{q},\boldsymbol{q}' ; \boldsymbol{k}, \boldsymbol{k}' ; t-t') =&\ nD_0
        v_{\boldsymbol{q}+\boldsymbol{q}'}(\boldsymbol{q},\boldsymbol{q}')R(|\boldsymbol{k}+\boldsymbol{k}'|, t-t')   v_{\boldsymbol{k}+\boldsymbol{k}'}(\boldsymbol{k},\boldsymbol{k}') \\
        &\ + nD_0 \boldsymbol{v}(\boldsymbol{q},\boldsymbol{q}')\cdot\left( \mathcal{I} - \frac{(\boldsymbol{q}+\boldsymbol{q}')}{|\boldsymbol{q}+\boldsymbol{q}'|} \otimes\frac{(\boldsymbol{k}+\boldsymbol{k}')}{|\boldsymbol{k}+\boldsymbol{k}'|} \right) 
        R_\perp(|\boldsymbol{k}+\boldsymbol{k}'|, t-t')
        \boldsymbol{v}(\boldsymbol{k},\boldsymbol{k}').
    \end{split}
    \label{eq:v22_reg_long}
    \end{equation}
In all three regularized vertices, the time indices $t\ge t'$ are associated with the extremal times of the vertex (and must therefore be appropriately integrated over in each diagram).  
\end{widetext}

\section{A Brief Account of the mean-Field Scenario}\label{sec:brief_mean_field_scenario}

Before proceeding, we briefly summarize the mean-field mode-coupling scenario for the dynamical glass transition, both to establish the notation used throughout and to define the reference solution around which the fluctuation expansion is organized. 
A more detailed discussion of the mean-field scenario is provided in Appendix \ref{sec:mean_field_scenario}.

\subsection{The Mode-Coupling Theory}
The most well known example of a resummation mentioned at the beginning of subsection \ref{sec:regularization} is provided by the mode-coupling theory \cite{leutheusser1984dynamical, bengtzelius1984dynamics, gotze2009complex}. 
Traditionally, in the context of systems evolving with Brownian dynamics this approximate
theory is introduced as follows \cite{szamel1991mode}. First, one notices that memory function $M(k;t)$ can be in a natural way expressed in terms of a simpler object, the so-called irreducible memory function, $M^\mathrm{irr}(k;t)$, through the following equation \cite{cichocki1987memory},
    \begin{equation}
    \begin{split}
        M(k;t) =&\ M^{\mathrm{irr}}(k;t)  - \int_0^t\mathrm{d}\tau M^{\mathrm{irr}}(k;t-\tau)M(k;\tau).      
    \end{split}
    \label{eq:relation_M_Mirr}
    \end{equation}  
It was argued in Ref. \cite{szamel2007dynamics} that the diagrammatic expansion for the irreducible memory function consists only of diagrams in the diagrammatic series for $M(k;t)$ that do not separate into disconnected components upon removal of a single
$\mathcal{V}_{22}$ vertex. Diagrammatically, the irreducible memory function $M^\mathrm{irr}(k;t)$ is therefore a simpler object than the memory function $M(k;t)$. Second, to arrive at the MCT one approximates the irreducible memory function in a self-consistent one-loop fashion \cite{szamel2007dynamics}, giving
\begin{widetext}
    \begin{equation}
    \begin{split}
        M_{\mathrm{MCT}}^{\mathrm{irr}}(k;t) = \frac{nD_0}{2}\int \frac{\mathrm{d}\boldsymbol{p}}{(2\pi)^d} v_{\boldsymbol{k}}(\boldsymbol{p}, \boldsymbol{k}-\boldsymbol{p}) F(p;t)F(|\boldsymbol{k}-\boldsymbol{p}|;t)v_{\boldsymbol{k}}(\boldsymbol{p}, \boldsymbol{k}-\boldsymbol{p}).   
    \end{split}
    \label{eq:Mirr_MCT}
    \end{equation}
In this approximation, it is then possible to show that the intermediate scattering function satisfies the following integro-differential equation
    \begin{equation}
        \frac{\partial F(k;t)}{\partial t} + \frac{D_0k^2}{S(k)} F(k;t) + \frac{n D_0}{2} \int_0^t\mathrm{d}\tau \int \frac{\mathrm{d}\boldsymbol{p}}{(2\pi)^d}\ v_{\boldsymbol{k}}(\boldsymbol{p}, \boldsymbol{k}-\boldsymbol{p})^2 F(p;t-\tau)F(|\boldsymbol{k}-\boldsymbol{p}|; t-\tau) \frac{\partial F(k; \tau)}{\partial \tau}=0.
    \label{eq:MCT}
    \end{equation}      
\begin{figure}
    \centering
    \includegraphics[width=0.75\linewidth]{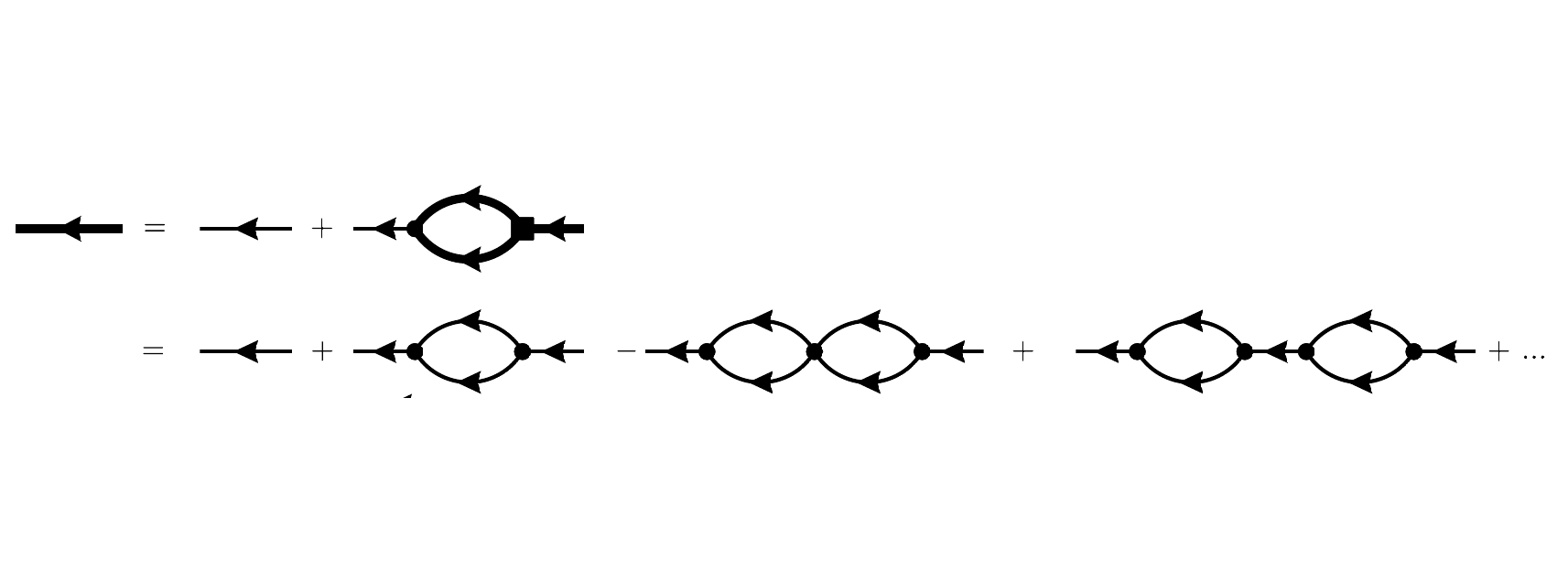}
    \vspace{-40pt}
    \caption{Diagrammatic form of the Dyson equation for the intermediate scattering function $F(k;t)$ using the mode-coupling approximation for the self-energy. The thick line represents the full propagator $F(k;t)$, and the thinner one denotes the bare propagator $F_0(k,t)$. The second row shows the first few bare diagrams that make up the MCT.}
\label{fig:MCT_contributions_propagator}
\end{figure}

\end{widetext}
Diagrammatically, this equation is equivalent to the resummation of the class of bare diagrams shown in Fig.~\ref{fig:MCT_contributions_propagator}.

Equation \eqref{eq:MCT} is the fundamental starting point of the celebrated mode-coupling theory (MCT) of the glass transition. While the physics encoded in Eq.~\eqref{eq:MCT} is by now well established\footnote{We refer the interested reader to Ref.~\cite{janssen2018mode} for a recent review, \citet{gotze2009complex} for an in-depth treatment and \ref{app:asymptotics_MCT} for a discussion of the asymptotic aspects of the theory.}, the present work builds upon and requires knowledge of the scenario that unfolds within MCT. We therefore recapitulate its principal results in the following section.

\subsection{Scaling Behavior of the Order Parameter}\label{sec:scaling_order_param}

Taking the long-time limit of Eq.~\eqref{eq:MCT} one gets a self-consistent equation for the Debye–Waller factor $F(k) = \lim_{t\to\infty} F(k;t)$. A nonzero solution of this equation indicates the emergence of a non-ergodic arrested phase which we refer to as a glass. The glass transition is thus characterized by the bifurcation of $F(k)$ from zero at a critical density $n_c$, marking the transition between liquid and dynamically arrested dynamics. The Debye–Waller factor therefore serves as the order parameter of the transition. 

In the supercooled regime, the intermediate scattering function $F(k;t)$ exhibits the hallmark two-step relaxation of glassy dynamics, decaying through an extended plateau corresponding to the Debye–Waller factor. This phenomenology is well described by MCT. To determine the scaling of the order parameter near the transition, we expand $F(k;t)$ around its critical plateau value $F_c(k)$
    \begin{equation}
        F(k;t) = F_c(k) + H(k;t;\varepsilon)
    \end{equation}
introducing a small deviation $H(k;t;\varepsilon)$ controlled by the distance $\varepsilon \equiv (n-n_c)/n_c$ from the critical point. Linearization of the MCT equations allows us to express $H(k;t ; \varepsilon)$ in terms of an eigenvalue problem involving a stability operator $C^{(1)}(\boldsymbol{k},\boldsymbol{p})$, whose largest eigenvalue $E_0(\varepsilon)$ approaches unity at criticality. 
The right eigenvector of the stability operator evaluated at the critical point defines the critical mode $h_0^{\mathrm{R}}(k)$. Expanding the deviations $H(k; t ; \varepsilon)$ in powers of the small parameter, one finds that the leading correction scales as $H(k;t ; \varepsilon) = \sqrt{|\varepsilon|} S(k)h_0^{\mathrm{R}}(k) g_\pm^{(1)}(\hat{t}) + \mathcal{O}(\varepsilon)$. This square-root dependence represents the generic scaling of the order parameter near the bifurcation. Here, $\hat{t} \equiv t / t_*(\varepsilon)$ is a rescaled time variable defined in terms of the diverging critical timescale $t_*(\varepsilon)$ that characterizes relaxation near the transition. The two branches $g_+^{(1)}$ and $g_-^{(1)}$ correspond, respectively, to the non-ergodic ($\varepsilon > 0$) and ergodic ($\varepsilon < 0$) sides of the transition.

The temporal part of the correction, encoded in the scaling function $g_\pm^{(1)}(\hat{t})$, obeys the so-called Götze $\beta$-scaling equations [see Eqs.~\eqref{eq:gotze_beta_scaling_non_ergodic} and \eqref{eq:gotze_beta_scaling}], which govern the slow dynamics in the vicinity of the plateau on either side of the transition. These functions describe the approach to and, in the ergodic phase, the departure from the plateau value of $F(k;t)$. Furthermore, the functions $g_\pm^{(1)}(\hat{t})$ display distinct dynamical regimes controlled by the exponent parameter $\lambda$, which can be computed directly from the equilibrium fluid structure. This parameter fixes the material dependent power-law exponents $a$ and $b$ describing the early $(\hat{t}\ll1)$ and late $(\hat{t}\gg1)$ $\beta$-relaxation regimes, and thus also fixes the two diverging timescales\textemdash{}$t_*(\varepsilon) \propto |\varepsilon|^{-1/2a}$ for the $\beta$-regime and $\tau_\alpha(\varepsilon) \propto |\varepsilon|^{-\gamma}$ with $\gamma = 1/2a+1/2b$ for the structural $\alpha$-relaxation\footnote{The $\alpha$-relaxation timescale is only well-defined in the ergodic phase, for $\varepsilon<0$.}\textemdash{}at the mode-coupling glass transition. 

\begin{figure}
    \centering
\includegraphics[width=0.9\linewidth]{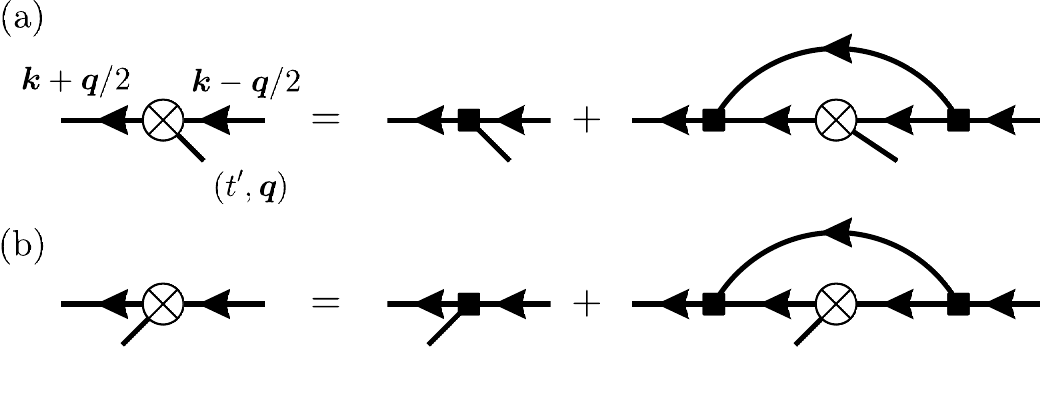}
    \vspace{-10pt}
    \caption{Diagrammatic representation of the resummation of rainbow diagrams yielding the dynamical susceptibilities $\chi_{\boldsymbol{q}}^+(\boldsymbol{k}; t ; t')$ [panel (a)] and $\chi_{\boldsymbol{q}}^-(\boldsymbol{k}; t ; t')$ [panel (b)]. Dangling ends at time $t'$ indicate points at which a propagator carrying momentum $\boldsymbol{q}$ can be attached; solid lines represent full propagators $F(k;t)$, and square vertices correspond to regularized mode-coupling vertices $\mathcal{v}^{\mathrm{reg}}_{12},\ \mathcal{v}^{\mathrm{reg}}_{21}$.}
    \label{fig:rainbow_insertions}
\end{figure}

\subsection{Scaling Behavior of Critical Fluctuations}\label{sec:scaling_critical_fluctuations}

As discussed above, the fluctuations of the order parameter are quantified by four-point functions, whose critical behaviour near the dynamical transition is governed by the underlying three-point susceptibilities. In the present framework we use the susceptibilities that were identified by one of us in a diagrammatic analysis of a four-point correlation function~\cite{szamel2008divergent}. These susceptibilities arise naturally from the resummation of specific class of diagrams: the so-called \textit{rainbow diagrams}. They are very closely related to the susceptibility that describes the change of $F(k;t)$ due to a static periodic potential that was introduced within inhomogeneous mode-coupling theory (IMCT) of Biroli \textit{et al.} \cite{biroli2006inhomogeneous}. In the following subsection, we summarize the asymptotic behaviour of three point susceptibilities used in our approach around the critical point.

The resummation of the rainbow diagrams, shown in Fig.~\ref{fig:rainbow_insertions}, naturally gives rise to two distinct three-point response functions, reflecting the presence of two inequivalent three-legged vertices in the theory [left- and right-handed, Eqs.~\eqref{eq:bare_left_handed_vertex}–\eqref{eq:bare_right_handed_vertex}]. We denote these by $\chi_{\boldsymbol{q}}^{\pm}(\boldsymbol{k};t;t')$, where the ``$+$" function corresponds to the right-handed vertex that injects momentum $\boldsymbol{q}$ into the propagator, and the ``$-$" function to the left-handed vertex that removes it. In this notation, $(\boldsymbol{k},t)$ labels the internal propagator, while $(\boldsymbol{q},t')$ denotes the external perturbation applied through the conjugate field. It is convenient to introduce the Fourier transforms with respect to the time at which the perturbation is applied, $t'$, 
    \begin{equation} \chi_{\boldsymbol{q}}^\pm(\boldsymbol{k}; t ; \omega) = \int_0^t \mathrm{d}t' e^{i\omega t'} \chi_{\boldsymbol{q}}^\pm(\boldsymbol{k}; t ; t').
    \label{eq:susceptibility_frequency_sppace}
    \end{equation}
With this definition, the two susceptibilities are related by a simple spatio-temporal transformation 
    \begin{equation}
        \chi^{-}_{\boldsymbol{q}}(\boldsymbol{k}; t ; \omega) = e^{i\omega t}\chi_{-\boldsymbol{q}}^{+}(\boldsymbol{k}; t ; -\omega).
    \end{equation}

Close to the mode-coupling transition and focusing on the leading contributions, the three-point susceptibilities obey a linear integral equation of the form (written here in Laplace and frequency space)
\begin{widetext}
    \begin{equation}
    \begin{split}
        \int \frac{\mathrm{d}\boldsymbol{p}}{(2\pi)^d} &\left[ (2\pi)^d\delta(\boldsymbol{k}-\boldsymbol{p}) - \mathcal{M}_{\boldsymbol{q}}(\boldsymbol{k}, \boldsymbol{p}) \right]\chi^{\pm}_{\boldsymbol{q}}(\boldsymbol{p}; z ; \omega) = \chi_{\boldsymbol{q}}^{(0\pm)}(\boldsymbol{k}; \hat{z} ; \omega)        
    \end{split}   
    \label{eq:dynamic_eigenvalue_susceptibility}
    \end{equation}    
\end{widetext}
with source terms $\chi_{\boldsymbol{q}}^{(0\pm)}(\boldsymbol{k}; \hat{z} ; \omega)$ that are regular in all arguments. Note that in the vicinity of the critical point, the source terms inherit the diverging timescale associated with the order parameter, which is why we have introduced the rescaled frequency variable $\hat{z}\equiv z t_*(\varepsilon)$ . The explicit expressions for the mass operator $\mathcal{M}_{\boldsymbol{q}}(\boldsymbol{k}, \boldsymbol{p})$ and the sources $\chi_{\boldsymbol{q}}^{(0\pm)}(\boldsymbol{k}; \hat{z} ; \omega)$ are given in Appendix \ref{app:critical_fluct_asymptotics}.

The solution to Eq.~\eqref{eq:dynamic_eigenvalue_susceptibility} is obtained by inversion of the mass operator which can be done perturbatively for long wavelengths ($q\ll1$) around the stability operator of the mode-coupling theory. This gives
    \begin{equation}
     \chi^{\pm}_{\boldsymbol{q}}(\boldsymbol{k}; z ; \omega) = \frac{1}{|\varepsilon|^{1/2}}\frac{\mathcal{B}_{\boldsymbol{q}}^{\pm}( \hat{z} ; \omega) S(k)h_0^{\mathrm{R}}(k)}{1 + (|\varepsilon|^{-1/4}\xi_0q)^2}   
     \label{eq:susceptibility_scaling}
    \end{equation}
where $\xi_0$ is a bare microscopic length determined by the liquid structure, and $\mathcal{B}_{\boldsymbol{q}}^{\pm}(\hat{z};\omega)$ is a scaling function\footnote{In the companion paper, we used $b_\pm$ to denote the the limit of the scaling function, $b_\pm \equiv \lim_{\hat{z},\omega\to0}\lim_{\boldsymbol{q}\to\boldsymbol{0}} \mathcal{B}^{\pm}_{\boldsymbol{q}}(\hat{z} ; \omega).$} that can be systematically expanded in powers of $q$.\footnote{To obtain this result, we have assumed that the responses $\chi_{\boldsymbol{q}}^\pm(\boldsymbol{k}; z ; \omega)$ depend only on the moduli $k$ and $q$ and not the angle between them. We cannot exclude that non-isotropic solutions also exist, as four-point functions have been shown to be anisotropic functions of wavevector arguments \cite{flenner2009anisotropic, berthier2011overview}.} From this form we identify the diverging dynamic correlation length $\xi_{\mathrm{d}} = \xi_0|\varepsilon|^{-1/4}$ that accompanies the dynamical arrest. More generally, one may define a frequency-dependent correlation length $\xi_{\mathrm{d}}(z)$\textemdash{}hence the term \textit{dynamic}\textemdash{}whose zero-frequency limit yields the expression above. A complete definition of $\xi_{\mathrm{d}}(z)$ would require a treatment of the frequency dependence of the mass operator, which lies beyond the scope of the present analysis. For a detailed discussion of this aspect, we refer the reader to earlier works by one of us on the time dependence of $\xi_{\mathrm{d}}(t)$ \cite{szamel2010diverging}. 

\begin{figure}[b]
    \centering   
    \includegraphics[width=\linewidth]{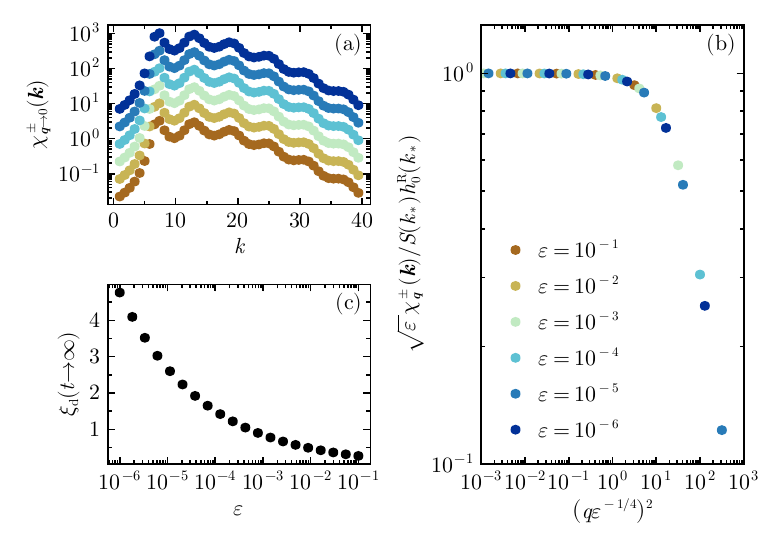}
    \caption{(a) Long wavelength ($q\to0$) profile of the susceptibilities $\chi^\pm_{\boldsymbol{q}\to0}(\boldsymbol{k})\equiv\lim_{\hat{z}\to0}\lim_{\omega\to0}\chi_{\boldsymbol{q}\to0}^\pm(k; \hat{z} ; \omega)$ for various relative distances $\varepsilon$ to the phase transition. (b) Finite-$q$ scaling of the susceptibility for various relative distances $\varepsilon$ to the phase transition. (c) Long-time limit of the correlation length $\xi_{\mathrm{d}}(t\to\infty) = \xi_0 |\varepsilon|^{-1/4}$. All panels show results for Percus-Yevick hard-spheres of unit diameter, for which all prefactors and constants are known.}
    \label{fig:susceptibility_scaling}
\end{figure}

To give the reader an impression of the typical behaviour of the susceptibilities, we show in Fig.~\ref{fig:susceptibility_scaling} the various scaling behaviours of the susceptibilities and associated correlation length in the non-ergodic phase for Percus-Yevick hard-spheres. 
The scaling function $\mathcal{B}^{\pm}_{\boldsymbol{q}}(\hat{t} ; \omega)$, in the limit $\omega\to0$, has been discussed in some detail in Refs.~\cite{biroli2006inhomogeneous, szamel2009three}. Briefly, much like the scaling function $g_{\pm}(t)$ of the propagator, the function $\mathcal{B}_{\boldsymbol{q}}^\pm(\hat{t} ; \omega)$ is also completely governed by the exponent parameter $\lambda$ and consequently by the dynamical exponents $a, b$ introduced in the previous section. It is possible to show that the scaling function of the integrated response follows a two-step temporal growth where, by virtue of the factorization theorem, we have
    \begin{equation}
    \label{eq:Xit1}
    \mathcal{B}^{\pm}_{\boldsymbol{q}}(\hat{t} ; \omega\to 0) 
    = \Xi\left((\xi_0 q)^2/\sqrt{|\varepsilon|}\right) \hat{t}^a \;\;\; \hat{t}\ll 1
    \end{equation}
    and
    \begin{equation}
    \label{eq:Xit2}
    \mathcal{B}^{\pm}_{\boldsymbol{q}}(\hat{t} ; \omega\to0) = \Xi\left((\xi_0 q)^2/\sqrt{|\varepsilon|}\right) \hat{t}^b \;\;\; \hat{t}\gg 1 .
    \end{equation}
Note that $\Xi(x)$ here is a regular function whose $x\gg1$ form is not known. The scaling behavior of the dynamical susceptibilities can, in principle, also be analyzed in the $\alpha$-regime. However, since these results are not directly relevant to the present discussion, we simply refer the reader to the detailed treatments of Refs.~\cite{biroli2006inhomogeneous, szamel2010diverging}. 

\subsection{Ginzburg Criterion}\label{sec:ginzburg_criterion}
\begin{figure}[t]
    \centering
    \includegraphics[width=0.75\linewidth]{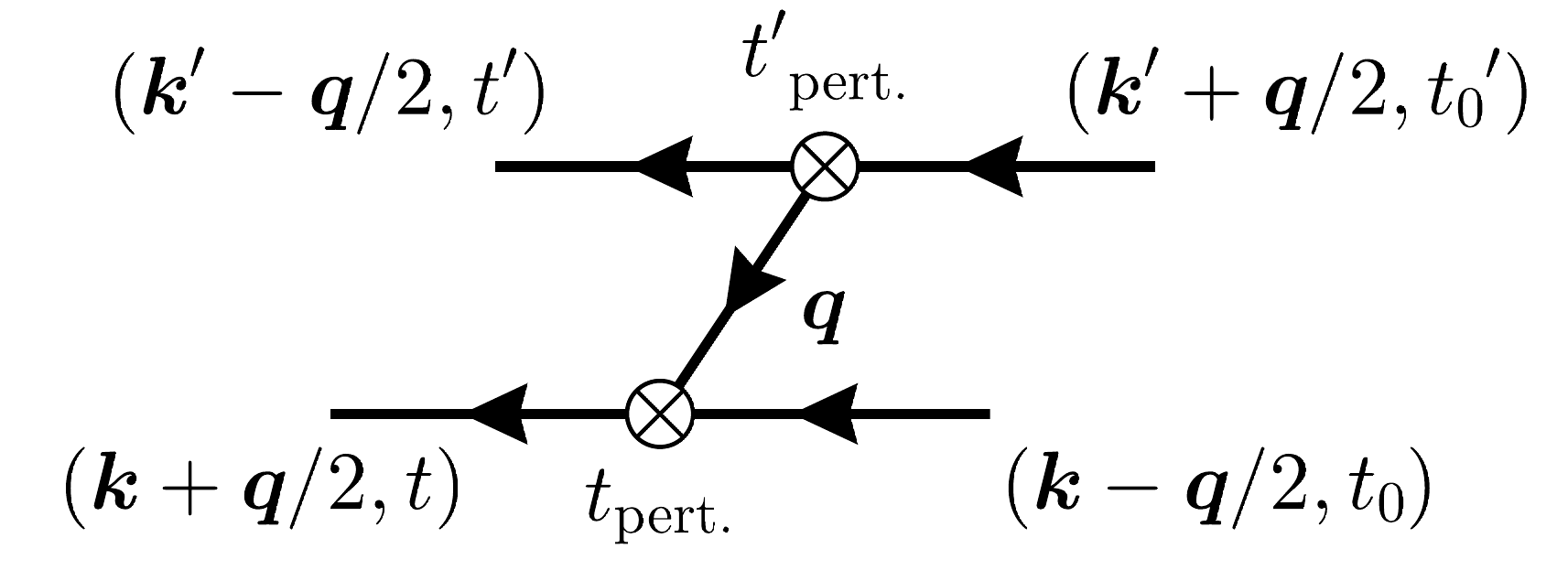}
    \caption{ Diagrammatic representation of Eq.~\eqref{eq:generic_susceptibility_product}. The external labels $(\boldsymbol{p},t)$ denote the incoming (or outgoing) momenta $\boldsymbol{p}$ at a given time $t$.
    }
\label{fig:general_junction_susceptibilities}
\end{figure}

Having established the critical behavior of three-point susceptibilities we can now use them to assess the consistency of mean-field theory via the Ginzburg criterion. We recall that \citet{goldenfeld1992lectures} defines the Ginzburg ratio as the ratio between the correlation function of the order parameter fluctuations and the order parameter squared, both integrated over the same correlation volume. To adapt this definition to the present case we consider the ratio of the four-point structure factor at a wavevector of the size of the inverse of the dynamic correlation length, $S_4(\boldsymbol{q}_*;\boldsymbol{k};t)$ at $|\boldsymbol{q}_*|\sim \xi_d^{-1}$, to the square of the change of the order paramater due to being $\varepsilon$ away from the transition point integrated over the volume of linear size $\xi_d$. Explicitly, we define the Ginzburg ratio $E_G$ as 
    \begin{equation}    
        E_G = \frac{S_4(\boldsymbol{q}_*;\boldsymbol{k};t)}{H(k;t;\varepsilon)^2(\xi_{\mathrm{d}})^d}.
    \label{eq:ginzburg_criteria}
    \end{equation}
Since we know from Ref.~\cite{szamel2008divergent} that 
    \begin{equation}
        S_4(\boldsymbol{q};\boldsymbol{k}; t) \propto \lim_{\omega\to 0} \left(\chi_{\boldsymbol{q}}^+(-\boldsymbol{k}; t ; -\omega)\chi_{\boldsymbol{q}}^-(\boldsymbol{k}; t ; \omega)\right),
    \end{equation}
we find that the numerator in Eq.~\eqref{eq:ginzburg_criteria} scales as $S_4(\boldsymbol{q}_*;\boldsymbol{k};t) \sim \varepsilon^{-1}$. Furthermore, recalling that $H(k;t;\varepsilon)\sim\varepsilon^{1/2}$ and that the correlation length diverges as $\xi_{\mathrm{d}}\sim\varepsilon^{-1/4}$, we find that the denominator scales as $\varepsilon^{1-d/4}$. Consequently, we find $E_G \sim |\varepsilon|^{(d - 8)/4}$ near the transition. This result shows that critical fluctuations dominate for $d < 8$. The Ginzburg criterion thus points to $d_c = 8$ as the upper critical dimension at which the mean-field description breaks down. This is in agreement with earlier field-theoretic considerations \cite{biroli2004diverging, biroli2007critical, franz2011field}.

This concludes our brief discussion of the mean-field scenario of the dynamical glass transition. While we have only discussed a mode-coupling-like treatment of critical fluctuations, we emphasize that a static mean-field approach using replicas leads to the same asymptotic description \cite{rizzo2013supercooled}.

\section{Leading Order Divergences Around the Mean-Field}\label{sec:leading_divergences_diagrammatic}

We aim to systematically compute critical fluctuation-dominated corrections to the mode-coupling approximation for the intermediate scattering function. Focusing on the vicinity of the mode-coupling transition, at small but finite distance $\varepsilon$ from criticality, we can resum these corrections and express them in terms of the mode-coupling propagators $F_{\mathrm{MCT}}(q; t)$ which are solutions to Eq.~\eqref{eq:MCT}. The diagrammatic rules of Sec.~\ref{sec:diagrammatic_rules} then provide the basis for constructing a perturbative expansion around the mode-coupling solution. Concretely, we write 
    \begin{equation}
        F(k;t) = F_{\mathrm{MCT}}(k;t) + \Delta F(k;t)
    \end{equation}
where $\Delta F(k;t)$ encodes the dominant (perturbative) corrections to the MCT solution, which we next describe.

The contributions to $\Delta F(k;t)$ arise from a class of diagrams in which each term represents a mode-coupling contribution whose propagators are systematically dressed by $n$ pairs of rainbow insertions (one $+$ type and one $-$ type). These insertions are connected via intermediate lines carrying the mode-coupling propagator $F_{\mathrm{MCT}}(q;t)$. As in any mean-field theory (and as discussed above), MCT already produces diverging susceptibilities near the transition, but these divergences do not couple back to the order parameter and thus leave the mean-field prediction unchanged. Dressing mode-coupling diagrams with rainbow insertions restores this coupling: after resummation of the insertions we get the diverging susceptibilities, which in this way re-enter the equations for the propagator, producing critical fluctuation-related corrections. In short, the resulting series perturbatively captures corrections beyond mean-field theory, as we consider non-mode-coupling diagrammatic contributions assembled from elements calculated in the mode-coupling approximation.

\begin{figure*}
    \centering
    \includegraphics[width=0.9\linewidth]{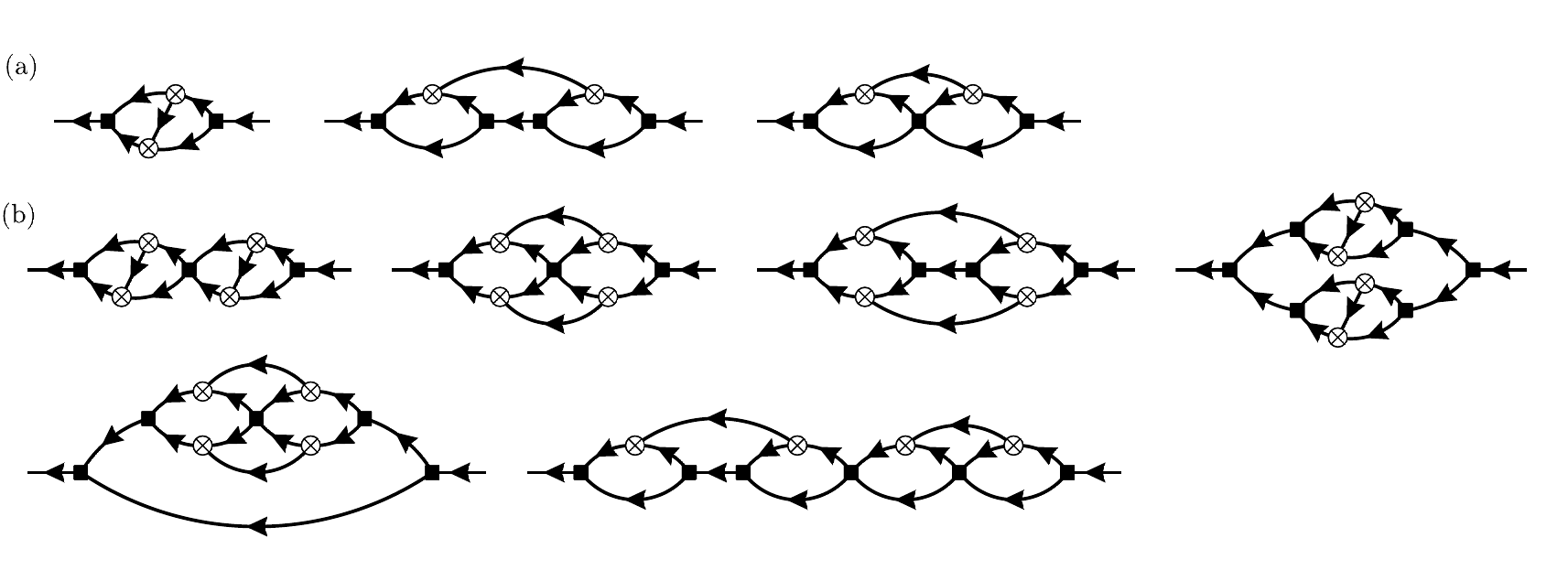}
    \vspace{-15pt}
    \caption{(a) Diagrammatic representations of the leading-order contributions involving a single pair of connected rainbow insertions. These capture the simplest topological structures relevant at this order. When evaluated near the mode-coupling transition, this class of diagrams scales as $\sim \varepsilon^{(d-4)/4}$. (b) A representative, though not exhaustive, set of diagrams featuring two pairs of rainbow insertions. Beyond the planar diagrams shown (i.e., diagrams that can be drawn on a plane without line crossings), the full set includes non-planar connections between susceptibility insertions as well as more complex topologies, all of which contribute at the same order. We recall that non-rainbow dressed vertices are regularized ones, given by Eqs.~\eqref{eq:regularization_left_handed_vertex},\ \eqref{eq:regularization_right_handed_vertex} and \eqref{eq:regularization_four_vertex}.}
    \label{fig:diagrammatic_most_diverging_2nd_4th_order}    
\end{figure*}

As an illustration, Fig.~\ref{fig:diagrammatic_most_diverging_2nd_4th_order}(a) shows the topologically relevant diagrams from this class containing a single pair of susceptibilities. In this way, the diagrammatic definition of susceptibility-dressed diagrams becomes explicit: the diagrams in Fig.~\ref{fig:diagrammatic_most_diverging_2nd_4th_order}(a) originate directly from the last three mode-coupling diagrams of the second line in Fig.~\ref{fig:MCT_contributions_propagator}, now written with renormalized propagators and vertices to account for all possible insertions. Similarly, Fig.~\ref{fig:diagrammatic_most_diverging_2nd_4th_order}(b) presents a selection of contributions with two pairs of insertions. 

\subsection{Asymptotic Behavior of Rainbow Insertions}

We now analyze the asymptotic scaling of the diagram class described above in the vicinity of the mode-coupling transition. Because every such diagram contains connected pairs of dynamical susceptibilities, any fluctuation-dominated contribution with $n$ pairs of susceptibilities necessarily involves $n$ factors of the form
    \begin{equation}
    \begin{split}
        \int \frac{\mathrm{d}\boldsymbol{q}}{(2\pi)^d}\int \frac{\mathrm{d}\omega}{2\pi} & e^{-i\omega(t_0-t_0')} \chi_{\boldsymbol{q}}^+(\boldsymbol{k}; t-t_0 ; -\omega) \\
        &\ \ \times F_{\mathrm{MCT}}(q;\omega) \chi_{\boldsymbol{q}}^-(\boldsymbol{k}'; t'-t_0' ; \omega)
    \end{split}    \label{eq:generic_susceptibility_product}
    \end{equation}
for some external wavevectors $\boldsymbol{k}\pm\boldsymbol{q}/2, \boldsymbol{k}'\mp\boldsymbol{q}/2$ and time slices $(t,t_0),\ (t',t_0')$ for the $+$ and $-$ susceptibilities respectively. The generic form of such contributions is shown diagrammatically in Fig.~\ref{fig:general_junction_susceptibilities}. Note that to obtain Eq.~\eqref{eq:generic_susceptibility_product}, we have written the connecting propagator in frequency space as
    \begin{equation}
        F_{\mathrm{MCT}}(q; t) = \int \frac{\mathrm{d}\omega}{2\pi} F_{\mathrm{MCT}}(q;\omega) e^{-i\omega t}
    \end{equation}
and have integrated over the internal time-slices ($t_{\mathrm{pert}}$ and $t_{\mathrm{pert}}'$ in the diagram of Fig.~\ref{fig:general_junction_susceptibilities}). The phase factors involving $\omega$ in expression \eqref{eq:generic_susceptibility_product} arise from the specific time-slices associated with the susceptibilities. These factors are, however, irrelevant for the scaling analysis. Near criticality, we expand around the mode-coupling solution and approximate $F_{\mathrm{MCT}}(q;\omega) = F_c(q)(2\pi)\delta(\omega) + \mathcal{O}(\sqrt{|\varepsilon}|)$. As a result, all such phase factors reduce to unity to leading order in $\varepsilon$.

Substituting this result in Eq.~\eqref{eq:generic_susceptibility_product} yields 
\begin{widetext}
    \begin{equation}
    \begin{split}
        \lim_{\omega\to0}\int \frac{\mathrm{d}\boldsymbol{q}}{(2\pi)^d}\chi_{\boldsymbol{q}}^+(\boldsymbol{k}; t-t_0 ; -\omega) F_{c}(q) \chi_{\boldsymbol{q}}^-(\boldsymbol{k}'; t'-t_0' ; \omega) =&\ |\varepsilon|^{(d-4)/4} \mathcal{m}_d(\boldsymbol{k}, \boldsymbol{k}' ; t-t_0, t'-t_0') + o(\varepsilon^{(d-4)/4})
    \end{split}   \label{eq:generic_susceptibility_product_result}
    \end{equation}
\end{widetext}
where to obtain this result we have used the low-$q$ scaling form Eq.~\eqref{eq:susceptibility_scaling} to identify the leading contributions in $\varepsilon$. In principle, the function $\mathcal{m}_d(\boldsymbol{k}, \boldsymbol{k}' ; t-t_0, t'-t_0')$ is finite and can be computed with knowledge of the large-$q$ behaviour of the susceptibilities [\textit{i.e.}, of the function $\Xi(x)$ implicitly defined in Eqs.~\eqref{eq:Xit1}-\eqref{eq:Xit2}]. Direct generalization gives that terms involving $n$ such insertions will scale as $\sim |\varepsilon|^{n(d-4)/4}$. We stress that the above analysis holds regardless of whether one considers the long-time limit or finite-time dynamics within the $\beta$-regime by virtue of the results of Sec.~\ref{sec:brief_mean_field_scenario}. To make the discussion of this section more concrete, we show below an explicit calculation for the computation of the simplest diagram with one pair of insertions. Throughout, we denote a particular diagram contribution with $A_i^{(n)}$, with $n$ corresponding to the number of rainbow insertions contained in the diagrams, and $i$ denotes the $i$-th contribution at this order. We use scripted letters $\mathcal{A}_i^{(n)}$ to refer to the leading (divergent) contribution to $A_i^{(n)}$ when evaluated near the mode-coupling transition.

\emph{Explicit Calculation}: We consider the first diagram shown in Fig.~\ref{fig:diagrammatic_most_diverging_2nd_4th_order}(a), which we denote by $A_1^{(2)}(k;t)$. The expression corresponding to the diagram is 
    \begin{widetext}
        \begin{equation}
        \begin{split}
            A_1^{(2)}(k;t)  =&\ \int_0^t\mathrm{d}\tau_4 \int_0^{\tau_4}\mathrm{d}\tau_3 \int_0^{\tau_3}\mathrm{d}\tau_2 \int_0^{\tau_2}\mathrm{d}\tau_1 \int \frac{\mathrm{d}\boldsymbol{q}}{(2\pi)^d}\frac{\mathrm{d}\boldsymbol{p}}{(2\pi)^d} F(k, t-\tau_4)\mathcal{v}_{12}^{\mathrm{reg}}(\boldsymbol{k} ; \boldsymbol{k}-\boldsymbol{p}+\boldsymbol{q}, \boldsymbol{p}-\boldsymbol{q} ; \tau_4-\tau_3) \\
            &\ \times \int_{\tau_2}^{\tau_3}\mathrm{d}t'\mathrm{d}t'' \chi_{\boldsymbol{q}}^{+}(\boldsymbol{k}-\boldsymbol{p}, \tau_3-\tau_2 ; t''-\tau_2) F(q, t''-t') \chi_{\boldsymbol{q}}^{-}(\boldsymbol{p}, \tau_3-\tau_2 ; t'-\tau_2) \mathcal{v}_{21}^{\mathrm{reg}}(\boldsymbol{p}, \boldsymbol{k}-\boldsymbol{p} ; \boldsymbol{k} ; \tau_2-\tau_1) F(k,\tau_1).
        \end{split}
        \label{eq:diagram1}
        \end{equation}
By writing the connecting propagator $F(q, t''-t')$ in frequency space, it is straightforward to show that Eq.~\eqref{eq:diagram1} can be rewritten as
    \begin{equation}
    \begin{split}
        ...\ \int\frac{\mathrm{d}\boldsymbol{q}}{(2\pi)^d}\int_{\tau_2}^{\tau_3}\mathrm{d}t'\mathrm{d}t'' &\chi_{\boldsymbol{q}}^{+}(\boldsymbol{k}-\boldsymbol{p}, \tau_3-\tau_2 ; t''-\tau_2) F(q, t''-t') \chi_{\boldsymbol{q}}^{-}(\boldsymbol{p}, \tau_3-\tau_2 ; t'-\tau_2)\ ...\\
        &= ...\ \int\frac{\mathrm{d}\boldsymbol{q}}{(2\pi)^d} \int \frac{\mathrm{d}\omega}{2\pi}\ \chi_{\boldsymbol{q}}^{+}(\boldsymbol{k}-\boldsymbol{p}, \tau_3-\tau_2 ; -\omega) F(q, \omega) \chi_{\boldsymbol{q}}^{-}(\boldsymbol{p}, \tau_3-\tau_2 ; \omega)\ ... \ 
    \end{split}
    \end{equation}
which agrees with the general form Eq.~\eqref{eq:generic_susceptibility_product} presented above.
\end{widetext}
Next, we evaluate $A_1^{(2)}(k;t)$ near the mode-coupling transition. This requires focusing on the $\beta$-regime, where dominant contributions arise from replacing $F(k;t) \to F_c(k)$ and from the small-$q$ contributions of the susceptibilities $\chi_{\boldsymbol{q}}^{\pm}(\boldsymbol{k} ; t)$, given by Eq.~\eqref{eq:susceptibility_scaling}. Writing $A_1^{(2)}(k;t)  = \mathcal{A}_1^{(2)}(k;\hat{t})\left[1+ \mathcal{O}(\sqrt{|\varepsilon|})\right]$, we find that the leading contribution reads
    \begin{equation}
        \mathcal{A}_1^{(2)}(k;\hat{t}) = a_1^{(2)}(k; \hat{t})S(k) |\varepsilon|^{(d-4)/4} + o(\varepsilon^{(d-4)/4})
    \label{eq:result_diagram_1}
    \end{equation}
where the amplitude $a_1^{(2)}(k; \hat{t})$ is regular in $k$ and is completely expressed in terms of (integrals over) the structure factor, the DW factor and the susceptibility scaling functions $\mathcal{B}_{\boldsymbol{q}}^\pm(\hat{t} ; \omega\to0)$. The leading contribution $\mathcal{A}_1^{(2)}(k;t) \sim |\varepsilon|^{(d-4)/4}$ is consistent with the generic discussion above and with previous results in $d = 3$ by \citet{szamel2013breakdown}. In this particular case, the function $\mathcal{m}_d$ of Eq.~\eqref{eq:generic_susceptibility_product_result} can be manipulated further and the total amplitude of this diagram is approximatively given by
\begin{widetext}
    \begin{equation}
        a_1^{(2)}(k;\hat{t}) \approx \mathcal{C}^{(2)}_d n_c S(k)(1-f_c(k))^2F_c(0)\left(\hat{t}^{x}\right)^2\int\frac{\mathrm{d}\boldsymbol{p}}{(2\pi)^d} \tilde{v}_{\boldsymbol{k}}(\boldsymbol{p}, \boldsymbol{k}-\boldsymbol{p})^2 S(p)h_0^{\mathrm{R}}(p)S(|\boldsymbol{k}-\boldsymbol{p}|)h_0^{\mathrm{R}}(|\boldsymbol{k}-\boldsymbol{p}|)   
    \end{equation}    
with $x = a,b$ depending on whether $\hat{t}\ll1$ or $\hat{t}\gg 1$ and the proportionality constant $\mathcal{C}^{(2)}_d$ is given by 
    \begin{equation}
        \mathcal{C}^{(2)}_d = \frac{1}{(\xi_0)^d}\frac{s_d}{(2\pi)^d}\int_0^{\infty} dy \frac{y^{d-1}\Xi(y)^2}{(1+y^2)^2}
    \end{equation}
which we expect to be finite on physical grounds.
\end{widetext}

This result extends naturally to the other diagrams with a single pair of rainbow insertions, though we omit their explicit evaluation here for brevity given the algebraic complexity. Summing the three diagrams shown in Fig.~\ref{fig:diagrammatic_most_diverging_2nd_4th_order}(a) yields the second–order source term, $A^{(2)}(k;t)$, whose evaluation near the mode-coupling transition gives $A^{(2)}(k;t) = \mathcal{A^{(2)}}(k;t)\left[1 + \mathcal{O}(\sqrt{|\varepsilon|}) \right]$ with
    \begin{equation}
        \mathcal{A}^{(2)}(k;t) = a^{(2)}(k;\hat{t})S(k)|\varepsilon|^{(d-4)/4} + o(\varepsilon^{(d-4)/4})
    \end{equation}
in which $a^{(2)}(k; \hat{t})$ is obtained from the sum of the amplitudes from all individual diagrams shown in Fig.~\ref{fig:diagrammatic_most_diverging_2nd_4th_order}(a). 

We can generalize the above reasoning to arbitrary even order $2n$. In this case, the contribution takes the form
    \begin{equation}
        A^{(2n)}(k;t) = \mathcal{A}^{(2n)}(k;\hat{t})\left[1 + \mathcal{O}(\sqrt{|\varepsilon|}) \right].
    \end{equation}
where the leading contribution scales as $\mathcal{A}^{(2n)}(k;\hat{t}) \sim a^{(2n)}(k; \hat{t})S(k)|\varepsilon|^{n(d-4)/4} + o(\varepsilon^{n(d-4)/4})$ with some generalized amplitudes $a^{(2n)}(k;\hat{t})$. Example diagrams at $n=2$ are shown in Fig.~\ref{fig:diagrammatic_most_diverging_2nd_4th_order}(b). Only even orders contribute, because all such corrections can be viewed as adding extra internal lines to mode-coupling contributions; the two vertices at the ends of each added line then renormalize into three-point susceptibilities. Structures that would correspond to odd orders therefore do not appear.

At this stage, one might be tempted to resum the entire family of $A^{(2n)}(k;t)$ diagrams into a closed expression. However, an important subtlety arises: these are not the most divergent contributions in the perturbative series. There exists another class of diagrams, denoted $F^{(2n)}(k;t)$, that can be constructed from the already divergent contributions $A^{(2n)}(k;t)$. As we discuss next, these more divergent contributions dominate the asymptotic behavior and satisfy their own self-consistent integral equation. 

\subsection{Identification of Most Dominant Diagrams}
\begin{figure}
    \centering
    \includegraphics[width=\linewidth]{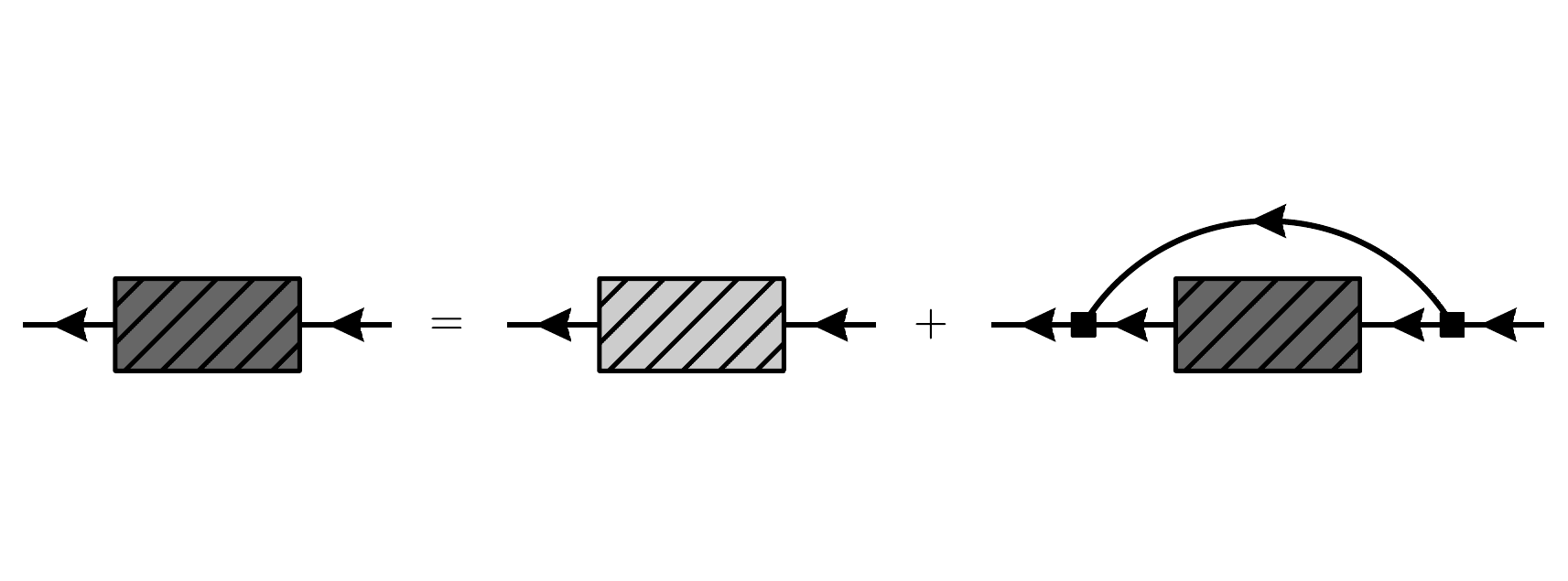}
    \vspace{-40pt}
    \caption{Diagrammatic representation of Eq.~\eqref{eq:overarching_rainbows_general} for the contributions $F^{(2n)}(k;t)$ (dark grey hatched) obtained by dressing $A^{(2n)}(k;t)$ (light grey hatched) with additional overarching rainbow insertions. Solid lines denote mode-coupling propagators and the light grey hatched block corresponds to the $A^{(2n)}$ insertion. This structure is formally analogous to the integral equation for the dynamical susceptibilities, with the overarching rainbows generating a self-consistent dressing of the $A^{(2n)}$ contributions.}
    \label{fig:generic_verarching_rainbow}
\end{figure}

The contributions $A^{(2n)}(k;t)$, while themselves divergent, can be dressed by additional overarching rainbow-like structures, generating the more divergent family of terms $F^{(2n)}(k;t)$ that contribute to the full correlation function. This construction is illustrated diagrammatically in Fig.~\ref{fig:generic_verarching_rainbow}, and is formally analogous to the resummation that yields the dynamical susceptibilities [compare with Fig.~\ref{fig:rainbow_insertions}]. The diagrammatic  relation between $A^{(2n)}(k;t)$ and $F^{(2n)}(k;t)$ contributions can be expressed by an integral equation of the form
\begin{widetext}
    \begin{equation}
    \begin{split}
        F^{(2n)}(k;t) = A^{(2n)}(k;t) + \int_0^t\mathrm{d}\tau_4...\int_0^{\tau_2}\mathrm{d}\tau_1 \int \frac{\mathrm{d}\boldsymbol{p}}{(2\pi)^d}& F(k,t-\tau_4)\mathcal{v}_{12}^{\mathrm{reg}}(\boldsymbol{k} ; \boldsymbol{p}, \boldsymbol{k}-\boldsymbol{p} ; \tau_4-\tau_3) F(|\boldsymbol{k}-\boldsymbol{p}|, \tau_3-\tau_2) \\
        &\ \times F^{(2n)}(p, \tau_3-\tau_2)\mathcal{v}_{21}^{\mathrm{reg}}(\boldsymbol{p}, \boldsymbol{k}-\boldsymbol{p} ; \boldsymbol{k} ; \tau_2-\tau_1)F(k, \tau_1).
    \end{split}
    \label{eq:overarching_rainbows_general}
    \end{equation}
This equation is formally analogous to the linear integral equation that governs the dynamical susceptibilities, discussed in Appendix~\ref{app:critical_fluct_asymptotics}. Close to the  mode-coupling transition, this allows us to re-express Eq. \eqref{eq:overarching_rainbows_general} in terms of the long-wavelength limit of the mass operator $\mathcal{M}_{\boldsymbol{0}}(\boldsymbol{k}, \boldsymbol{p})$ acting on $F^{(2n)}(k;t)$, just as is the case for the susceptibilities in Eq.~\eqref{eq:dynamic_eigenvalue_susceptibility}. Specifically, we get
    \begin{equation}
    \begin{split}
        \int \frac{\mathrm{d}\boldsymbol{p}}{(2\pi)^d} &\left[(2\pi)^d\delta(\boldsymbol{k}-\boldsymbol{p}) - \mathcal{M}_{\boldsymbol{0}}(\boldsymbol{k}, \boldsymbol{p}) \right] F^{(2n)}(p;t) = A^{(2n)}(k;t).        
    \end{split}
    \label{eq:overarching_rainbows_beta_regime}
    \end{equation}
Following the standard procedure to invert the mass operator around the mode-coupling transition reveals that we can write the solution as
    \begin{equation}
        F^{(2n)}(k;t) = f_0^{(2n)}(k;\hat{t})S(k)h_0^{\mathrm{R}}(k),
    \end{equation}
where to leading order in $\varepsilon$ we have
    \begin{equation}
        f_0^{(2n)}(k;\hat{t}) = \frac{1}{2g(1-\lambda)\sqrt{|\varepsilon|}} \int \frac{\mathrm{d}\boldsymbol{k}}{s_d} \left(\frac{h_0^{\mathrm{L}}(k)}{k^{d-1}} \right)\frac{\mathcal{A}^{(2n)}(k;\hat{t})}{S(k)}.
    \label{eq:solution_diagrammatic_overarching}
    \end{equation}
Here, the quantity $g$ is a positive constant that can be computed within MCT. We conclude that the additional rainbow dressing enhances the singular behavior of the contributions $A^{(2n)}(k; \hat{t})$ by an order $|\varepsilon|^{-1/2}$, leading to
    \begin{equation}
        F^{(2n)}(k;t) = f^{(2n)}(k; \hat{t})S(k) |\varepsilon|^{n(d-4)/4 - 1/2} + \mathcal{O}(\varepsilon^{n(d-4)/4})
    \end{equation}
\end{widetext}
in the $\beta$-regime, where $f^{(2n)}(k; \hat{t})$ is a renormalized amplitude that can be computed using the procedures outlined above. Summing over all such contributions, the fluctuation dominated corrections from rainbow-dressed diagrams yield an asymptotic series of the form 
    \begin{equation}
        \Delta F(k;t) = \sum_{n=1}^{\infty} f^{(2n)}(k; \hat{t})S(k) |\varepsilon|^{n(d-4)/4 - 1/2}
    \label{eq:asymptotic_series}
    \end{equation}
around the mode-coupling transition. 

The asymptotic result Eq.~\eqref{eq:asymptotic_series} shows that fluctuation corrections near the transition are organized as a hierarchy of terms whose singularity depends on spatial dimension. In fact, below four dimensions ($d < 4$), the asymptotic expansion breaks down as each successive term becomes increasingly singular. Importantly for our purposes, the scaling of the leading contributions provides a useful cross-check of the upper critical dimension determined from a Ginzburg criterion in Sec.~\ref{sec:ginzburg_criterion}. To determine the upper critical dimension here, we compare the leading fluctuation contributions $\Delta F(k)$ to the mean-field correction $H(k; t ; \varepsilon) \sim \sqrt{|\varepsilon|}$ near the transition \cite{gotze1985properties}. Inspection of Eq.~\eqref{eq:asymptotic_series} reveals that the fluctuation terms scale as $|\varepsilon|^{(d - 4)/4 - 1/2}$ and, therefore, dominate the mean-field contribution (which scales as $\sim \sqrt{|\varepsilon|}$) for all dimensions $d < 8$, identifying $d_c = 8$ as the upper critical dimension, in agreement with the result presented earlier.

The asymptotic series Eq.~\eqref{eq:asymptotic_series} constitutes the principal result of the first part of our work. The diagrams discussed here provide the most divergent corrections that can be systematically constructed from the dressing of mode-coupling contributions. This is why we consider them as natural diagrams that arise from critical fluctuations. 

We note, however, that other classes of diagrams, which are not directly related to the critical fluctuations of the order parameter, may also give rise to divergences at the mode-coupling transition. We do not believe that their existence affects the present conclusions. This point will be further discussed in Sec.~\ref{sec:conclusions}.

\section{Mapping to a Stochastic Process}\label{sec:mapping_stochastic_process}

Having identified and characterized the leading divergent corrections to MCT, the next step is to resum them. A direct resummation of the asymptotic series, Eq.~\eqref{eq:asymptotic_series}, is hindered by two factors: (i) in physically relevant dimensions there is no characteristic scale beyond which higher-order terms become parametrically subleading, and (ii) the amplitudes of the leading corrections are not available in closed form for generic wavevectors. These obstacles make a brute-force resummation unwieldy.

To circumvent these difficulties, we reformulate the problem in terms of an effective stochastic process whose perturbative solution reproduces the dominant divergent contributions identified above. This mapping provides a compact and physically transparent method of resumming fluctuation dominated contributions, and is in spirit closely related to approaches successfully applied to disordered critical phenomena \cite{parisi1979random, parisi1981critical} as well as to replica-based field-theoretic treatments of the glass transition \cite{franz2011field, franz2012quantitative}. In what follows, we construct this effective theory starting from the full microscopic dynamical problem.

\subsection{Postulated Stochastic Equation}

To proceed, we introduce two random auxiliary fields $u^\pm_{\boldsymbol{p}}(t)$ which inject ($+$) or remove ($-$) momentum $\boldsymbol{p}$ from the order parameter. In their presence, the order parameter is generalized to a two–wavevector, two–time object, $F_u(\boldsymbol{k}_+, \boldsymbol{k}_-; t,0)$, with an off-diagonal representation $\boldsymbol{k}_\pm = \boldsymbol{k} \pm \boldsymbol{q}/2$, that makes the net momentum transfer $\boldsymbol{q}$ from the external fields explicit. We then postulate that the time evolution of this two-point function is governed by a nonlinear integro-differential equation, structurally analogous to mode-coupling theory, but now coupled to the random external fields:
    \begin{widetext}
    \begin{equation}
    \begin{split}
        &\frac{\partial }{\partial t}F_u(\boldsymbol{k}_+, \boldsymbol{k}_-; t, 0) + D_0 k_+ \int \frac{\mathrm{d}\boldsymbol{p}}{(2\pi)^d}\int_0^t \mathrm{d}\tau R_u(\boldsymbol{k}_+, \boldsymbol{p} ; t, \tau) \frac{p}{S(p)} F_u(\boldsymbol{p}, \boldsymbol{k}_-  ; \tau, 0) = \mathcal{S}^{+}_u(\boldsymbol{k}_+,\boldsymbol{k}_- ; t, 0) + \mathcal{S}^{-}_u(\boldsymbol{k}_+,\boldsymbol{k}_- ; t, 0).
    \end{split}
    \label{eq:postulate_microscopic_stochastic}
    \end{equation}
The two source terms on the right-hand-side of Eq.~\eqref{eq:postulate_microscopic_stochastic} couple the random fields to the two-point functions via
    \begin{equation}
     \mathcal{S}_u^{+}(\boldsymbol{k}, \boldsymbol{k}' ; t) = D_0k \int \frac{\mathrm{d}\boldsymbol{p}}{(2\pi)^d}\frac{\mathrm{d}\boldsymbol{p}'}{(2\pi)^d}  \int_0^t\mathrm{d}\tau R_u(\boldsymbol{k}, \boldsymbol{p} ; t, \tau) v_{\boldsymbol{p}}(\boldsymbol{p}-\boldsymbol{p}', \boldsymbol{p}') u_{\boldsymbol{p}'}^+(\tau)F_u(\boldsymbol{p}-\boldsymbol{p}', \boldsymbol{k}'  ; \tau, 0),    \label{eq:pos_source}
    \end{equation}
and
     \begin{equation}
         \mathcal{S}_u^{-}(\boldsymbol{k}, \boldsymbol{k}' ; t) = nD_0 \int \frac{\mathrm{d}\boldsymbol{p}}{(2\pi)^d}\frac{\mathrm{d}\boldsymbol{p}'}{(2\pi)^d} \int_0^t\mathrm{d}\tau S(k) u_{\boldsymbol{p}}^-(t) v_{\boldsymbol{k}+\mathbf{p}}(\boldsymbol{p},\boldsymbol{k})R_u(\boldsymbol{k}+\boldsymbol{p}, \boldsymbol{p}' ; t,\tau) \frac{p'}{S(p')} F_u(\boldsymbol{p}', \boldsymbol{k}'  ; \tau, 0).
     \label{eq:neg_source}
     \end{equation}
The time evolution of the sources and the two-point function is mediated by an inhomogeneous resolvent operator
    \begin{equation}
    \begin{split}
    R_u&(\boldsymbol{k}, \boldsymbol{k}'; t, t')\equiv \left[(2\pi)^d\delta(\boldsymbol{k}-\boldsymbol{k}')\delta(t-t') + M_u^{\mathrm{irr}}(\boldsymbol{k}, \boldsymbol{k}'; t, t') \right]^{-1},
    \end{split}
    \end{equation}
with the inhomogeneous irreducible memory kernel
    \begin{equation}
    \begin{split}
        &M_u^{\mathrm{irr}}(\boldsymbol{k}, \boldsymbol{k}' ; t, t')
        = \frac{nD_0}{2}\int \frac{\mathrm{d}\boldsymbol{p}}{(2\pi)^d} \frac{\mathrm{d}\boldsymbol{p}'}{(2\pi)^d} v_{\boldsymbol{k}}(\boldsymbol{p}, \boldsymbol{k}-\boldsymbol{p}) F_u(\boldsymbol{p}, \boldsymbol{p}' ; t, t')F_u(\boldsymbol{k}-\boldsymbol{p}, \boldsymbol{k}'-\boldsymbol{p}' ; t, t') v_{\boldsymbol{k}'}(\boldsymbol{p}', \boldsymbol{k}'-\boldsymbol{p}')
    \end{split}
    \label{eq:inhomogeneous_Mirr}
    \end{equation}
\end{widetext}
taken in a mode-coupling approximation. Denoting $\llbracket ... \rrbracket$ the disorder average, the random fields are taken to be colored Gaussian processes with zero mean $\llbracket u_{\boldsymbol{q}}^{\pm}(t) \rrbracket = 0$ and variances satisfying
    \begin{equation}
        \llbracket u_{\boldsymbol{q}}^+(t) u_{\boldsymbol{q}'}^-(t') \rrbracket = F_{\mathrm{MCT}}(q, t-t')(2\pi)^d\delta(\boldsymbol{q}-\boldsymbol{q}')\Theta(t-t')
    \label{eq:noise_noise_correlator}
    \end{equation}
and $\llbracket u_{\boldsymbol{q}}^+(t) u_{\boldsymbol{q}'}^+(t')\rrbracket = \llbracket u_{\boldsymbol{q}}^-(t) u_{\boldsymbol{q}'}^-(t')\rrbracket = 0$. These last two conditions along with the Heaviside function $\Theta(x)$ in Eq.~\eqref{eq:noise_noise_correlator} are necessary to enforce causal responses after disorder averaging. The physical correlation function is then obtained by considering the disordered averaged order-parameter 
    \begin{equation}
        F(k;t) = \llbracket F_u(\boldsymbol{k}_+, \boldsymbol{k}_- ; t, 0) \rrbracket.
    \end{equation}

While the statistics of the random fields Eq.~\eqref{eq:noise_noise_correlator} suggest an analogy of $u^\pm_{\boldsymbol{q}}(t)$ with density fluctuations $\delta\varrho(\boldsymbol{q},t) = \sum_i \exp[i\boldsymbol{q}\cdot\boldsymbol{r}_i]$, they cannot be identified with them. To see this, decompose $u^\pm_{\boldsymbol{q}}(t) = a_{\boldsymbol{q}}(t) \pm ib_{\boldsymbol{q}}(t)$ where the required statistics imply that $a_{\boldsymbol{q}}(t)$ and $b_{\boldsymbol{q}}(t)$ are independent Gaussian processes of zero mean and variance $\llbracket a_{\boldsymbol{q}}(t) a_{\boldsymbol{q}'}(t') \rrbracket = \llbracket b_{\boldsymbol{q}}(t) b_{\boldsymbol{q}'}(t') \rrbracket = \frac{1}{2}\times \llbracket u_{\boldsymbol{q}}^-(t) u_{\boldsymbol{q}'}^+(t') \rrbracket$. This specific decomposition is not obeyed by density fluctuations $\delta\varrho(\boldsymbol{q},t)$ under (canonical) statistical averaging. The random fields $u^\pm_{\boldsymbol{q}}(t)$ are therefore best understood as a purely mathematical construction whose properties are dictated by physical features of the liquid. 

\subsection{Generalized Dynamical Susceptibilities}
We now introduce the notation and definitions needed for the perturbative solution of Eqs.~\eqref{eq:postulate_microscopic_stochastic}–\eqref{eq:inhomogeneous_Mirr}. Our goal is to expand the stochastic order parameter $F_u(\boldsymbol{k}_+, \boldsymbol{k}_-; t, 0)$, which evolves under the full microscopic dynamics in the presence of stochastic fields $u^{\pm}$, as a series in powers of these fields. Throughout, we use $\boldsymbol{q}$ to denote the total momentum transferred to the propagator over the time interval $(0,t)$. We find that the zeroth-order term corresponds to the unperturbed, spatially homogeneous two-point function $F_{\mathrm{MCT}}(k;t)$ of MCT, while higher-order terms capture the linear and nonlinear responses to the perturbations:
\begin{widetext}
    \begin{equation}
    \begin{split}
        F_u(\boldsymbol{k}_+, \boldsymbol{k}_-; t, 0) =&\ F_{\mathrm{MCT}}(k;t) (2\pi)^d\delta(\boldsymbol{q}) + F_u^{(1)}(\boldsymbol{k}_+, \boldsymbol{k}_- ; t, 0) + F_u^{(2)}(\boldsymbol{k}_+, \boldsymbol{k}_-; t, 0) + \ldots      
    \end{split}
    \label{eq:expansion_form_Fu}
    \end{equation}
At first order, the correction can be written in a linear-response–like form. Specifically, we introduce response functions $\mathcal{X}^\pm$ such that
    \begin{equation}
        F_u^{(1)}(\boldsymbol{k}_+, \boldsymbol{k}_- ; t, t') \equiv \sum_{\nu=\pm}\int \mathrm{d}t_1 \int \frac{\mathrm{d}\boldsymbol{q}_1}{(2\pi)^d}  \mathcal{X}_{\boldsymbol{q}_1}^{(\nu)}(\boldsymbol{k}; t, t' ;t_1)u_{\boldsymbol{q}_1}^{(\nu)}(t_1) \mathds{1}_{(t,t')}(t_1) (2\pi)^d\delta(\boldsymbol{q}_1-\boldsymbol{q}).
    \label{eq:Fu1_susceptibility}
    \end{equation}
Here, the indicator function $\mathds{1}_{(t,t')}(t_{\mathrm{pert}}) \equiv \Theta(t-t_{\mathrm{pert}})\Theta(t_{\mathrm{pert}}-t')$ where the Heaviside functions $\Theta(t)$ ensures that the system only responds to causal perturbations [\textit{i.e.}\ those that occur in the time interval $(t,t')$]. The functions $\mathcal{X}^{\pm}$ are formally defined through functional differentiation of the order parameter 
    \begin{equation}
        \frac{\delta F_u(\boldsymbol{k}_+, \boldsymbol{k}_- ; t, t')}{\delta u^{(\pm)}_{\boldsymbol{q}_{\mathrm{pert}}}(t_{\mathrm{pert}})}\Bigg\vert_{u^{(\pm)}=0} = \mathcal{X}_{\boldsymbol{q}_{\mathrm{pert}}}^{(\pm)}(\boldsymbol{k}; t, t' ; t_{\mathrm{pert}}) \mathds{1}_{(t,t')}(t_{\mathrm{pert}}) \delta(\boldsymbol{q}_{\mathrm{pert}}-\boldsymbol{q}).
    \label{eq:variational_def_X1}
    \end{equation}
In the limit $u^{(\pm)}_{\boldsymbol{q}_{\mathrm{pert}}}(t_{\mathrm{pert}})\to0$, time- and space-translation invariance are restored, restricting the time dependences of the response functions to $(t-t')$ and $(t_{\mathrm{pert}}-t')$ and ensuring momentum conservation in Eq.~\eqref{eq:variational_def_X1}: $\mathcal{X}_{\boldsymbol{q}_{\mathrm{pert}}}^{(\pm)}(\boldsymbol{k}; t, t' ; t_{\mathrm{pert}}) \to \mathcal{X}_{\boldsymbol{q}_{\mathrm{pert}}}^{(\pm)}(\boldsymbol{k}; t- t' ; t_{\mathrm{pert}}-t')$. We will show later that the responses $\mathcal{X}^{(\pm)}_{\boldsymbol{q}}$ introduced here are closely related to the microscopic susceptibilities $\chi^{\pm}_{\boldsymbol{q}}$ introduced earlier.

The generalization to higher-order response functions follows in complete analogy, with causality always enforced by the explicit time ordering encoded in the Heaviside functions. To make the structure transparent, we represent these generalized response functions diagrammatically. At order $n$, the contribution takes the form
    \begin{equation}
    \begin{split}
        F^{(n)}_u(\boldsymbol{k}_+, \boldsymbol{k}_- ; t, 0)  =&\raisebox{-1.33ex}{\includegraphics[height=10ex,page=1]{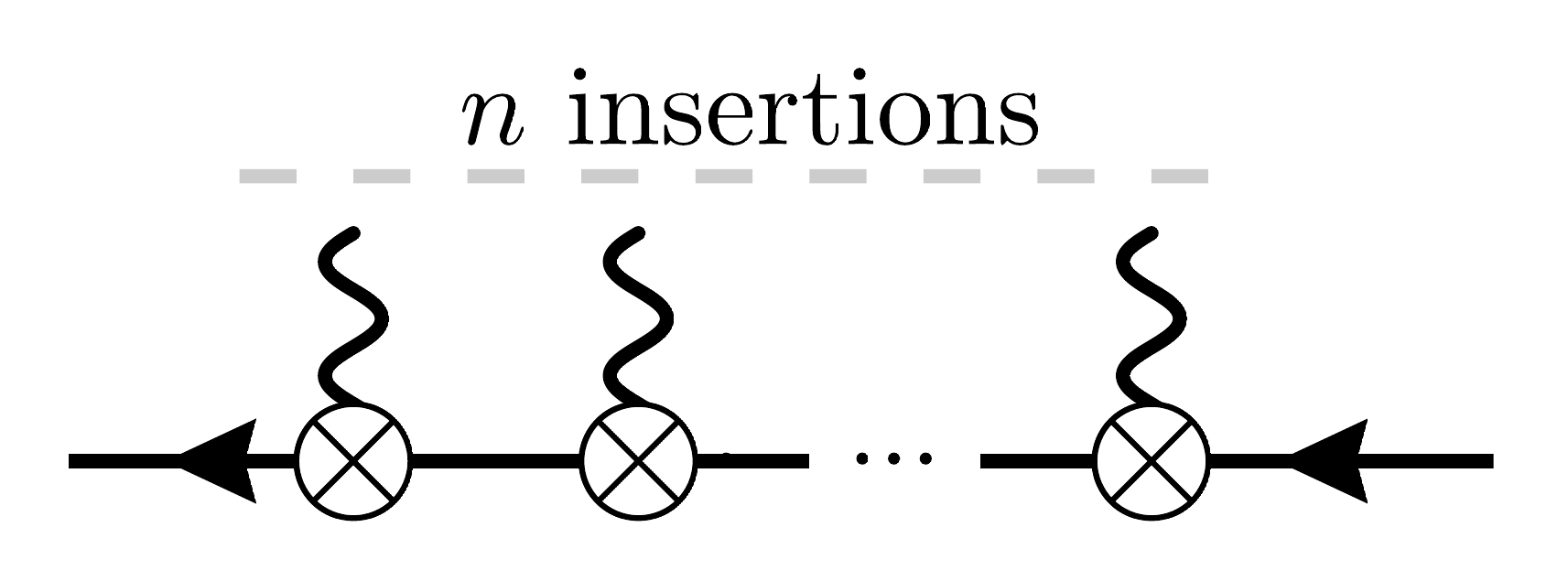}}
    \end{split}
    \end{equation}
where the wavy line \raisebox{-0.25ex}{\includegraphics[height=2ex,page=1]{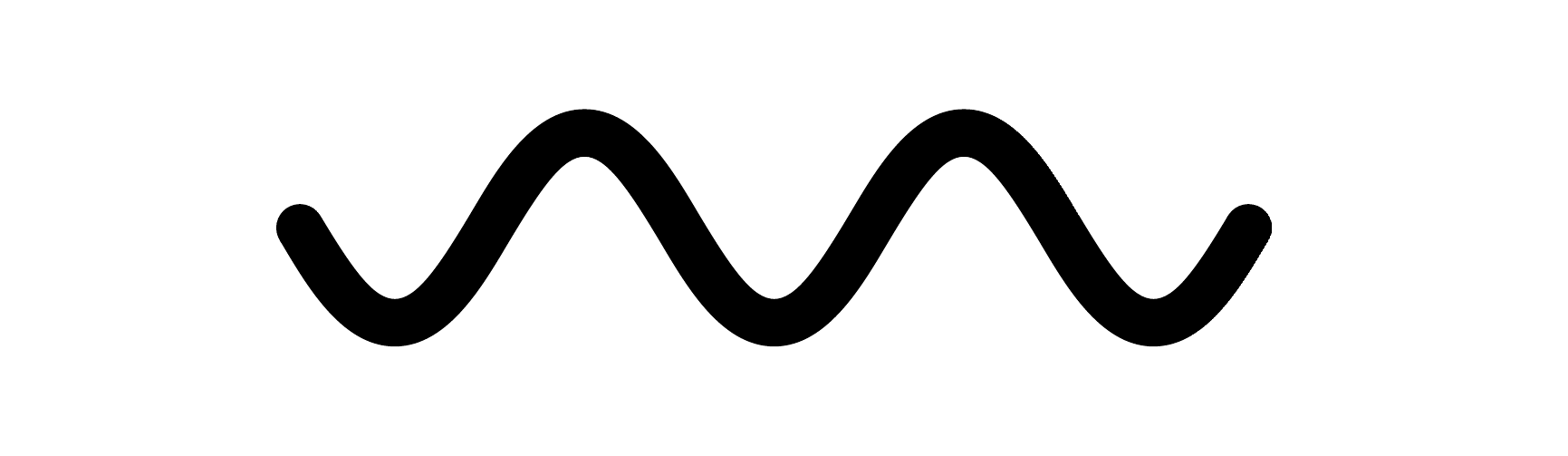}} indicates a summation over the $\pm$ insertions at that time point. We also importantly note that the initial condition reads
    \begin{equation}
        F^{(n)}_u(\boldsymbol{k}_+, \boldsymbol{k}_- ; 0, 0) = 0
   \label{eq:generalized_susceptibility_init}   
    \end{equation}
for $n \geq 1$.
\end{widetext}

\subsection{Perturbative Analysis of the Stochastic Process}

\begin{figure*}
    \centering
    \includegraphics[width=0.7\linewidth]{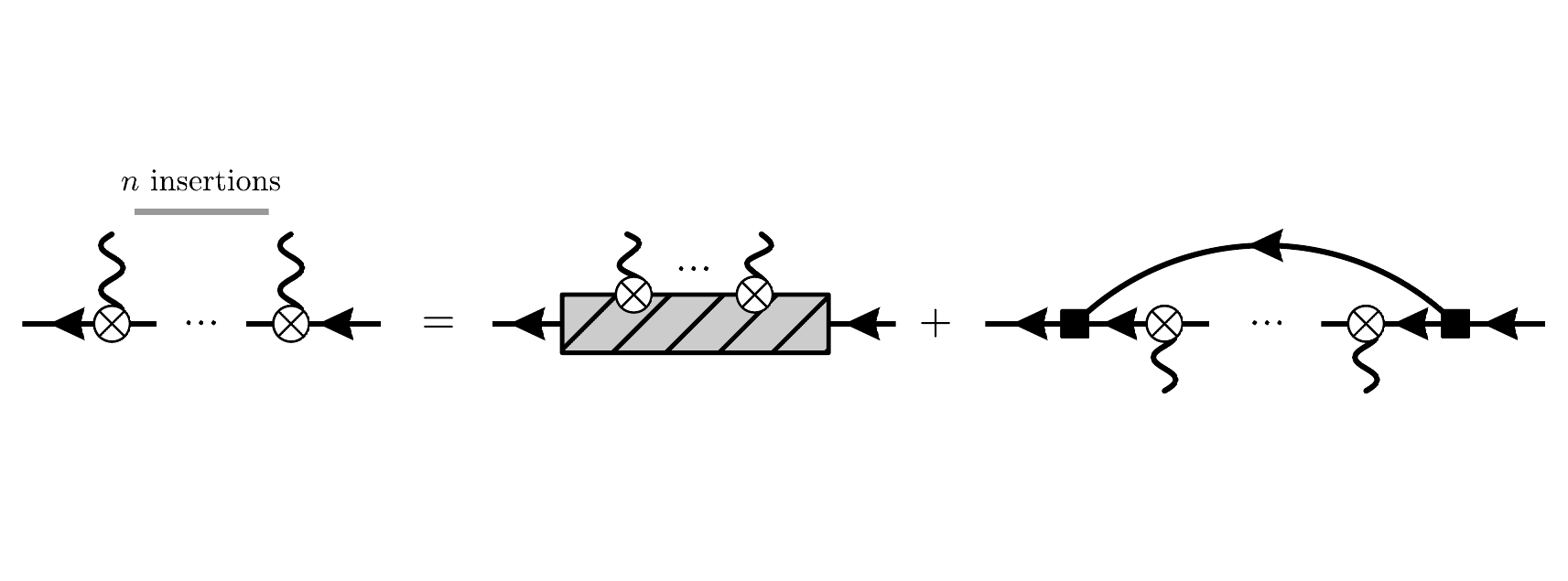}
        \vspace{-40pt}
\caption{Diagrammatic structure of Eq.~\eqref{eq:overarching_rainbows_general_stochastic}. The grey hatched box corresponds to the effective source terms $A_u^{(n)}(\boldsymbol{k}_+, \boldsymbol{k}_- ; t, 0)$.}
    \label{fig:overarching_rainbows_general_stochastic}
\end{figure*}

We now turn to a systematic perturbative analysis of Eqs.~\eqref{eq:postulate_microscopic_stochastic}–\eqref{eq:inhomogeneous_Mirr}, which describe the evolution of the order parameter under the influence of a stochastic field. Specifically, our goal is to expand the two-point function $F_u(\boldsymbol{k}_+, \boldsymbol{k}_-; t, 0)$ order by order in the random fields $u_{\boldsymbol{p}}^{\pm}(t)$. As mentioned above, the zeroth order solution recovers the MCT solution. The higher-order terms describe the response of the system to the stochastic perturbations and, as we will show, can be matched with the dominant diagrammatic contributions identified in Sec.~\ref{sec:leading_divergences_diagrammatic}.

We examine the generic structure of the perturbative expansion in the fields $u_{\boldsymbol{q}}^\pm(t)$. Substituting Eq.~\eqref{eq:expansion_form_Fu} in Eq.~\eqref{eq:postulate_microscopic_stochastic}, the $n$-th order contribution will satisfy 
    \begin{widetext}
        \begin{equation}
        \begin{split}
            \frac{\partial }{\partial t}F_u^{(n)}(\boldsymbol{k}_+, \boldsymbol{k}_-; t, 0) &+ D_0 k_+ \int \frac{\mathrm{d}\boldsymbol{p}}{(2\pi)^d} \int_0^t \mathrm{d}\tau\sum_{m=0}^n \begin{pmatrix}
                n \\ m
            \end{pmatrix} R_u^{(m)}(\boldsymbol{k}_+, \boldsymbol{p} ; t, \tau) \frac{p}{S(p)} F_u^{(n-m)}(\boldsymbol{p}, \boldsymbol{k}_-  ; \tau, 0)\\
            &= \mathcal{S}^{+,(n)}_u(\boldsymbol{k}_+,\boldsymbol{k}_- ; t, 0) + \mathcal{S}^{-,(n)}_u(\boldsymbol{k}_+,\boldsymbol{k}_- ; t, 0)  
        \end{split}
        \label{eq:postulate_microscopic_stochastic_expansion}
        \end{equation}
where $R_u^{(n)}$, $\mathcal{S}_u^{\pm,(n)}$ denote the contribution to the resolvent and the sources at order $n$ in the random fields respectively. It is convenient to single out the $m=0$ contribution, and by using Dyson's equation [Eq.~\eqref{eq:dyson}], the result Eq.~\eqref{eq:postulate_microscopic_stochastic_expansion} can be rewritten in an integral form (see Appendix~\ref{app:derivation_stochastic_integral_eq}):
    \begin{equation}
    \begin{split}
        F_u^{(n)}(\boldsymbol{k}_+, \boldsymbol{k}_-; t, 0) =&\ \int_0^t\mathrm{d}\tau  \frac{F_{\mathrm{MCT}}(k_+, t-\tau)}{S(k_+)}\left(\mathcal{S}^{+,(n)}_u(\boldsymbol{k}_+,\boldsymbol{k}_- ; \tau, 0) + \mathcal{S}^{-,(n)}_u(\boldsymbol{k}_+,\boldsymbol{k}_- ; \tau, 0)\right)\\
        &-\ \int_0^t \mathrm{d}\tau\int_0^{\tau} \mathrm{d}\tau' \int \frac{\mathrm{d}\boldsymbol{p}}{(2\pi)^d} \frac{F(k_+, t-\tau)}{S(k_+)} D_0k_+ \sum_{m=1}^n \begin{pmatrix}
                n \\ m
            \end{pmatrix} R_u^{(m)}(\boldsymbol{k}_+, \boldsymbol{p} ; \tau, \tau') \frac{p}{S(p)} F_u^{(n-m)}(\boldsymbol{p}, \boldsymbol{k}_-  ; \tau', 0).
    \end{split}
    \label{eq:stochastic_expansion_integral_form}
    \end{equation}
The first line contains the contributions from the stochastic sources directly at order $n$, while the second line accounts for processes in which the perturbation propagates through higher–order corrections to the resolvent. To proceed, we must evaluate the $n$-th order contribution to the resolvent $R_u^{(n)}$. We recall that the resolvent $R_u$ can also be written in terms of the memory function $M_u$ (here generalized to the presence of external fields) as
    \begin{equation}
    \begin{split}
        R_u(\boldsymbol{k}, \boldsymbol{k}'; t,t') =&\ (2\pi)^d\delta(\boldsymbol{k}-\boldsymbol{k}')\delta(t-t') - M_u(\boldsymbol{k}, \boldsymbol{k}'; t,t') 
    \end{split}
    \end{equation}
from which we deduce that the $n$-th order contribution to the resolvent is given by $R_u^{(n)}(\boldsymbol{k}, \boldsymbol{k}'; t,t') = -M_u^{(n)}(\boldsymbol{k}, \boldsymbol{k}'; t,t')$. The $n$-th order memory kernels satisfy a natural recursion
    \begin{equation}
    \begin{split}
        M_u^{(n)}(\boldsymbol{k}, \boldsymbol{k}' ; t, t') =&\ \int_{t'}^{t} \mathrm{d}\tau_2 \int_{t'}^{\tau_2}\mathrm{d}\tau_1 R(k, t-\tau_2) M_u^{\mathrm{irr},(n)}(\boldsymbol{k}, \boldsymbol{k}' ; \tau_2, \tau_1)R(k', \tau_1-t')\\
        &- \sum_{m=1}^{n-1} \begin{pmatrix}
            n \\ m
        \end{pmatrix} \int_{t'}^{t} \mathrm{d}\tau_3 ...\int_{t'}^{\tau_2}\mathrm{d}\tau_1 \int \frac{\mathrm{d}\boldsymbol{p}}{(2\pi)^d} R(k, t-\tau_3) M_u^{\mathrm{irr}, (m)}(\boldsymbol{k}, \boldsymbol{p} ; \tau_3, \tau_2) M_u^{(n-m)}(\boldsymbol{p}, \boldsymbol{k}' ; \tau_1, t')
    \end{split}
    \label{eq:relation_M_Mirr_stochastic}
    \end{equation}
for $n\geq2$. This relation is a direct generalization of the decomposition Eq.~\eqref{eq:relation_M_Mirr} to the presence of spatio-temporally varying external fields. Importantly, the relation Eq.~\eqref{eq:relation_M_Mirr_stochastic} expresses the reducible memory at order $n$ as the sum of a direct irreducible insertion at order $n$, plus all possible decompositions into lower–order irreducible and reducible pieces. The recursion is initialized by the $n=1$ case:
    \begin{equation}
        M_u^{(1)}(\boldsymbol{k}, \boldsymbol{k}'; t,t') = \int_{t'}^{t} \mathrm{d}\tau_2 \int_{t'}^{\tau_2}\mathrm{d}\tau_1 R(k, t-\tau_2) M_u^{\mathrm{irr},(1)}(\boldsymbol{k}, \boldsymbol{k}' ; \tau_2, \tau_1)R(k', \tau_1-t')
    \label{eq:DeltaMu}
    \end{equation}
with the zeroth–order resolvent $R(k,t)$ given in Eq.~\eqref{eq:resolvent}. Using the results above, and after some algebraic manipulations, we obtain a compact integral equation for the $n$-th–order contribution to the order parameter in terms of the regularized vertices $\mathcal{v}_{12}^{\mathrm{reg}},\ \mathcal{v}_{21}^{\mathrm{reg}}$ Eqs.~\eqref{eq:regularization_left_handed_vertex}-\eqref{eq:regularization_right_handed_vertex}:
    \begin{equation}
    \begin{split}        
        F^{(n)}_u(\boldsymbol{k}_+, \boldsymbol{k}_- ; t,0) =&\ A_u^{(n)}(\boldsymbol{k}_+, \boldsymbol{k}_-; t, 0)\\
        &\ +  \int_0^t \mathrm{d}\tau_4...\int_0^{\tau_2} \mathrm{d}\tau_1 \int \frac{\mathrm{d}\boldsymbol{p}}{(2\pi)^d}F(k_+, t-\tau_4) \mathcal{v}_{12}^{\mathrm{reg}}(\boldsymbol{k}_+ ; \boldsymbol{k}-\boldsymbol{p}, \boldsymbol{p}+\tfrac{\boldsymbol{q}}{2} ; \tau_4-\tau_3)\\
        &\ \times F(|\boldsymbol{k}-\boldsymbol{p}|, \tau_3-\tau_2)  F_u^{(n)}(\boldsymbol{p}+\tfrac{\boldsymbol{q}}{2}, \boldsymbol{p}-\tfrac{\boldsymbol{q}}{2} ; \tau_3,\tau_2)\mathcal{v}_{21}^{\mathrm{reg}}(\boldsymbol{k}-\boldsymbol{p}, \boldsymbol{p}-\tfrac{\boldsymbol{q}}{2} ; \boldsymbol{k}_- ; \tau_2-\tau_1)F(k_-, \tau_1).
    \end{split}
    \label{eq:overarching_rainbows_general_stochastic}
    \end{equation}
A derivation is outlined in Appendix \ref{app:derivation_overarching_rainbows_general_stochastic_eq}. 
In Eq.~\eqref{eq:overarching_rainbows_general_stochastic}, $A^{(n)}(\boldsymbol{k}_+, \boldsymbol{k}_- ; t, 0)$ collects the effective source terms at order $n$. Although their explicit algebraic form is cumbersome, their diagrammatic content is transparent. At a given order $n$, one starts from the standard mode-coupling diagrams for the memory kernel $M_{\mathrm{MCT}}(k,t)$ [Eqs.~\eqref{eq:relation_M_Mirr}–\eqref{eq:Mirr_MCT}] and generates all new topologies by replacing a subset of the propagators $F_u$ with their $l$-th order variations $\mathcal{X}^{(\nu_1\ldots\nu_l)}_{\boldsymbol{q}_1\ldots\boldsymbol{q}_l}$. The only requirement is that the inserted vertices together carry a total order $\sum_i l_i = n$ to respect the order of the perturbative contribution. Note that the case with a single insertion of order $n$ is excluded because it has already been isolated in the previous steps to write Eq.~\eqref{eq:overarching_rainbows_general_stochastic}. To clarify this construction, we explicitly work through the cases $n=1$ and $n=2$ in the following subsections.

The structure of Eq.~\eqref{eq:overarching_rainbows_general_stochastic} mirrors that of Eq.~\eqref{eq:overarching_rainbows_general} and is shown diagrammatically in Fig.~\ref{fig:overarching_rainbows_general_stochastic}, confirming that the perturbative solution of the stochastic equation automatically regenerates the expected series of overarching rainbow diagrams. Evaluating this expression in the $\beta$-regime near the mode-coupling transition, we can recast it in terms of the mass operator in its zero frequency limit, yielding the leading contributions (written in Laplace frequency space)
\begin{equation}
            F_u^{(n)}(\boldsymbol{k}_+, \boldsymbol{k}_- ; z, 0) = A_u^{(n)}(\boldsymbol{k}_+, \boldsymbol{k}_- ; z, 0) + \int \frac{\mathrm{d}\boldsymbol{p}}{(2\pi)^d}\mathcal{M}_{\boldsymbol{q}}(\boldsymbol{k}, \boldsymbol{p})F_u^{(n)}(\boldsymbol{p}_+, \boldsymbol{p}_-; z, 0)  
    \end{equation}
which coincides with Eq.~(9) of the companion paper \cite{companionpaper} in the long-time limit ($z\to0$), and, after disorder averaging, with Eq.~\eqref{eq:overarching_rainbows_beta_regime} of Sec.~\ref{sec:leading_divergences_diagrammatic}. Here we have taken the Laplace transform with respect to the observation time $t$, holding the initial time fixed at zero.  What remains is to verify that the stochastic process reproduces not only the recursive rainbow structure but also the correct building blocks from which these diagrams are assembled. Here, this role is played by the effective source terms $A_u^{(n)}$, which encode how insertions of generalized susceptibilities generate the rainbow series. We now turn to their explicit structure.

\subsubsection*{Calculation at $1^{\mathrm{st}}$ Order: The Dynamical Susceptibilities}
To gain intuition for the general recursive structure, it is instructive to begin with the lowest nontrivial order of the expansion. For the proposed stochastic process to asymptotically capture the correct physics, the first-order contribution must reproduce the three-point dynamical susceptibilities introduced in Sec.~\ref{sec:scaling_critical_fluctuations}. Taking the $n=1$ case of Eq.\ \eqref{eq:overarching_rainbows_general_stochastic} gives 
    \begin{equation}
    \begin{split}
    F^{(1)}_u(\boldsymbol{k}_+, \boldsymbol{k}_- ; t,0) =&\ A_u^{(1)}(\boldsymbol{k}_+, \boldsymbol{k}_-; t, 0)\\
        &\ +  \int_0^t \mathrm{d}\tau_4...\int_0^{\tau_2} \mathrm{d}\tau_1 \int \frac{\mathrm{d}\boldsymbol{p}}{(2\pi)^d}F_{\mathrm{MCT}}(k_+, t-\tau_4) \mathcal{v}_{12}^{\mathrm{reg}}(\boldsymbol{k}_+ ; \boldsymbol{k}-\boldsymbol{p}, \boldsymbol{p}+\tfrac{\boldsymbol{q}}{2} ; \tau_4-\tau_3)\\
        &\ \times F_{\mathrm{MCT}}(|\boldsymbol{k}-\boldsymbol{p}|, \tau_3-\tau_2)  F_u^{(1)}(\boldsymbol{p}+\tfrac{\boldsymbol{q}}{2}, \boldsymbol{p}-\tfrac{\boldsymbol{q}}{2} ; \tau_3,\tau_2)\mathcal{v}_{21}^{\mathrm{reg}}(\boldsymbol{k}-\boldsymbol{p}, \boldsymbol{p}-\tfrac{\boldsymbol{q}}{2} ; \boldsymbol{k}_- ; \tau_2-\tau_1)F_{\mathrm{MCT}}(k_-, \tau_1),   
    \end{split}
    \label{eq:stochastic_eq_1st_order}
    \end{equation}
The explicit structure of the source terms can now be analyzed. To leading order in the external fields, we find
    \begin{equation}
    \begin{split}
        A^{(1)}_u(\boldsymbol{k}_+, \boldsymbol{k}_- ; t, 0) =&\ \int_0^t\mathrm{d}\tau \frac{F_{\mathrm{MCT}}(k_+, t-\tau)}{S(k_+)} \left(\mathcal{S}_u^{+,(1)}(\boldsymbol{k}_+, \boldsymbol{k}_- ; \tau, 0) + \mathcal{S}_u^{-,(1)}(\boldsymbol{k}_+, \boldsymbol{k}_- ; \tau, 0)  \right) \\
        =&\ \int_0^t\mathrm{d}\tau \frac{F_{\mathrm{MCT}}(k_+, t-\tau)}{S(k_+)} \left( D_0k_+ \int_0^{\tau}\mathrm{d}\tau' R(k_+, \tau-\tau') v_{\boldsymbol{k}_+}(\boldsymbol{k}_-, \boldsymbol{q})u^+_{\boldsymbol{q}}(\tau') F_{\mathrm{MCT}}(k_-, \tau')  \right) \\
        &\ + \int_0^t\mathrm{d}\tau \frac{F_{\mathrm{MCT}}(k_+, t-\tau)}{S(k_+)}\left(nD_0 S(k_+) \int_0^{\tau}\mathrm{d}\tau' u^-_{-\boldsymbol{q}}(\tau)v_{\boldsymbol{k}_-}(\boldsymbol{k}_+, -\boldsymbol{q})R(k_+, \tau-\tau') \frac{k_-}{S(k_-)} F_{\mathrm{MCT}}(k_-, \tau')  \right) \\
        =&\ \int_0^t\mathrm{d}t'\left( \chi_{\boldsymbol{q}}^{(0+)}(\boldsymbol{k}; t ; t') u_{\boldsymbol{q}}^+(t') +  \chi_{-\boldsymbol{q}}^{(0-)}(\boldsymbol{k}; t ; t') u_{-\boldsymbol{q}}^-(t')\right)
    \end{split}
    \end{equation}
in which we have recognized the bare sources $\chi_{\boldsymbol{q}}^{(0\pm)}(\boldsymbol{k}; t ; t')$ [see Eqs.~\eqref{eq:bare_source_Xq0+}-\eqref{eq:bare_source_Xq0-} in Appendix~\ref{app:critical_fluct_asymptotics}]. This establishes the formal equivalence between the first perturbative correction to the correlator and the dynamical susceptibility. More generally, substituting the definition of $F_u^{(1)}(\boldsymbol{k}_+, \boldsymbol{k}_- ; t, 0)$ into Eq.~\eqref{eq:stochastic_eq_1st_order} and taking a variation of the latter with respect to the fields $u^{\pm}_{\boldsymbol{q}_{\mathrm{pert}}}(t_{\mathrm{pert}})$ shows that the generalized response functions $\mathcal{X}^{\pm}_{\boldsymbol{q}_{\mathrm{pert}}}(\boldsymbol{k}; t ; t_{\mathrm{pert}})$ obey equations that are strictly equivalent to those satisfied by the susceptibilities $\chi^{\pm}_{\boldsymbol{q}_{\mathrm{pert}}}(\boldsymbol{k}; t ; t_{\mathrm{pert}})$, see Eq.~\eqref{eq:eom_susceptibilities_diagrammatic} in Appendix~\ref{app:critical_fluct_asymptotics}, which reduces to Eq.~\eqref{eq:dynamic_eigenvalue_susceptibility} when evaluated near the mode-coupling transition. In other words, the perturbative expansion of the effective stochastic problem naturally reproduces the rainbow diagram resummation at first order in the external fields. 

\subsubsection*{Calculation at $2^{\mathrm{nd}}$ Order}

The calculation at second order is already considerably more involved. Recall that we are concerned only with the dominant contributions near the mode-coupling transition. At higher orders (including second order), the contributions to $A_u^{(n)}$ involving $\mathcal{S}_u^{\pm, (n)}$ are subleading. To see this, we realize that these contributions involve at most $n-1$ susceptibility insertions, compared to the $n$ insertions contained in the terms on the second line of Eq.~\eqref{eq:stochastic_expansion_integral_form}. Consequently, the perturbative contributions emanating from $\mathcal{S}_u^\pm$ in Eq.~\eqref{eq:postulate_microscopic_stochastic} can be safely neglected. The relevant second–order source can then be decomposed into three distinct contributions $A^{(2)}_u(\boldsymbol{k}_+, \boldsymbol{k}_- ; t,0) = A^{(2)}_{u,1}(\boldsymbol{k}_+, \boldsymbol{k}_- ; t,0) + A^{(2)}_{u,2}(\boldsymbol{k}_+, \boldsymbol{k}_- ; t,0) + A^{(2)}_{u,3}(\boldsymbol{k}_+, \boldsymbol{k}_- ; t,0)$. 

The first contribution emerges from the variation of the term of form $(R * M_u^{\mathrm{irr},(2)} * R)$ which is part of the $m=2$ contribution in Eq.~\eqref{eq:stochastic_expansion_integral_form}, and can be written as 
    \begin{equation}
    \begin{split}
        A^{(2)}_{u,1}(\boldsymbol{k}_+, \boldsymbol{k}_- ; t,0) =&\ \frac{1}{2}\int_0^t\mathrm{d}\tau_4...\int_0^{\tau_2}\mathrm{d}\tau_1 \int \frac{\mathrm{d}\boldsymbol{p}}{(2\pi)^d}\frac{\mathrm{d}\boldsymbol{q}_1}{(2\pi)^d}\frac{\mathrm{d}\boldsymbol{q}_2}{(2\pi)^d}  F_{\mathrm{MCT}}(k_+, t-\tau_4) \mathcal{v}_{12}^{\mathrm{reg}}(\boldsymbol{k}_+ ; \boldsymbol{k}-\boldsymbol{p}+\boldsymbol{q}_1, \boldsymbol{p}+\boldsymbol{q}_2-\tfrac{\boldsymbol{q}}{2} ; \tau_4-\tau_3) \\
        &\ \times F_u^{(1)}(\boldsymbol{k}-\boldsymbol{p}+\boldsymbol{q}_1, \boldsymbol{k}-\boldsymbol{p} ; \tau_3, \tau_2) F_u^{(1)}(\boldsymbol{p}+\boldsymbol{q}_2-\tfrac{\boldsymbol{q}}{2}, \boldsymbol{p}-\tfrac{\boldsymbol{q}}{2} ; \tau_3, \tau_2) \mathcal{v}_{21}^{\mathrm{reg}}(\boldsymbol{k}-\boldsymbol{p}, \boldsymbol{p}-\tfrac{\boldsymbol{q}}{2} ; \tau_2-\tau_1) F_{\mathrm{MCT}}(k_-, \tau_1) \\
        &\ \times (2\pi)^d \delta(\boldsymbol{q}_1+\boldsymbol{q}_2-\boldsymbol{q})
    \end{split}
    \label{eq:Au2_1}
    \end{equation}
and is diagrammatically represented by the first diagram of Fig.~\ref{fig:stochastic_sources_2nd_order_dynamic}. 

The second term originates from the perturbative contribution of form $(R * M_u^{\mathrm{irr},(1)} * R * M_u^{\mathrm{irr},(1)} * R)$ [again, part of the $m=2$ contribution in Eq.~\eqref{eq:stochastic_expansion_integral_form}] , and can be written as
    \begin{equation}
    \begin{split}
        A^{(2)}_{u,2}(\boldsymbol{k}_+, \boldsymbol{k}_- ; t,0) =\ -\int_0^t\mathrm{d}\tau_6...\int_0^{\tau_2}\mathrm{d}\tau_1 &\int \frac{\mathrm{d}\boldsymbol{p}_1}{(2\pi)^d} \frac{\mathrm{d}\boldsymbol{p}_2}{(2\pi)^d}\frac{\mathrm{d}\boldsymbol{q}_1}{(2\pi)^d}\frac{\mathrm{d}\boldsymbol{q}_2}{(2\pi)^d} F(k_+, t-\tau_6) \mathcal{v}_{12}^{\mathrm{reg}}(\boldsymbol{k}_+ ; \boldsymbol{k}-\boldsymbol{p}_2+\boldsymbol{q}_2, \boldsymbol{p}_2+\boldsymbol{q}_1-\tfrac{\boldsymbol{q}}{2} ; \tau_6-\tau_5) \\
        &\ \times F_u^{(1)}(\boldsymbol{k}-\boldsymbol{p}_2+\boldsymbol{q}_2, \boldsymbol{k}-\boldsymbol{p}_2 ; \tau_5,\tau_4) F_{\mathrm{MCT}}(|\boldsymbol{p}_2+\boldsymbol{q}_1-\tfrac{\boldsymbol{q}}{2}|, \tau_5-\tau_4)\\
        &\ \times \mathcal{v}_{22}^{\mathrm{reg}}(\boldsymbol{k}-\boldsymbol{p}_2, \boldsymbol{p}_2+\boldsymbol{q}_1-\tfrac{\boldsymbol{q}}{2} ; \boldsymbol{k}-\boldsymbol{p}_1+\boldsymbol{q}_1, \boldsymbol{p}_1-\tfrac{\boldsymbol{q}}{2} ; \tau_4-\tau_3) \\
        &\ \times F_u^{(1)}(\boldsymbol{k}-\boldsymbol{p}_1+\boldsymbol{q}_1, \boldsymbol{k}-\boldsymbol{p} ; \tau_3, \tau_2)F_{\mathrm{MCT}}(|\boldsymbol{p}_1-\tfrac{\boldsymbol{q}}{2}|, \tau_3-\tau_2) \\
        &\ \times \mathcal{v}_{21}^{\mathrm{reg}}(\boldsymbol{k}-\boldsymbol{p} , \boldsymbol{p}_1-\tfrac{\boldsymbol{q}}{2} ; \boldsymbol{k}_- ; \tau_2-\tau_1)F_{\mathrm{MCT}}(k_-, \tau_1) (2\pi)^d \delta(\boldsymbol{q}_1+\boldsymbol{q}_2-\boldsymbol{q}).
    \end{split}
    \label{eq:Au2_2}
    \end{equation}
Equation \eqref{eq:Au2_2} is diagrammatically represented by the second diagram in Fig.~\ref{fig:stochastic_sources_2nd_order_dynamic}. 

The third contribution comes from the perturbative contribution of form $(R * M_u^{\mathrm{irr},(1)} * F_u^{(1)})$, contained in the $m=1$ contribution to Eq.~\eqref{eq:stochastic_expansion_integral_form}. After appropriate algebraic manipulations and a rewriting of the right-most susceptibility insertion diagrammatically illustrated in Fig.~\ref{fig:iteration_trick_diagrammatic}, its dominant component around the mean-field transition can be expressed as
    \begin{equation}
    \begin{split}
        A^{(2)}_{u,3}(\boldsymbol{k}_+, \boldsymbol{k}_- ; t,0) =\int_0^t\mathrm{d}\tau_8...\int_0^{\tau_2}\mathrm{d}\tau_1 &\int \frac{\mathrm{d}\boldsymbol{p}_1}{(2\pi)^d} \frac{\mathrm{d}\boldsymbol{p}_2}{(2\pi)^d}\frac{\mathrm{d}\boldsymbol{q}_1}{(2\pi)^d}\frac{\mathrm{d}\boldsymbol{q}_2}{(2\pi)^d} F_{\mathrm{MCT}}(k_+, t-\tau_8) \mathcal{v}_{12}^{\mathrm{reg}}(\boldsymbol{k}_+ ; \boldsymbol{k}-\boldsymbol{p}_2+\boldsymbol{q}_2, \boldsymbol{p}_2+\boldsymbol{q}_1-\tfrac{\boldsymbol{q}}{2} ; \tau_8-\tau_7) \\
        &\ \times F_u^{(1)}(\boldsymbol{k}-\boldsymbol{p}_2+\boldsymbol{q}_2 , \boldsymbol{k}-\boldsymbol{p}_2 ; \tau_7, \tau_6) F_{\mathrm{MCT}}(|\boldsymbol{p}_2+\boldsymbol{q}_1-\tfrac{\boldsymbol{q}}{2}|, \tau_7-\tau_6)\\
        &\ \times \mathcal{v}_{21}^{\mathrm{reg}}(\boldsymbol{k}-\boldsymbol{p}_2, \boldsymbol{p}_2+\boldsymbol{q}_1-\tfrac{\boldsymbol{q}}{2} ; \boldsymbol{k}+\boldsymbol{q}_1-\tfrac{\boldsymbol{q}}{2} ; \tau_6-\tau_5) F_{\mathrm{MCT}}(|\boldsymbol{k}+\boldsymbol{q}_1-\tfrac{\boldsymbol{q}}{2}|, \tau_5-\tau_4)\\
        &\ \times \mathcal{v}_{12}^{\mathrm{reg}}(\boldsymbol{k}+\boldsymbol{q}_1-\tfrac{\boldsymbol{q}}{2} ; \boldsymbol{k}-\boldsymbol{p}_1+\boldsymbol{q}_1, \boldsymbol{p}_1-\tfrac{\boldsymbol{q}}{2} ; \tau_4-\tau_3) F_{\mathrm{MCT}}(|\boldsymbol{p}_1-\tfrac{\boldsymbol{q}}{2}|, \tau_3-\tau_2) \\
        &\ \times F_u^{(1)}(\boldsymbol{k}-\boldsymbol{p}_1+\boldsymbol{q}_1 , \boldsymbol{k}-\boldsymbol{p}_1 ; \tau_3, \tau_4) \mathcal{v}_{21}^{\mathrm{reg}}(\boldsymbol{k}-\boldsymbol{p}_1, \boldsymbol{p}_1-\tfrac{\boldsymbol{q}}{2} ; \tau_2-\tau_1) F_{\mathrm{MCT}}(k_-, \tau_1)\\
        &\ \times (2\pi)^d\delta(\boldsymbol{q}_1+\boldsymbol{q}_2-\boldsymbol{q}).
    \end{split}
    \label{eq:iterated_source3_second_order}
    \end{equation}
\begin{figure}
    \centering
    \includegraphics[width=0.9\linewidth]{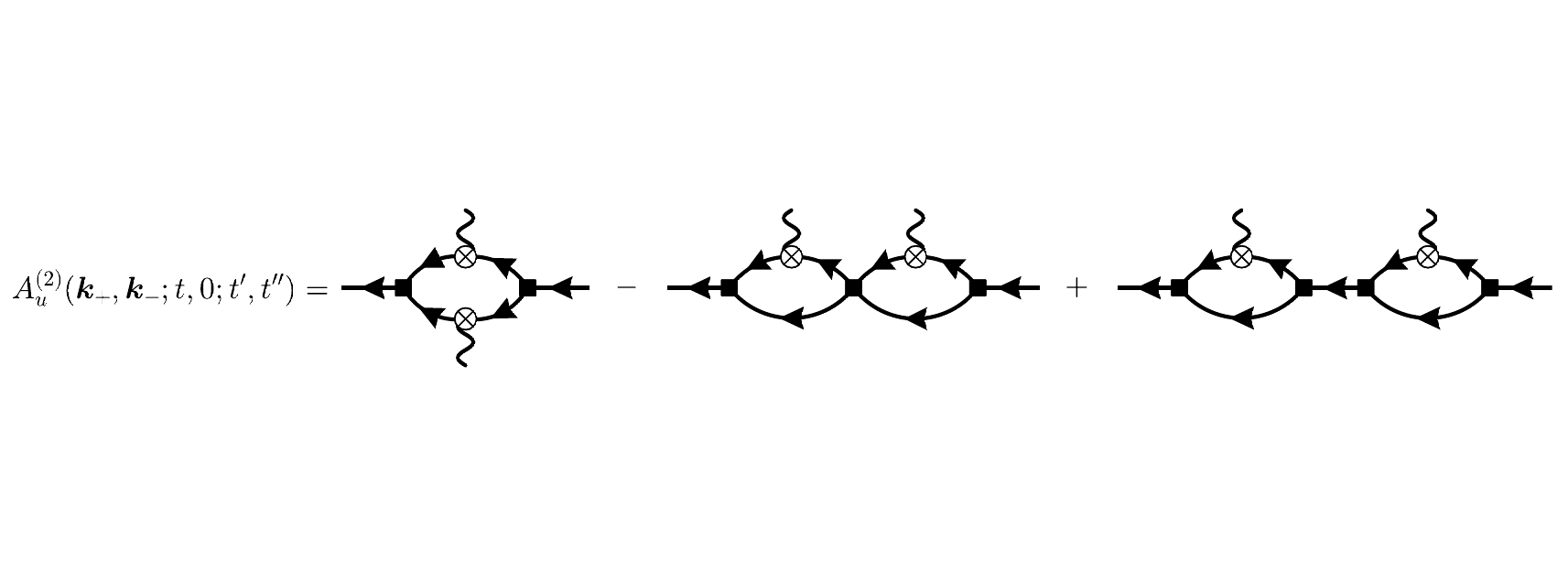}
    \vspace{-60pt}
    \caption{Second-order effective sources $A_u^{(2)}(\boldsymbol{k}_+, \boldsymbol{k}_-, t ; t', t'')$ arising from the perturbative expansion of Eqs.~\eqref{eq:postulate_microscopic_stochastic}–\eqref{eq:inhomogeneous_Mirr}. These terms represent the second-order response of the system to external stochastic fields. The total momentum transferred from the fields is $\boldsymbol{q}$, reflecting the net intake of momentum through the pair of insertions at intermediate times.}
    \label{fig:stochastic_sources_2nd_order_dynamic}
\end{figure}
The equivalence between the topology of these contributions and the original diagrams identified within the diagrammatic theory [Fig.~\ref{fig:diagrammatic_most_diverging_2nd_4th_order} (a)] is now evident. Substituting the definitions of $F_u^{(1)}$ in terms of the generalized susceptibilities and performing the disorder average, this correspondence becomes exact, yielding specifically 
    \begin{equation}
        \llbracket A_u^{(2)}(\boldsymbol{k}_+, \boldsymbol{k}_- ; t, 0) \rrbracket = a^{(2)}(k; \hat{t})S(k)|\varepsilon|^{(d-4)/4} + o(\varepsilon^{(d-4)/4})
    \end{equation}
around the mode-coupling transition.
\end{widetext}

\begin{figure}
    \centering
    \includegraphics[width=0.9\linewidth]{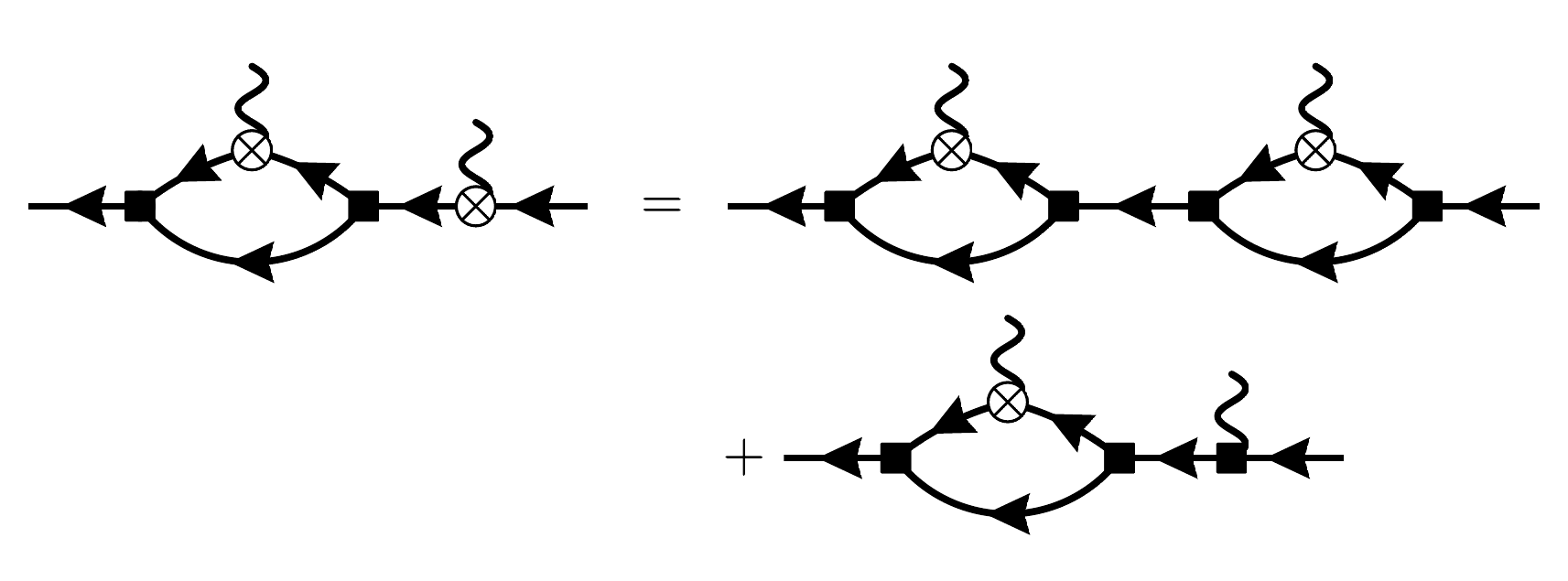}
    \caption{Iteration scheme for obtaining the contribution $A_{u,3}^{(2)}(\boldsymbol{k}_+, \boldsymbol{k}_- ; t,0)$ [Eq.~\eqref{eq:iterated_source3_second_order}]. The diagram on the left is generated directly from the perturbative expansion [Eq.\eqref{eq:overarching_rainbows_general_stochastic}]. To arrive at the right-hand side, the right-most insertion is rewritten using Eq.~\eqref{eq:stochastic_eq_1st_order}. We have defined the ``bare insertion" \raisebox{-4ex}{\includegraphics[height=10ex,page=1]{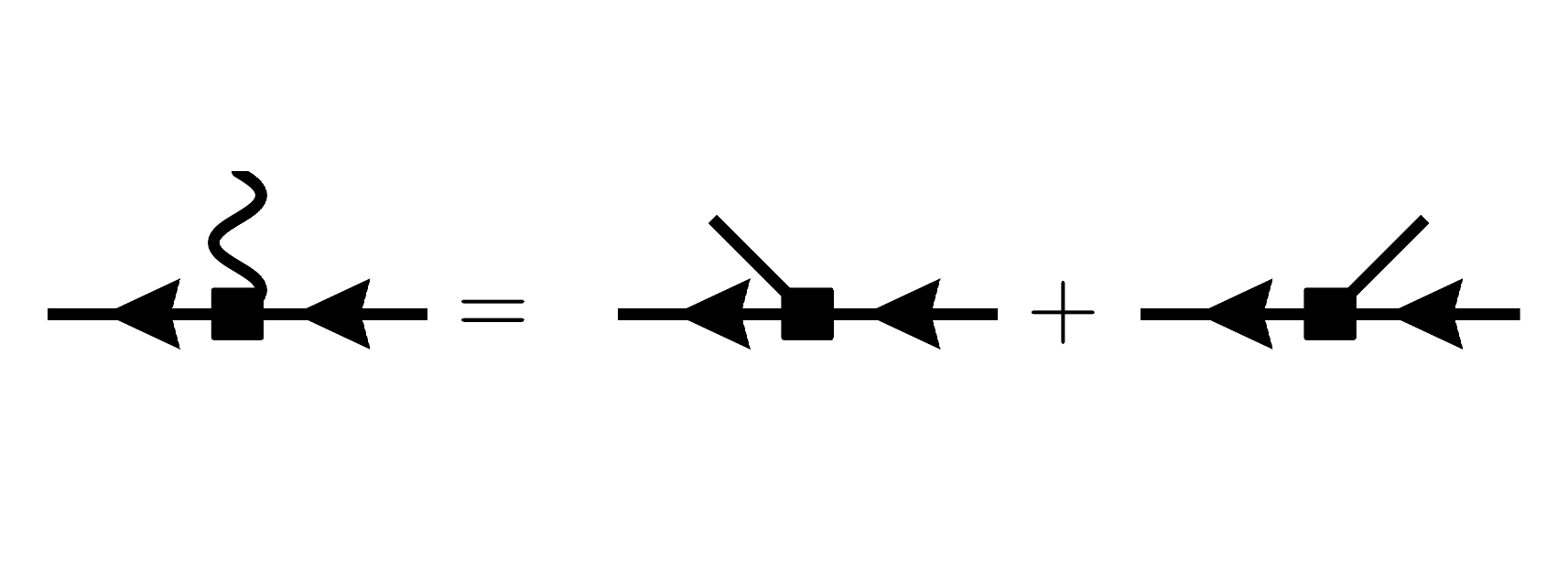}} for notational convenience. The second term on the right-hand side is subdominant after disorder averaging and evaluation near the mode-coupling contribution, as it scales only as $\varepsilon^{(d-2)/4}$, compared to the leading term that scales as $\varepsilon^{(d-4)/4}$ as discussed in the main text.}
    \label{fig:iteration_trick_diagrammatic}
\end{figure}

\subsection{Resummation of Fluctuations Beyond Mean-Field}

Looking at the diagrammatic structure of the generated source terms $A_u^{(n)}$ at arbitrary order, it is not difficult to convince oneself that the perturbative scheme reproduces order by order the class of diagrams identified in Sec.~\ref{sec:leading_divergences_diagrammatic}. This allows us to formally write 
    \begin{equation}
        \llbracket \sum_{n=0}^{\infty} F^{(n)}_u(\boldsymbol{k}_+, \boldsymbol{k}_- ; t, 0) \rrbracket = F_{\mathrm{MCT}}(k; t) + \Delta F(k;t) 
    \label{eq:equivalence_stochastic_diagrammatics}
    \end{equation}
to leading order in $\varepsilon$. This establishes full equivalence between the stochastic framework [left-hand side of Eq.~\eqref{eq:equivalence_stochastic_diagrammatics}] and the diagrammatic theory [right-hand side of Eq.~\eqref{eq:equivalence_stochastic_diagrammatics}]. While the leading-order contributions coincide exactly with those of the diagrammatic theory, the stochastic process also generates a hierarchy of subdominant divergences. These are however formally unimportant in the asymptotic limit $\varepsilon\to0$. We emphasize that the equivalence holds both in the long-time limit and extends dynamically throughout the critical $\beta$-regime.

\section{Asymptotic Dynamics Beyond Mean-Field}\label{sec:asymptotic_dynamics_beyond_mean-field}

Having established that the stochastic process captures the dominant fluctuations around the mean-field dynamics, we now turn to the physical interpretation of this process. We focus on the slow, glassy dynamics characteristic of the $\beta$-regime, where the perturbative corrections under consideration become most relevant. We recall that the density is assumed to be close to the presumed critical density $n_c$, with $n = n_c(1 + \varepsilon)$ and $|\varepsilon|\ll 1$. In this regime, the intermediate scattering function exhibits a characteristic two-step decay, which motivates the asymptotic decomposition around a plateau value $F_p(k)$:
    \begin{equation}
        F_u(\boldsymbol{k}_+, \boldsymbol{k}_-; t, 0) = F_p(k) (2\pi)^d\delta(\boldsymbol{q}) + H_u(\boldsymbol{k}_+, \boldsymbol{k}_-; \hat{t})
    \label{eq:generalized_MCT_ansatz}
    \end{equation}
where here the rescaled time variable is $\hat{t} \equiv t / t_u(\varepsilon)$ with the critical timescale $t_u(\varepsilon)$ to be fixed by matching asymptotic regimes. Note that the critical timescale in principle depends on the stochastic variables. Taking a Laplace transform, 
    \begin{equation}
    zF_u(\boldsymbol{k}_+, \boldsymbol{k}_-; z, 0) = F_p(k) (2\pi)^d\delta(\boldsymbol{q}) + \hat{z}H_u(\boldsymbol{k}_+, \boldsymbol{k}_-; \hat{z})
    \label{eq:generalized_MCT_ansatz_frequency}
    \end{equation}
with $\hat{z}\equiv t_u(\varepsilon)z$. Note that in Eq.~\eqref{eq:generalized_MCT_ansatz} [Eq.~\eqref{eq:generalized_MCT_ansatz_frequency}] we have omitted the explicit two-time [two-frequency] notation for the deviation $H_u$ since in the limit $|\varepsilon|\ll1$, the noise–noise correlator Eq.~\eqref{eq:noise_noise_correlator} becomes approximately time-independent. As a consequence, the spatio-temporal stochastic fields $u^\pm_{\boldsymbol{q}}(t)$ can, to leading order, be represented by an effective quenched process (\textit{i.e.}\ a static realization of random fields) that we will specify below. This approximation restores time-translation invariance, thereby justifying the use of a single time argument in the representation of $H_u$. 

The plateau is expressed as a perturbation of the mean-field value $F_p(k) = F_c(k) + \delta F(k)$ where $F_c(k)$ satisfies its own self-consistent equation [see Eq.~\eqref{eq:MCT_EOS}] and $\delta F(k)$ denotes a non-trivial correction from the resummation. We assume that the corrections around the plateau are small $H_u(\boldsymbol{k}_+, \boldsymbol{k}_-, \hat{t}) \ll F_p(k)$, and consequently that $\mathcal{L}\{ H_u(\boldsymbol{k}_+, \boldsymbol{k}_-; \hat{t})^{n+1}\}(\hat{z}) \ll  \mathcal{L}\{H_u(\boldsymbol{k}_+, \boldsymbol{k}_-; \hat{t})^{n}\}(\hat{z})$ \cite{gotze1989beta}. The microscopic correction $\delta F(k)$ is also expected to be weak, since our resummation only addresses long-wavelength critical fluctuations. We therefore neglect $\delta F(k)$, and work with the approximation $F_p(k) \approx F_c(k)$. 

By substituting the expansion Eq.~\eqref{eq:generalized_MCT_ansatz} in the equation of motion Eq.~\eqref{eq:postulate_microscopic_stochastic}, we can write down an integral equation of the following form
\begin{widetext}
    \begin{equation}
        \frac{1}{f_c(k_+)}\int \frac{\mathrm{d}\boldsymbol{p}}{(2\pi)^d} \big((2\pi)^d\delta(\boldsymbol{k}-\boldsymbol{p}) - \mathcal{M}_{\boldsymbol{q}}(\boldsymbol{k}, \boldsymbol{p}) \big) \hat{z}H_u(\boldsymbol{p}_+, \boldsymbol{p}_-; \hat{z}) = I_u(\boldsymbol{k}_+, \boldsymbol{k}_-; \hat{z})
    \label{eq:stochastic_eigen_problem}
    \end{equation}
\end{widetext}
where the left-hand side contains the mass operator of the dynamical susceptibility $\mathcal{M}_{\boldsymbol{q}}(\boldsymbol{k}, \boldsymbol{p})$ (see Appendix \ref{app:critical_fluct_asymptotics}, Eq.~\eqref{eq:mass_operator_stability} for its definition) acting linearly on $H_u$. We stress that the mass operator appears here after taking the zero-frequency limit, which is justified by our restriction to the critical $\beta$-regime. All remaining contributions, \textit{i.e.} those involving $\varepsilon$ explicitly and the non-linear terms in $H_u$, are collected on the right-hand side in the function $I_u$. Ideally, one would attempt the standard route of performing a systematic expansion of the deviations $H_u$ in powers of $\varepsilon$, \textit{i.e.} $H_u(\boldsymbol{k}_+, \boldsymbol{k}_-; \hat{t} ; \varepsilon) = \sum_{n=1}^{\infty} r(\varepsilon)^n H_u^{(n)}(\boldsymbol{k}_+, \boldsymbol{k}_-; \hat{t})$ with $r(\varepsilon)$ some monomial in $|\varepsilon|$ as is done in the standard mean-field treatment \cite{andreanov2009mode}. In the present case, however, such an expansion is not feasible: this generates a hierarchy of coupled equations that cannot be truncated in a controlled way at fixed order in $\varepsilon$ unlike in MCT. This makes it necessary to resum the deviations from the plateau to all orders in $\varepsilon$. To this end, we propose the Ansatz
    \begin{equation}
    H_u(\boldsymbol{k}_+, \boldsymbol{k}_-; \hat{t}) = H(k)g_u(\boldsymbol{q};\hat{t}),    \label{eq:generalized_factorization_theorem}
    \end{equation}
which cleanly separates the short-wavelength structural fluctuations in $H(k)$, inherited from the underlying mode-coupling structure, from the long wavelength dynamical fluctuations described by $g_u(\boldsymbol{q};\hat{t})$. 

To make progress, we further assume the decomposition of the amplitude around its mean-field value $H(k) = S(k) \left(h_0^{\mathrm{R}}(k) + \delta h_0^{\mathrm{R}}(k) \right)$ where we recall that $h_0^{\mathrm{R}}(k)$ is the principal eigenfunction of the stability matrix of MCT and we denote $\delta h_0^{\mathrm{R}}(k)$ the static corrections from the resummation of the leading order divergences. By the same reasoning as for the plateau value $F_p(k)$ these short-distance corrections are expected to be weak, and we therefore approximate $H(k)$ by the mean-field result: $H(k) \approx S(k)h_0^{\mathrm{R}}(k)$. Carrying out a low-$q$ expansion we can write down a dynamical equation for $g_u(\boldsymbol{q};\hat{t})$, as we outline next. In summary, our strategy is to extract simultaneously (a) the long-wavelength physics ($q\ll1$) and (b) the asymptotics near the transition $(|\varepsilon|\ll1)$.

Focusing on the left-hand side of Eq.~\eqref{eq:stochastic_eigen_problem} and applying the approximations outlined above together 
we eventually obtain 
\begin{widetext}
    \begin{equation}
    \begin{split}
        \int \frac{\mathrm{d}\boldsymbol{p}}{(2\pi)^d} \big((2\pi)^d\delta(\boldsymbol{k}-\boldsymbol{p}) - \mathcal{M}_{\boldsymbol{q}}(\boldsymbol{k}, \boldsymbol{p}) \big) \hat{z}H_u(\boldsymbol{p}_+, \boldsymbol{p}_-; \hat{z}) =&\ -\frac{S(k)h_0^{\mathrm{R}}(k)}{f_c(k)}\Gamma q^2 \hat{z}g_u(\boldsymbol{q};\hat{z}) + \mathcal{O}(q^4g_u, \sqrt{\varepsilon}g_u).
    \end{split}
    \label{eq:SBR_mass_operator}
    \end{equation}
We note that the additional contribution $\sim \mathcal{O}\left(\sqrt{\varepsilon}g_u\right)$ from the mass-operator is subleading, since $g_u$ already resums contributions at all orders in $\varepsilon$, and any additional prefactors of form $|\varepsilon|^\alpha g_u(\boldsymbol{q};\hat{z})$ with positive $\alpha$ are suppressed in the small-$\varepsilon$ limit. This constitutes the first key difference from the mean-field analysis, where the same term is not subleading in the non-ergodic phase ($\varepsilon<0$).

As mentioned above, the inhomogeneity on the right-hand side of Eq.~\eqref{eq:stochastic_eigen_problem} can be written as a series expansion in $\varepsilon$ and $H_u$:
    \begin{equation}
        I_u(\boldsymbol{k}_+, \boldsymbol{k}_-; \hat{z}) = I_u^{(0)}(\boldsymbol{k}_+, \boldsymbol{k}_- ; \hat{z}) + I_u^{(1)}(\boldsymbol{k}_+, \boldsymbol{k}_-; \hat{z}) + \mathcal{O}(uH_u, H_u^3, \varepsilon H_u^2, \varepsilon^2),
    \end{equation}
paralleling the standard mean-field treatment discussed in detail in Appendix \ref{sec:mean_field_scenario}. The term $I^{(0)}_u$ consists of two contributions. The first is the frequency-independent mean-field term familiar from the standard MCT treatment, which involves the function $C^{(0)}(k)$ discussed in Appendix~\ref{app:asymptotics_MCT}. The second arises from the stochastic fields and is absent in the mean-field treatment. Explicitly,
    \begin{equation}
        I_u^{(0)}(\boldsymbol{k}_+, \boldsymbol{k}_- ; \hat{z}) = - \frac{S(k)}{f_c(k)}\varepsilon C^{(0)}(k) - \frac{S(k)}{f_c(k)}s_u(\boldsymbol{k}_+, \boldsymbol{k}_- ; \hat{z}) + \mathcal{O}(\varepsilon g_u, ug_u) 
    \label{eq:I_u0_final}
    \end{equation}
with 
    \begin{equation}
    \begin{split}
        s_u(\boldsymbol{k}_+, \boldsymbol{k}_-; \hat{z}) \approx&\ S(k)c(k) f_c(k)(1-f_c(k)) \left( u^+_{\boldsymbol{q}}(\hat{z}) + nu^-_{\boldsymbol{q}}(\hat{z}) \right) + \mathcal{O}\left(q u^\pm, g_uu^\pm\right). 
    \end{split}
    \label{eq:s_u}
    \end{equation}
Close to the transition, the random sources $u^{\pm}_{\boldsymbol{q}}(\hat{z})$ only enter through the linear combination in Eq.~\eqref{eq:s_u} above. It is convenient to introduce a single effective stochastic process $u_{\boldsymbol{q}}(t) = u^+_{\boldsymbol{q}}(t) + n_cu^-_{\boldsymbol{q}}(t)$. By construction, we have $\llbracket u_{\boldsymbol{q}}(t) \rrbracket = 0$, and $\llbracket u_{\boldsymbol{q}}(t) u_{\boldsymbol{q}'}(t') \rrbracket = n_cF_c(q)\left( \Theta(t-t') + \Theta(t'-t)\right)(2\pi)^d\delta(\boldsymbol{q}-\boldsymbol{q}') + \mathcal{O}(\sqrt{|\varepsilon|}) = n_cF_c(q)(2\pi)^d\delta(\boldsymbol{q}-\boldsymbol{q}') + \mathcal{O}(\sqrt{|\varepsilon}|)$ ; Hence, the Heaviside functions cancel and, to leading order in $\varepsilon$, the process $u_{\boldsymbol{q}}(t) \approx u_{\boldsymbol{q}}$ is effectively quenched, such that Eq.~\eqref{eq:s_u} can be rewritten as
    \begin{equation}
        s_u(\boldsymbol{k}_+, \boldsymbol{k}_-) \approx S(k)c(k) f_c(k)(1-f_c(k)) u_{\boldsymbol{q}} + \mathcal{O}\left(q u_{\boldsymbol{q}}, \sqrt{\varepsilon}u_{\boldsymbol{q}}\right).
    \end{equation}
With this additional approximation, the contribution $I_u^{(0)}$ becomes frequency independent. The terms that make up $I^{(1)}_u$ depend on $H_u$ but not on $\varepsilon$ explicitly. Their structure is analogous to that encountered in the mean-field treatment, except that the scaling function $g_u(\boldsymbol{q},t)$ is now also wavenumber dependent. There are three such $H_u$-dependent contributions, so that $I^{(1)}_u(\boldsymbol{k}_+, \boldsymbol{k}_-; \hat{z}) = I^{(1)}_{u,1}(\boldsymbol{k}_+, \boldsymbol{k}_-; \hat{z}) + I^{(1)}_{u,2}(\boldsymbol{k}_+, \boldsymbol{k}_-; \hat{z}) + I^{(1)}_{u,3}(\boldsymbol{k}_+, \boldsymbol{k}_-; \hat{z})$. Explicitly, one finds for the first two
    \begin{equation}
    \begin{split}
        I^{(1)}_{u,1}(\boldsymbol{k}_+, \boldsymbol{k}_-; \hat{z}) \approx&\ -\frac{S(k)h_0^{\mathrm{R}}(k)^2}{f_c(k)^2} \hat{z}^2\int \frac{\mathrm{d}\boldsymbol{p}}{(2\pi)^d} g_u(\boldsymbol{p};\hat{z}) g_u(\boldsymbol{q}-\boldsymbol{p};\hat{z})
    \end{split}
    \label{eq:Iu_1}
    \end{equation}
and
    \begin{equation}
    \begin{split}
        I^{(1)}_{u,2}(\boldsymbol{k}_+, \boldsymbol{k}_-; \hat{z}) \approx&\ \frac{S(k)h_0^{\mathrm{R}}(k)^2}{f_c(k)^2(1-f_c(k))} \hat{z}^2\int \frac{\mathrm{d}\boldsymbol{p}}{(2\pi)^d} g_u(\boldsymbol{p};\hat{z}) g_u(\boldsymbol{q}-\boldsymbol{p};\hat{z}).
    \end{split}
    \label{eq:Iu_2}
    \end{equation}
Lastly, for the third contribution we get
    \begin{equation}
    \begin{split}
        I^{(1)}_{u,3}(\boldsymbol{k}_+, \boldsymbol{k}_-; \hat{z}) \approx&\ -\frac{S(k)}{f_c(k)} \int \frac{\mathrm{d}\boldsymbol{p}}{(2\pi)^d} C^{(2)}(\boldsymbol{k}, \boldsymbol{p})h_0^{\mathrm{R}}(p)h_0^{\mathrm{R}}(|\boldsymbol{k}-\boldsymbol{p}|) \hat{z}\mathcal{L}\left\{ \int\frac{\mathrm{d}\boldsymbol{p}'}{(2\pi)^d}g_u(\boldsymbol{p}';\hat{t})g_u(\boldsymbol{q}-\boldsymbol{p}';\hat{t})\right\}(\hat{z}).
    \end{split}
    \label{eq:Iu_3}
    \end{equation}
Detailed derivations of Eqs.~\eqref{eq:Iu_1}-\eqref{eq:Iu_3} are provided in Appendix~\ref{app:derivation_stochastic_inhomogeneities}. We can then assemble the results of Eqs.~\eqref{eq:Iu_1}-\eqref{eq:Iu_3} to give for $I_u^{(1)}$:
    \begin{equation}
    \begin{split}
       I_u^{(1)}(\boldsymbol{k}_+, \boldsymbol{k}_-; \hat{z}) \approx&\ \frac{S(k)}{f_c(k)} \frac{h_0^{\mathrm{R}}(k)^2}{(1-f_c(k))} \hat{z}^2 \int \frac{\mathrm{d}\boldsymbol{p}}{(2\pi)^d} g_u(\boldsymbol{p};\hat{z}) g_u(\boldsymbol{q}-\boldsymbol{p};\hat{z}) \\
       &\ - \frac{S(k)}{f_c(k)} \int \frac{\mathrm{d}\boldsymbol{p}}{(2\pi)^d} C^{(2)}(\boldsymbol{k}, \boldsymbol{p})h_0^{\mathrm{R}}(p)h_0^{\mathrm{R}}(|\boldsymbol{k}-\boldsymbol{p}|) \hat{z}\mathcal{L}\left\{ \int\frac{\mathrm{d}\boldsymbol{p}'}{(2\pi)^d}g_u(\boldsymbol{p}'; \hat{t})g_u(\boldsymbol{q}-\boldsymbol{p}'; \hat{t})\right\}(\hat{z}) + \mathcal{O}(q^2g_u, \sqrt{\varepsilon}g_u).
    \end{split}
    \label{eq:I_u1_final}
    \end{equation}
By collecting the results of Eqs.~\eqref{eq:SBR_mass_operator}, \eqref{eq:I_u0_final} and \eqref{eq:I_u1_final} and contracting the resulting equation with the left eigenfunction of the stability operator, we finally obtain
    \begin{equation}
       -\Gamma q^2 \hat{z}g_s(\boldsymbol{q};\hat{z}) = -\sigma - s(\boldsymbol{q}) - \lambda \hat{z}\mathcal{L} \left\{ \int \frac{\mathrm{d}\boldsymbol{p}}{(2\pi)^d} g_s(\boldsymbol{p}, \hat{t})g_s(\boldsymbol{q}-\boldsymbol{p}; \hat{t}) \right\}(\hat{z}) + \hat{z}^2\int \frac{\mathrm{d}\boldsymbol{p}}{(2\pi)^d} g_s(\boldsymbol{p}; \hat{z})g_s(\boldsymbol{q}-\boldsymbol{p}; \hat{z})
    \label{eq:SBR_Fourier_space}
    \end{equation}
in which $s(\boldsymbol{q}) \equiv \mu u_{\boldsymbol{q}}$ is a rescaled stochastic process with prefactor
    \begin{equation}
        \mu \equiv \int \frac{\mathrm{d}\boldsymbol{k}}{s_d} \left( \frac{h_0^{\mathrm{L}}(k)}{k^{d-1}}\right) S(k)c(k)f_c(k)(1-f_c(k))
    \label{eq:prefactor_noise_SBR}
    \end{equation}
and where we have also introduced the mean-field separation parameter 
    \begin{equation}
        \sigma \equiv \varepsilon \int \frac{\mathrm{d}\boldsymbol{k}}{s_d} \left(\frac{h_0^{\mathrm{L}}(k)}{k^{d-1}}\right)C^{(0)}(k).
    \label{eq:separation_parameter}
    \end{equation}
Note that in Eq. \eqref{eq:SBR_Fourier_space} we have also changed the notation $g_u \to g_s$ to emphasize the dependence on the stochastic process $s(\boldsymbol{q})$. Finally, the rescaled stochastic process $s(\boldsymbol{q})$ is defined through $\llbracket s(\boldsymbol{q}) \rrbracket = 0$ and $\llbracket s(\boldsymbol{q}) s(\boldsymbol{q}')\rrbracket = \Delta\sigma^2 (2\pi)^d\delta(\boldsymbol{q}-\boldsymbol{q}')$ with $\Delta\sigma^2\equiv \mu^2 n_c F_c(0)$. We can we can write Eq.~\eqref{eq:SBR_Fourier_space} in direct space as 
    \begin{equation}
        s(\boldsymbol{x}) + \sigma = -\Gamma \nabla^2 g_s(\boldsymbol{x};\hat{t}) - \lambda g_s(\boldsymbol{x};\hat{t})^2 + \frac{\mathrm{d}}{\mathrm{d}\hat{t}}\int_0^{\hat{t}}\mathrm{d}\tau g_s(\boldsymbol{x};\hat{t}-\tau)g_s(\boldsymbol{x};\tau).
    \label{eq:kinetic_SBR}
    \end{equation}
In this representation, the Gaussian process $s(\boldsymbol{x})$ satisfies $\llbracket s(\boldsymbol{x}) \rrbracket = 0$ and $\llbracket s(\boldsymbol{x}) s(\boldsymbol{x}')\rrbracket = \Delta\sigma^2 \delta(\boldsymbol{x}-\boldsymbol{x}')$. 
\end{widetext}

Remarkably, we recognize in Eq.~\eqref{eq:kinetic_SBR} the equation of stochastic beta-relaxation (SBR) initially derived using a combination of replica and supersymmetric field-theoretic techniques. This equation is the second important result of the present work. We emphasize that all phenomenological constants that parametrize Eq.~\eqref{eq:kinetic_SBR} can be computed from liquid structure alone, through Eqs.~\eqref{eq:MCT_exponent_parameter}, \eqref{eq:prefactor_noise_SBR} and \eqref{eq:Gamma_def_Qtensor}. We have evaluated these constants for the canonical hard-sphere system in the Percus-Yevick approximation and report $\lambda = 0.735$, $\Gamma = 0.0708$ and $\Delta\sigma^2 = 0.0045$.\footnote{Note that the value of the exponent parameter $\lambda$ is well known, and the value of $\Gamma$ was computed microscopically in Refs.~\cite{biroli2006inhomogeneous,szamel2013breakdown}.} These values are in line with a recent study where SBR was fitted to simulation data \cite{laudicina2025nonmonotonic}. 

\section{Properties of the Effective Theory}\label{sec:properties_SBR}
Having established an appropriate effective theory for the dynamics of liquids in the $\beta$-regime, we next discuss properties of the solutions to the theory.

\subsection{Destabilization of the Mean-Field Critical Point}
Taking the long-time limit of Eq.~\eqref{eq:kinetic_SBR} and writing $g_s(\boldsymbol{x}) \equiv \lim_{t\to\infty}g_s(\boldsymbol{x},t)$ leads to the stationary equation
    \begin{equation}
        \sigma + s(\boldsymbol{x}) = - \Gamma \nabla^2 g_s(\boldsymbol{x}) + (1-\lambda) g_s(\boldsymbol{x})^2.
    \label{eq:stationary_SBR}
    \end{equation}
Equation \eqref{eq:stationary_SBR} can be written as the Euler–Lagrange condition of a Lyapunov functional:
        \begin{eqnarray}
        \label{eq:lyapunov_SBR} 
        \mathfrak{F}[g_s] = \int \mathrm{d}\boldsymbol{x} \bigg( \frac{\Gamma}{2}(\nabla g_s(\boldsymbol{x}))^2 +  
        \frac{(1-\lambda)}{3}g_s(\boldsymbol{x})^3 
         \nonumber \\ -  (\sigma + s(\boldsymbol{x})) g_s(\boldsymbol{x})\bigg).
    \end{eqnarray}
This representation makes it immediately clear that the functional does not possess a finite global minimizer. Recall that at the mean-field glass transition the coefficient $(1-\lambda)>0$ is positive \cite{gotze1985properties}. Consequently, the coefficient of the cubic term in the Lapunov functional Eq.~\eqref{eq:lyapunov_SBR} is positive, and thus the solutions to Eq.~\eqref{eq:stationary_SBR} systematically tend towards $g_s(\boldsymbol{x})\to-\infty$, signalling a departure from the plateau and the final structural relaxation. The random field $s(\boldsymbol{x})$ enters only linearly in $g_s(\boldsymbol{x},t)$, and therefore affects the dynamics merely transiently. It locally accelerates or delays the decay of $g_s(\boldsymbol{x},t)$ regions with $\sigma+s(\boldsymbol{x})<0$ relax more rapidly, while positive fluctuations slow down the decay. Nevertheless, such spatial heterogeneities cannot stabilize any nontrivial stationary profile, since the positive cubic term enforces a net drift of $g_s(\boldsymbol{x},t)$ towards $-\infty$, ultimately driving the system away from the plateau and into structural relaxation at sufficiently large times. 

Interestingly, and as noted in Ref.~\cite{franz2011field}, Eq.~\eqref{eq:lyapunov_SBR} is formally equivalent to the action describing the spinodal of the random-field Ising model. In recent years, strong connections have been drawn between the spinodal of the RFIM and effective theories describing the behavior of structural glasses \cite{franz2013static, biroli2014random, nandi2014critical, biroli2018random, biroli2018randomii}. The present dynamical derivation opens the door for further exploration of this intriguing connection.

\begin{figure}
    \centering
    \includegraphics[width=0.8\linewidth]{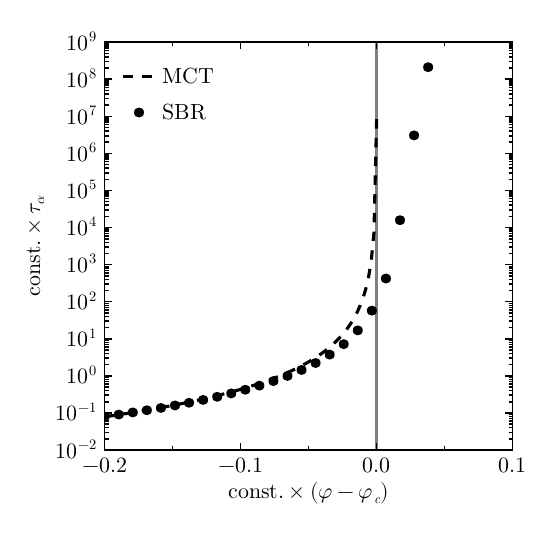} 
    \caption{Structural relaxation time $\tau_\alpha$ as a function of packing fraction for Percus-Yevick hard spheres of unit diameter, obtained by numerically solving the SBR equations beyond the mean-field MCT transition. For each packing fraction, 10 independent noise realizations were generated on a cubic grid of $40^3$ sites with lattice spacing equal to the hard sphere diamater, $\Delta x = 1$. In contrast to MCT, $\tau_\alpha$ remains finite at all packing fractions, with the mean-field divergence replaced by a smooth crossover. The sample-to-sample stastistical variation of $\tau_{\alpha}$ is smaller than the scatter size, and thus not shown.
    }
    \label{fig:relax_time_PY_HS_SBR}
\end{figure}

To illustrate that ergodicity is restored, we numerically solve the SBR equations using the parameters computed for Percus-Yevick hard-spheres. The results are shown in Fig.~\ref{fig:relax_time_PY_HS_SBR}. Following standard mode-coupling analyses, we define the structural relaxation time as the point at which the deviations from the plateau become significant, i.e. when $F(k,\tau_{\alpha}) - F_c(k) = H(k)\llbracket g_s(\boldsymbol{x}, \tau_{\alpha}) \rrbracket \sim \mathcal{O}(1)$. The signs of the crossover are now evident: rather than diverging at $\varphi_c$, $\tau_{\alpha}$ remains finite for all packing fractions, growing steeply but smoothly through the mean-field transition region. The putative MCT singularity is thus replaced by a rapid but continuous crossover, consistent with the absence of a true dynamical arrest in simulations and experiments.

\subsection{Scaling Laws Beyond Mean-Field}
We now examine whether the characteristic mean-field scaling laws survive once quenched fluctuations are included. The starting point is the SBR equation in direct space Eq.~\eqref{eq:kinetic_SBR}. To proceed analytically, we adopt a separation-of-variables ansatz, $g_s(\boldsymbol{x}; t) = B_s(\boldsymbol{x})g_{\mathrm{scal.}}(t)$ where $g_{\mathrm{scal.}}(t)$ carries the temporal scaling and is assumed, to leading order, to be independent of the stochastic contributions. This allows us to check explicitly whether the mean-field asymptotics are retained.

\paragraph*{The first scaling law.} Suppose $g_{\mathrm{scal.}}(t)\propto \hat{t}^{-a}$, whose Laplace transform is $g_{\mathrm{scal.}}(\hat{z}) \propto \hat{z}^{a-1}\Gamma(1-a)$. At high frequencies (short times), the gradient term is subleading and one has 
    \begin{equation}
        B_s(\boldsymbol{x})^2 \left(\lambda\Gamma(1-2a)-\Gamma(1-a)^2 \right) = 0
    \end{equation}
which is solved by $\lambda = \Gamma(1-a)^2/\Gamma(1-2a)$. This is precisely the first MCT scaling law. At lower frequencies the solution acquires corrections of order $z^a$, which mix with the leading $z^{2a}$ terms and complicate a direct analytic treatment. These, however, do not directly affect the leading scaling.

\paragraph*{The second scaling law.} Now suppose $g_{\mathrm{scal.}}(t)\propto -\hat{t}^{b}$, with Laplace transform $g_{\mathrm{scal.}}(\hat{z})\propto -\hat{z}^{-b-1}\Gamma(1+b)$. In the long-time (low-frequency) limit the gradient term is again subleading, leaving
    \begin{equation}
        B_s(\boldsymbol{x})^2 \left(\lambda\Gamma(1+2b)-\Gamma(1+b)^2 \right) = 0
    \end{equation}
with solution $\lambda = \Gamma(1+b)^2/\Gamma(1+2b)$. This reproduces the second MCT scaling law. Subleading corrections appear at higher frequencies as terms of order $\hat{z}^{-b}$.

We thus recover both asymptotic scaling regimes, $g_{\mathrm{scal.}}(\hat{t})\sim \hat{t}^{-a}$ and $g_{\mathrm{scal.}}(\hat{t})\sim -\hat{t}^b$, together with the algebraic relations linking the exponents to the exponent parameter $\lambda$. These relations are therefore robust against critical fluctuations. This conclusion is consistent with earlier claims, and more generally the scaling relations are expected to persist under arbitrary loop corrections in the field-theoretic formulation of MCT \cite{andreanov2009mode}. In this sense, the temporal scaling laws are universal in the limits $\hat{z}\gg1$ and $\hat{z}\ll1$. SBR however produces strong corrections for $\hat{z}\sim \mathcal{O}(1)$, i.e. the critical $\beta$-regime. 

\paragraph*{Diverging timescale.} Having established that quenched fluctuations do not modify the relations determining the dynamical exponents $a$ and $b$, we now examine their effect on the diverging timescales associated with the critical scenario. The analysis is essentially local. At each spatial point, we may define a local effective separation parameter $\sigma_{\mathrm{eff.}}(\boldsymbol{x}) = \sigma + s(\boldsymbol{x})$. Since the diverging timescales only come from the temporal sector of the equations, and this sector is untouched by the presence of the quenched field, SBR preserves the mean-field divergences at least at a local level. We therefore find that the characteristic timescale $t_u(\sigma)$ introduced in Eq.~\eqref{eq:generalized_MCT_ansatz} scales as $t_u(\sigma) \propto |\sigma_{\mathrm{eff.}}(\boldsymbol{x})|^{-1/2a}$. 

\subsection{Interpretation of the Equation}

The SBR equation should be understood as an asymptotic effective theory that captures the effect of critical fluctuations beyond mean field in systems close to a dynamical transition of the mode-coupling type. Its range of validity is restricted to the vicinity of the avoided critical point, where $\varepsilon \ll 1$. Moreover, it is asymptotic in the spatial sense: the derivation retains only the leading long-wavelength contributions, represented by the isotropic Laplacian term. Continuing the expansion further, additional gradient terms appear, as well as higher-order nonlinearities (\textit{e.g.}\ cubic terms in $g_u$). These contributions are neglected in the present formulation because they are irrelevant in the asymptotic scaling regime of the singularity considered in this work. In this sense, SBR provides a minimal description for the inclusion of critical fluctuations around the mean-field dynamical glass transition.

The physical content of the theory is most clearly seen through the intermediate scattering function  $F(k;t)$ in the $\beta$-regime. For a given correlation function $F(k;t)$, the factorization theorem gives
    \begin{equation}
        F(k;t) = F_p(k) + H(k)G(t)
    \end{equation}
where $F_p(k)$ is the DW factor. This factorization has been extensively tested and is known to hold in the regime where MCT applies, with both $F_p(k)$ and the amplitudes $H(k)$ typically showing excellent agreement with data in a mode-coupling setting \cite{nauroth1997quantitative}. The universal scaling function $G(t)$ is then obtained as the 
disorder average of the SBR solution     
\begin{equation}
        G(t) = \llbracket g_s(\boldsymbol{x},t)\rrbracket.
\end{equation}
Furthermore, SBR suggests the existence of a self-consistent disorder in structural glasses that remains quenched and controls the relaxation in the $\beta$-regime. This provides a bridge to spin-glass theories: whereas in spin glasses, quenched disorder enters through random couplings or fields, in structural glasses the analogous disorder is \textit{self-induced}. The latter emerges from the system's own configuration and remains effectively frozen on $\beta$-relaxation timescales. The two systems therefore exhibit an effective equivalence over these timescales. 

\section{Discussion \& Outlook} \label{sec:conclusions}
\begin{figure}
    \centering
    \includegraphics[width=0.66\linewidth]{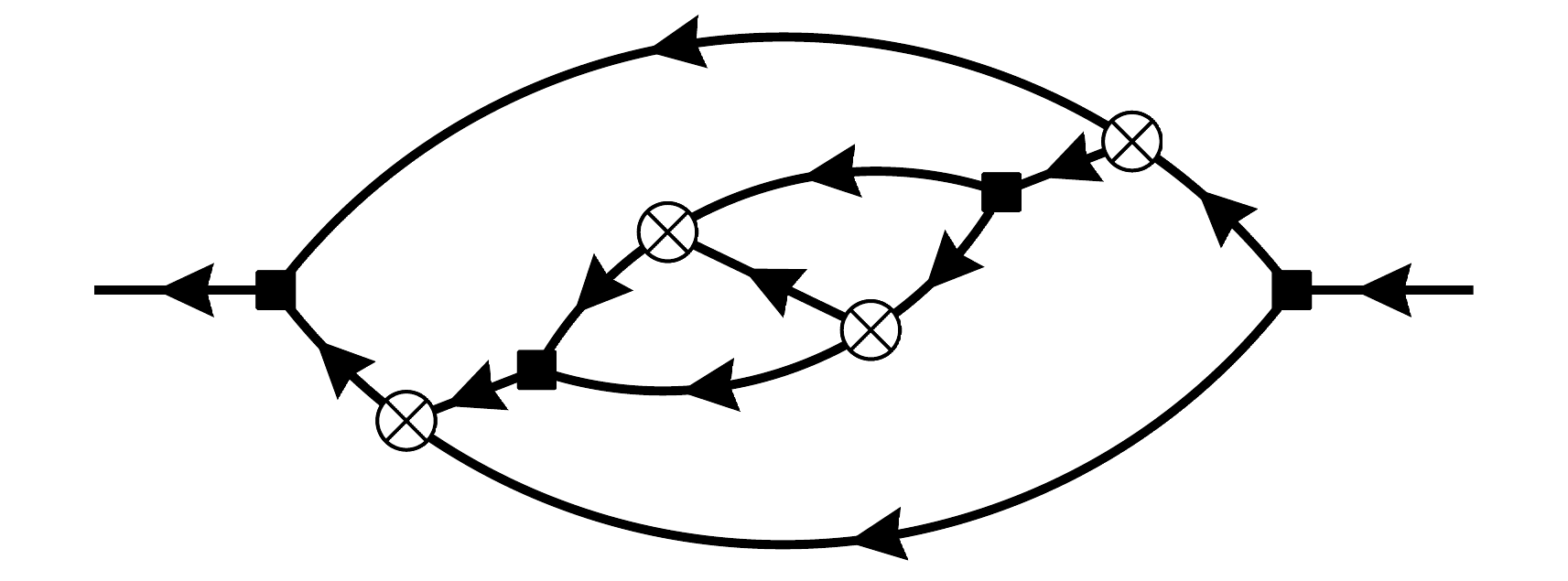}
    \caption{Example diagram that would be resummed if the stochastic process was completely self-consistent [\textit{i.e.}\ $\llbracket u^+_{\boldsymbol{q}}(t) u^-_{\boldsymbol{q}'}(t') \rrbracket = F(q,t-t')(2\pi)^d\delta(\boldsymbol{q}-\boldsymbol{q}')\Theta(t-t')$]. Perturbatively around MCT, this diagram scale as $\sim \varepsilon^{(d-4)/2}$ in the long-wavelength limit. Similar diagrams can be obtained by replacing the internal insertion with other diverging terms such as $A^{(2)}_2$ or $A^{(2)}_3$ from Fig.~\ref{fig:diagrammatic_most_diverging_2nd_4th_order}(a)}
    \label{fig:sample_diagram_not_generated}
\end{figure}
Starting from a diagrammatic formulation of the celebrated mode-coupling theory (MCT), we have analyzed the critical fluctuations that accompany the dynamical transition into a dynamically arrested state. At the diagrammatic level, we identified a class of critical fluctuation-dominated diagrams that contribute to the propagator. We studied the properties of this class of diagrammatic contributions in the vicinity of the transition. This allowed us to confirm the upper critical dimension of the theory, $d_c = 8$; a result consistent with the standard Ginzburg criterion and with prior works. To explicitly resum these dominant contributions, we introduced an effective stochastic process that asymptotically describes the fluctuation-dominated regime, \textit{i.e.}\ when critical fluctuations control the dynamics. Specifically, we have argued that the perturbative solution of the stochastic process reproduces, order by order, the diagrammatic contributions that we identified. Interestingly, the stochastic process is one with annealed disorder which is self-consistently determined by the mean-field order parameter. This naturally embodies the notion of dynamically self-induced disorder in structural glasses \cite{berthier2021self}.

We then analyzed the stochastic process in the $\beta$-regime of relaxation, where the fluctuation-dominated diagrams are the most dominant. This allowed us to show that the mean-field transition is completely destabilized for $d<d_c$ solely due to the inclusion of critical fluctuations. Furthermore, the analysis of the (pseudo-)critical dynamics provides a derivation of stochastic beta-relaxation (SBR) theory from microscopic dynamical considerations. Traditionally, SBR is derived from a combination of replica and supersymmetric field-theoretical techniques \cite{rizzo2013supercooled, rizzo2015qualitative, rizzo2016dynamical, rizzo2020solvable}. Here we demonstrated that it also emerges from the microscopic dynamics through a specific diagrammatic resummation, thereby providing a unification of static and dynamical approaches to the glass transition beyond mean-field. A further strength of our formulation is that all coupling constants of the effective theory can be expressed directly in terms of liquid structural correlations and mean-field quantities such as the non-ergodicity parameter. We have computed them for the paradigmatic Percus-Yevick hard-sphere system. This enables parameter-free predictions beyond mean-field theory.  Importantly, our work demonstrates that the observation of a dynamical crossover results solely from the self-consistent treatment of critical fluctuations, without invoking `activated processes'.

Despite these advances, several limitations of the present framework should be noted. The theory does not address the known shortcomings of MCT in the limit of infinite spatial dimensions, where dynamical mean-field theory provides the correct microscopic description \cite{maimbourg2016solution}. Likewise, while the fluctuation mechanism identified here applies for $d<8$, the situation for $d>8$ remains less clear: the corrections captured by Eq.~\eqref{eq:asymptotic_series} become regular, restoring the mean-field critical scenario, yet simulations continue to observe a dynamical crossover \cite{charbonneau2022dimensional, charbonneau2024dynamics}. We note further that our analysis is restricted to the asymptotic $\beta$-regime around the avoided critical point; the long-time $\alpha$-regime, likely dominated by non-perturbative processes such as `activated hopping' \cite{charbonneau2014hopping} (also sometimes referred to as instantons \cite{charbonneau2025rare}) and dynamic facilitation \cite{scalliet2022thirty}, lies beyond the reach of the present treatment. Clarifying these aspects, and particularly the microscopic mechanisms of the crossover in high dimensions and the non-perturbative dynamics in the $\alpha$-regime, remains an important direction for future work.

A related question concerns the completeness of the resummation. The present scheme resums \emph{all} diagrams arising from dressing the MCT skeleton diagrams with three-point susceptibilities associated with critical fluctuations of the order parameter. There are, however, other classes of divergent diagrams. A natural candidate arises from promoting Eq.~\eqref{eq:noise_noise_correlator} to a fully self-consistent treatment, \textit{i.e.}\, from replacing $F_{\mathrm{MCT}}(q;t)$ by the full intermediate scattering function $F(q;t)$ in the noise correlator. Diagrammatically, this corresponds to not only dressing MCT diagrams with susceptibilities but also replacing the connections between these susceptibilities by the full $F(q;t)$. An example diagram of this type is shown in Fig.~\ref{fig:sample_diagram_not_generated}. Preliminary calculations indicate that including such diagrams leads to the same effective theory, but with renormalized coupling constants. Whether these constants can be computed in closed form remains an open question. 

Another natural direction for future research is to extend the present analysis to higher-order glass-transition singularities predicted by MCT when more than one control parameter is involved. For example, \citet{nandi2014critical} have analyzed mean-field critical fluctuations near $A_3$-type critical points \cite{gotze1989logarithmic}. In agreement with their results, we find that the dominant contributions to four-point functions scale as $\varepsilon^{(d-4)/3}$ close to the transition, which implies an upper critical dimension of $d_c = 6$ from a Ginzburg-type argument given that the order parameter scales as $H(k ;t ; \varepsilon) \propto \varepsilon^{1/3}$ there. By analogy, we expect that the corresponding scaling equation for the fluctuating field $g_s(\boldsymbol{x},t)$ near an $A_3$ singularity could be derived along the same lines as in the present work, provided that higher-order terms in the expansion are retained. In such a scenario, the characteristic logarithmic decay of correlation functions should persist, much like the von-Schweidler law is preserved by the current calculations. 

Finally, we would like to emphasize the difference between the present approach and another extension of MCT that some of us developed earlier and that became known as the generalized mode-coupling theory (GMCT) \cite{szamel2003colloidal, wu2005high, janssen2015microscopic}. The latter theory relies upon a Mori continued fraction-like expression of the memory kernel to delay the factorization approximation of the original MCT. However, GMCT, in its current form, still includes only wavevector-diagonal contributions to higher-order functions, leaving the overall mean-field scenario unchanged \cite{luo2020generalized, laudicina2022dynamical}. In contrast, the present approach explicitly includes the wavevector-off-diagonal contributions that encode critical fluctuations of the order parameter. The two approaches are therefore diagrammatically distinct.

In summary, we provided here a unified framework that consistently extends MCT beyond the mean field picture and established a firm microscopic basis for stochastic beta-relaxation as an effective theory for supercooled liquids around the dynamical crossover point. While many open questions remain, the present work sets the stage for systematic, first-principles-based investigations of fluctuation effects in glassy dynamics beyond mean-field.

\section*{Acknowledgements} 

CCLL and LMCJ acknowledge the Dutch Research Council (NWO) for financial support through a Vidi grant. GS acknowledges the support of NSF Grant No. CHE 2154241. CCLL also warmly thanks CSU Chemistry for their hospitality during part of this work. We are indebted to Tommaso Rizzo for useful discussions.
\bibliography{apssamp}

\clearpage
\appendix

\begin{widetext}

\section{Detailed Account of the Mean-Field Scenario} \label{sec:mean_field_scenario}

In this section, we discuss how the MCT of the glass transition is obtained by the self-consistent ring approximation to the memory function $M(k;t)$, and discuss the critical properties of the theory.  

\subsection{Asymptotic Dynamics Around the Transition}\label{app:asymptotics_MCT}

Let us assume that there is a critical density $n_c$ at which a stable glass phase emerges. Observables evaluated at the mean-field critical point will be denoted by a subscript $\cdot\cdot\cdot_c$. We are interested in studying the dynamics in the vicinity of the critical point. We consider a small perturbation around the critical density $n = n_c(1 + \varepsilon)$, with $|\varepsilon|\ll 1$. Since we know that in the supercooled regime, $F(k;t)$ exhibits a two-step decay through a long-lived plateau, we write 
    \begin{equation}
        F(k;t) = F_c(k) + H(k; \hat{t} ; \varepsilon).
    \label{eq:MCT_asymptotic_expansion1}
    \end{equation}
with $F_c(k)$ the DW factor at the critical point. 

A key assumption is that the deviations from the plateau are small, and thus we have that $H(k; t ; \varepsilon)^{l+1} \ll H(k; t ; \varepsilon)^l$ for $l\geq 0$ \cite{gotze1989beta}.
It is convenient to proceed in Laplace space, where Eq.~\eqref{eq:MCT} reads 
    \begin{equation}
        F(k;z) = \frac{S(k)}{z + \tfrac{D_0k^2}{S(k)}(1+M^{\mathrm{irr}}_{\mathrm{MCT}}(k;z))^{-1}}
    \label{eq:MCT_freq_space}
    \end{equation}
which can be recast as 
    \begin{equation}
        0 = zF(k;z) - S(k) + D_0k \left[ zM_{\mathrm{MCT}}^{\mathrm{irr}}(k;z)\right]^{-1} \frac{k}{S(k)}zF(k;z)
    \label{eq:MCT_low_freq}
    \end{equation}
for small frequencies $z$ (\textit{i.e.} large-time physics). In deriving Eq.~\eqref{eq:MCT_low_freq}, we have assumed that $M_{\mathrm{MCT}}^{\mathrm{irr}}(k;z) \gg 1$ for $z\ll 1$, consistent with the expectation that the memory kernel admits a decomposition analogous to Eq.~\eqref{eq:MCT_asymptotic_expansion1}, with relaxation proceeding via a long-lived plateau. Under this assumption, the subsequent calculations remain valid only up to an overall microscopic timescale, which must be fixed by matching to the short-time behavior.  

Substituting the ansatz, Eq.~\eqref{eq:MCT_asymptotic_expansion1}, into Eq.~\eqref{eq:MCT_low_freq}, one obtains two coupled equations corresponding to the frequency-independent and frequency-dependent contributions. The static part yields the self-consistency condition that fixes the critical Debye–Waller factor,
    \begin{equation}
        \frac{F_c(k)}{S(k) - F_c(k)} = m_{c\mathrm{MCT}}^{\mathrm{irr}}(k)
    \label{eq:MCT_EOS}
    \end{equation}
where $m_{c\mathrm{MCT}}^{\mathrm{irr}}(k) = \left(D_0k^2/S(k) \right)^{-1}\lim_{t\rightarrow\infty} M_{c\mathrm{MCT}}^{\mathrm{irr}}(k, t)$ evaluated self-consistently at the critical density. The frequency-dependent contributions govern the critical dynamics in the vicinity of the ergodic–nonergodic transition, and are described by 
    \begin{equation}
        \int \frac{\mathrm{d}\boldsymbol{p}}{(2\pi)^d}\big[ (2\pi)^d\delta(\boldsymbol{k}-\boldsymbol{p}) - C^{(1)}(\boldsymbol{k}, \boldsymbol{p})\big] \frac{zH(p; z ; \varepsilon)}{S(p)}  = I(k;z ; \varepsilon)
    \label{eq:asymptotic_expansion_MCT_ergodic}
    \end{equation}
in which we have introduced the stability operator $C^{(1)}(\boldsymbol{k}, \boldsymbol{p})$ (\textit{i.e.}\ the Jacobian of the memory kernel) defined as 
    \begin{equation}
    \begin{split}
        C^{(1)}(\boldsymbol{k}, \boldsymbol{k}') \equiv&\ nS(k) (1-f_c(k))^2 \tilde{v}_{\boldsymbol{k}}(\boldsymbol{k}', \boldsymbol{k}-\boldsymbol{k}')^2 S(|\boldsymbol{k}-\boldsymbol{k}'|)f_c(|\boldsymbol{k}-\boldsymbol{k}'|) S(k'). 
    \end{split}
    \label{eq:MCT_stab_matrix_3D}
    \end{equation}
In the language of critical phenomena, $(\mathrm{Id.}-C^{(1)})$ in Eq.~\eqref{eq:asymptotic_expansion_MCT_ergodic} plays a role equivalent to that of a mass operator in an effective action, with the mode-coupling transition marked by the vanishing of its largest eigenvalue, as we will see below.

The right-hand side of Eq.~\eqref{eq:asymptotic_expansion_MCT_ergodic} can be expanded in powers of the small parameter $\varepsilon$ and of the small corrections $H$, yielding an $\varepsilon$-independent term, quadratic terms in $H$, and higher-order corrections:
    \begin{equation}
        I(k;z ; \varepsilon) = I^{(0)}(k; \varepsilon) + I^{(1)}(k;z ; \varepsilon) + \mathcal{O}(\varepsilon^2, \varepsilon H, H^3).
    \end{equation}
The $H$-independent contribution, $I^{(0)}(k; \varepsilon)$ reflects the variation of the mode-coupling vertices upon changing control parameters (\textit{i.e.}\ $\frac{\mathrm{d}}{\mathrm{d}n} [n\tilde{v}_{\boldsymbol{k}}(\boldsymbol{p}, \boldsymbol{k}-\boldsymbol{p})^2]$ in the case of the density). The next term, $I^{(1)}(k;z;\varepsilon)$, contains all contributions quadratic in $H$ and reads
    \begin{equation}
    \begin{split}
        I^{(1)}(k;z ; \varepsilon) =&\ -\frac{1}{(1-f_c(k))} \left[ \int \frac{\mathrm{d}\boldsymbol{p}}{(2\pi)^d} C^{(1)}(\boldsymbol{k}, \boldsymbol{p})\frac{zH(p; z ; \varepsilon)}{S(p)}\right]^2 + \int \frac{\mathrm{d}\boldsymbol{p}}{(2\pi)^d} C^{(1)}(\boldsymbol{k}, \boldsymbol{p}) \frac{zH(p; z ; \varepsilon)}{S(p)}\frac{zH(k; z ; \varepsilon)}{S(k)} \\
        & + \int \frac{\mathrm{d}\boldsymbol{p}}{(2\pi)^d}C^{(2)}(\boldsymbol{k}, \boldsymbol{p}) \frac{z\mathcal{L}\{ H(p; t ; \varepsilon) H(|\boldsymbol{k}-\boldsymbol{p}|; t ; \varepsilon)  \}(z)}{S(p)S(|\boldsymbol{k}-\boldsymbol{p}|)}.
    \end{split}
    \end{equation}
The precise form of the kernel $C^{(2)}(\boldsymbol{k}, \boldsymbol{p})$ is unimportant for the critical scenario.

Next, we expand the corrections as a power series in $\varepsilon$, giving
    \begin{equation}
        H(k; t ; \varepsilon) = \sum_{l=1}^{\infty} r(\varepsilon)^l S(k)h_{\pm}^{(l)}(k)g_{\pm}^{(l)}(\hat{t}),
    \label{eq:ergodic_MCT_ansatz}
    \end{equation}
where $\hat{t} \equiv t/t_*(\varepsilon)$ is a rescaled time variable by $t_*(\varepsilon)$ corresponding to the critical timescale associated with the relaxation near the transition. It must be determined self-consistently at the end of the computation. The ``$\pm$" notation refers to the sign of the perturbation in $\varepsilon$. Equivalently in frequency space
    \begin{equation}
        zH(k; z ; \varepsilon) = \sum_{l=1}^{\infty} r(\varepsilon)^l S(k)h_{\pm}^{(l)}(k) \hat{z}g_{\pm}^{(l)}(\hat{z}),
    \label{eq:ergodic_MCT_ansatz_frequency}
    \end{equation}
where $\hat{z} = z t_*(\varepsilon)$ denotes the corresponding rescaled frequency. The monomial $r(\varepsilon)$ serves as a small expansion parameter controlling the perturbative series in Eq.~\eqref{eq:ergodic_MCT_ansatz}. Note also that we have factorized wavevector and temporal dependences in the Ansatz Eq.~\eqref{eq:ergodic_MCT_ansatz}. This is known as the ``factorization theorem" in the MCT literature \cite{gotze1999recent}. \citet{gotze1985properties} and more recently \citet{andreanov2009mode} have argued that $r(\varepsilon) = \sqrt{\varepsilon}$ is the only prescription physically relevant prescription. The other possibility, $r(\varepsilon) = \varepsilon$, would lead to an absence of structural relaxation in the dynamical calculation, even in the ergodic phase. Focusing then on the equation at leading order, one obtains the following eigenvalue problem
    \begin{equation}
        \int \frac{\mathrm{d}\boldsymbol{p}}{(2\pi)^d} \left[(2\pi)^d\delta(\boldsymbol{k}-\boldsymbol{p}) - C^{(1)}(\boldsymbol{k},\boldsymbol{p}) \right] h_{\pm}^{(1)}(p) = 0 + \mathcal{O}(\varepsilon)
    \label{eq:MCT_eigenproblem}
    \end{equation}
which only fixes the amplitudes $h_{\pm}^{(1)}(p)$, and not the frequency dependent function $g_{\pm}^{(1)}(\hat{z})$. The latter is fixed by the next-leading order terms, as we illustrate further below.

In order to obtain closed expressions for the amplitudes $h_{\pm}^{(1)}(k)$, let us denote the spectrum of the stability operator by $E_{\alpha}(\varepsilon)$, such that right and left eigenfunctions can be defined via
    \begin{equation}
        \int \frac{\mathrm{d}\boldsymbol{p}}{(2\pi)^d}\ C^{(1)}(\boldsymbol{k}, \boldsymbol{p})h_{\alpha}^{\mathrm{R}}(p) = E_{\alpha}(\varepsilon)h_{\alpha}^{\mathrm{R}}(k)
    \end{equation}
and
    \begin{equation}        
        \int \frac{\mathrm{d}\boldsymbol{p}}{(2\pi)^d}\ \left(\frac{h_{\alpha}^{\mathrm{L}}(p)}{p^{d-1}}\right)C^{(1)}(\boldsymbol{p}, \boldsymbol{k}) = \left(\frac{h_{\alpha}^{\mathrm{L}}(k)}{k^{d-1}}\right) E_{\alpha}(\varepsilon).
    \end{equation}
It is clear that the solution to Eq.~\eqref{eq:MCT_eigenproblem} requires $h_{\pm}^{(1)}(k)$ to be a right eigenfunction with eigenvalue one: $h_{\pm}^{(1)}(k) = h_0^{\mathrm{R}}(k)$. To guarantee stability, this eigenvalue must be the largest one, which we denote $E_0(\varepsilon)$. Consequently, we have $E_0(0)=1$ at the critical point. 

\paragraph*{Analysis at long times.} In the long-time limit, the solution to leading order determined above is known up to a multiplicative constant, which we determine next. Focusing on the behavior of Eq.~\eqref{eq:asymptotic_expansion_MCT_ergodic} in the zero frequency limit ($z\to0$) leads to an expression for $g_\pm\equiv \lim_{\hat{z}\to0} \hat{z} g_{\pm}^{(1)}(\hat{z})$: 
    \begin{equation}
        g_\pm^2 = \frac{\pm 1}{1-\lambda} \int \frac{\mathrm{d}\boldsymbol{k}}{s_d} \left(\frac{h_0^{\mathrm{L}}(k)}{k^{d-1}}\right) C^{(0)}(k)
    \label{eq:MCT_g_coef_nonergodic}
    \end{equation}
where $s_d$ denotes the surface area of a $d$-dimensional unit sphere and we have introduced the exponent parameter
    \begin{equation}
        \lambda \equiv \int \frac{\mathrm{d}\boldsymbol{k}}{s_d}\int \frac{\mathrm{d}\boldsymbol{p}}{(2\pi)^d} \left(\frac{h_0^{\mathrm{L}}(k)}{k^{d-1}}\right)  C^{(2)}(\boldsymbol{k}, \boldsymbol{p}) h_0^{\mathrm{R}}(|\boldsymbol{k}-\boldsymbol{p}|)h_0^{\mathrm{R}}(p)
    \label{eq:MCT_exponent_parameter}
    \end{equation}
of the mode-coupling theory. All the results above depend on the form of the kernels $C^{(0)}$ and $C^{(2)}$. We emphasise that they only fix system-specific constants. Within the mode-coupling theory, they are computable from pair-structure alone and are given by 
    \begin{equation}
    \begin{split}
        C^{(0)}(k) =&\ (1-f_c(k))^2\Bigg[\left(\pm1 + \frac{n_c}{S(k)}\frac{\mathrm{d}S(k)}{\mathrm{d}n} \bigg\vert_{n=n_c}\right)m_c^{\mathrm{irr.}}(k) \\
        &\ \hspace{2cm} + n_cS(k)\int\frac{\mathrm{d}\boldsymbol{p}}{(2\pi)^d} \tilde{v}^c_{\boldsymbol{k}}(\boldsymbol{p}, \boldsymbol{k}-\boldsymbol{p})\delta\tilde{v}_{\boldsymbol{k}}(\boldsymbol{p}, \boldsymbol{k}-\boldsymbol{p})F_c(p)F_c(|\boldsymbol{k}-\boldsymbol{p}|)\Bigg]
    \end{split}
    \end{equation}
with
    \begin{equation}
    \delta\tilde{v}_{\boldsymbol{k}}(\boldsymbol{p}, \boldsymbol{p}') = k^{-1}\hat{\boldsymbol{k}}\cdot \left(\boldsymbol{p}\frac{\mathrm{d}c(p)}{\mathrm{d}n}\bigg\vert_{n=n_c} + \boldsymbol{p}'\frac{\mathrm{d}c(p')}{\mathrm{d}n}\bigg\vert_{n=n_c} \right)
    \end{equation}
and
    \begin{equation}
        C^{(2)}(\boldsymbol{k}, \boldsymbol{k}') = \frac{n_cS(k)}{2}(1-f_c(k))^2 \tilde{v}_{\boldsymbol{k}}(\boldsymbol{k}-\boldsymbol{k}', \boldsymbol{k}')^2 S(|\boldsymbol{k}-\boldsymbol{k}'|)S(k').
    \label{eq:def_C2_MF}
    \end{equation}
We note that in order to obtain the results Eqs.~\eqref{eq:MCT_g_coef_nonergodic}-\eqref{eq:MCT_exponent_parameter}, the following normalizations for the eigenfunctions were fixed \cite{gotze1985properties}
    \begin{equation}
        \int \frac{\mathrm{d}\boldsymbol{k}}{s_d} \left( \frac{h_0^{\mathrm{L}}(k)}{k^{d-1}}\right) h_0^{\mathrm{R}}(k) = 1
    \label{eq:normalization1}
    \end{equation}
and
    \begin{equation}
        \int \frac{\mathrm{d}\boldsymbol{k}}{s_d}\left(\frac{h_0^{\mathrm{L}}(k)}{k^{d-1}}\right)\frac{h_0^{\mathrm{R}}(k)^2}{(1-f_c(k))} = 1.
    \label{eq:normalization2}
    \end{equation}
    
From Eq.~\eqref{eq:MCT_g_coef_nonergodic}, we see that the only physical solution for $g$ is obtained if the positive branch is taken, which corresponds to $\varepsilon$ being positive. This is of course a reasonable result, as for a negative perturbation of the critical density, we expect to be below the transition and therefore in the ergodic regime where the correlation function decays to zero [\textit{i.e.}\ $F(k,t\to\infty) =0$]. In this case, the temporal deviations $g_-^{(1)}(\hat{t})$ should not bounded at large times. 

The linear stability of Eq.~\eqref{eq:MCT_eigenproblem} with respect to $\varepsilon$ (in the case $\varepsilon>0$) can be used to determine the behavior of the critical eigenvalue in the non-ergodic phase
    \begin{equation}
        E_0(\varepsilon) = 1 - 2g(1-\lambda)\sqrt{\varepsilon} + \mathcal{O}(\varepsilon)
    \label{eq:critical_eigval_MCT}
    \end{equation}
where we henceforth denote the constant $g = g_+$. In contrast, this calculation cannot be repeated in the ergodic regime ($\varepsilon<0$). Accordingly, in this case, we assume that the critical eigenvalue $E_0(-\varepsilon)$ is independent of $\varepsilon$ to leading order (or, at minimum, has no leading $\varepsilon$-dependence).

\paragraph*{Analysis at finite times.} The corresponding scaling functions $g_{\pm}^{(1)}(\hat{z})$ are fixed by the inhomogeneity contributions $I(k;z ; \varepsilon)$ in Eq.~\eqref{eq:asymptotic_expansion_MCT_ergodic}. There are two cases to consider. When $\varepsilon>0$, that is, in the non-ergodic phase, we find that $g_+^{(1)}(\hat{z})$ satisfies a scale-invariant equation of form
    \begin{equation}
        2g(1-\lambda) g_+^{(1)}(\hat{z}) + \frac{1}{\hat{z}} = \lambda \mathcal{L}\{g_+^{(1)}(t)^2\}(\hat{z}) - \hat{z} g_+^{(1)}(\hat{z})^2
    \label{eq:gotze_beta_scaling_nonergodic_frequency}
    \end{equation}
which, in the time-domain is given by 
    \begin{equation}
    \begin{split}
        2g(1-\lambda)& g_+^{(1)}(\hat{t}) + 1 =\ \lambda g_+^{(1)}(\hat{t})^2 - \frac{\mathrm{d}}{\mathrm{d}\hat{t}} \int_0^{\hat{t}}\mathrm{d}\tau g_+^{(1)}(\hat{t}-\tau)g_+^{(1)}(\tau).        
    \end{split}
    \label{eq:gotze_beta_scaling_non_ergodic}
    \end{equation}
There are two asymptotic regimes to consider. When $\hat{t}\ll 1$, we find that $g_+^{(1)}(\hat{t}) \propto \hat{t}^{-a}$ is a solution with a condition on the exponent $a$ 
    \begin{equation}
        \lambda = \frac{\Gamma(1-a)^2}{\Gamma(1-2a)}.
    \end{equation}
We can now determine the critical timescale $t_*(\varepsilon)$. Consider the regime in which the deviation of the correlation function from its plateau value $F_c(k)$ becomes of order unity. Since the wavenumber-dependent amplitude is itself of order one, this implies $\sqrt{\varepsilon} g_+^{(1)}(\hat{t}) \sim 1$. Using the short-time asymptotic form $g_+^{(1)}(\hat{t})\sim \hat{t}^{-a}$, this condition can be written as $\sqrt{\varepsilon}(t_0/t_*(\varepsilon))^{-a} \sim 1$ where $t_0$ is a microscopic timescale that matches the short-time non-critical behavior. Rearranging yields $t_*(\varepsilon) = t_0\varepsilon^{-1/2a}$. Thus, the characteristic timescale $t_{*}(\varepsilon)$ diverges upon approaching the transition point. In the opposite limit $\hat{t}\gg 1$, the system crosses over to the non-ergodic (frozen) solution discussed previously. 

In the ergodic phase, we get instead 
    \begin{equation}
        \frac{1}{\hat{z}} = \lambda \mathcal{L}\{g_-^{(1)}(\hat{t})^2\}(\hat{z}) - \hat{z} g_-^{(1)}(\hat{z})^2,
    \label{eq:gotze_beta_scaling_ergodic_frequency}
    \end{equation}
which, in the time-domain transforms to 
    \begin{equation}
        1 = \lambda g_-^{(1)}(\hat{t})^2 - \frac{\mathrm{d}}{\mathrm{d}\hat{t}} \int_0^{\hat{t}}\mathrm{d}\tau g_-^{(1)}(\hat{t}-\tau)g_-^{(1)}(\tau).
    \label{eq:gotze_beta_scaling}
    \end{equation}
This is the celebrated G\"otze $\beta$-scaling equation. At short rescaled times ($\hat{t}\ll1$), the solution is again given by $g_-(\hat{t}) \propto \hat{t}^{-a}$, akin to its non-ergodic counterpart. In the opposite limit, at late times ($\hat{t}\gg1$), the onset of structural relaxation is described by another power-law, this time $g_-(\hat{t}) \propto -\hat{t}^{b}$ where the dynamical exponent $b$ satisfies 
    \begin{equation}
        \lambda = \frac{\Gamma(1+b)^2}{\Gamma(1+2b)}.
    \end{equation}
The critical timescale in the $\beta$-regime is the same as in the non-ergodic phase. This side of the transition also allows us to compute the critical timescale in the $\alpha$-regime. For this, we consider the condition $|F(k,\tau_{\alpha}) - F_c(k)|\sim\mathcal{O}(1)$ in the second scaling window, which gives the condition $\tau_{\alpha}(\varepsilon) \sim \varepsilon^{-\gamma}$ with $\gamma = 1/2a + 1/2b$ following similar arguments as for the critical $\beta$-regime. Thus, on both sides of the transition, the theory predicts two intrinsic diverging timescales at $\varepsilon=0$ (\textit{i.e.}\ critical slowing down of the $\alpha$- and $\beta$-regime). Importantly, the complete scaling behavior for the critical scenario is governed by the sole exponent parameter $\lambda$, computable from liquid structure alone through Eq.~\eqref{eq:MCT_exponent_parameter}. 

\subsection{Critical Fluctuations in the Mean-Field} \label{app:critical_fluct_asymptotics}

In the present diagrammatic formulation, the dominant contributions to four-point functions arises from the resummation of rainbow diagrams which are shown in Fig.~\ref{fig:rainbow_insertions}. This resummation leads to the definition of two distinct three-point functions from the two types of three-legged vertices present in the theory [left- and right-handed vertices Eqs.~\eqref{eq:bare_left_handed_vertex}-\eqref{eq:bare_right_handed_vertex}]. We denote these two distinct response functions $\chi_{\boldsymbol{q}}^\pm(\boldsymbol{k}; t; t')$ (+ for the right-handed and - for the left-handed one). Here, $(\boldsymbol{k},t)$ labels the underlying propagator, while $(\boldsymbol{q},t')$ denotes the external perturbation induced by the conjugate field. Writing down the resummations of the rainbow diagrams gives
    \begin{equation}
    \begin{split}
        &\chi_{\boldsymbol{q}}^\pm(\boldsymbol{k}; t; t') = \chi_{\boldsymbol{q}}^{(0\pm)}(\boldsymbol{k}; t ; t') + \int_{t'}^{t}\mathrm{d}\tau_4\int_{t'}^{\tau_4}\mathrm{d}\tau_3 \int_0^{t'}\mathrm{d}\tau_2\int_0^{\tau_2}\mathrm{d}\tau_1 \int \frac{\mathrm{d}\boldsymbol{p}}{(2\pi)^d} F(|\boldsymbol{k}\pm\frac{\boldsymbol{q}}{2}|, t-\tau_4)\tilde{v}^{\mathrm{reg}}_{12}(\boldsymbol{k}\pm\frac{\boldsymbol{q}}{2} ; \boldsymbol{k}-\boldsymbol{p}, \boldsymbol{p}\pm\frac{\boldsymbol{q}}{2} ; \tau_4-\tau_3)\\
        & \hspace{5cm} \times F(|\boldsymbol{k}-\boldsymbol{p}|, \tau_3-\tau_2)\chi_{\boldsymbol{q}}^{\pm}(\boldsymbol{p}, \tau_3-\tau_2 ; t')  \tilde{v}^{\mathrm{reg}}_{21}(\boldsymbol{k}-\boldsymbol{p}, \boldsymbol{p}\mp\frac{\boldsymbol{q}}{2} ; \boldsymbol{k}\mp\frac{\boldsymbol{q}}{2} ; \tau_2-\tau_1) F(|\boldsymbol{k}\mp\frac{\boldsymbol{q}}{2}|, \tau_1)
    \end{split}
    \label{eq:eom_susceptibilities_diagrammatic}
    \end{equation}
with bare sources given by 
    \begin{equation}
        \chi_{\boldsymbol{q}}^{(0+)}(\boldsymbol{k}; t ; t') = \int_{0}^t \mathrm{d}\tau F(|\boldsymbol{k}+\frac{\boldsymbol{q}}{2}|, t-\tau) \tilde{v}_{12}^{\mathrm{reg}}(\boldsymbol{k}+\frac{\boldsymbol{q}}{2} ; \boldsymbol{k}-\frac{\boldsymbol{q}}{2}, \boldsymbol{q} ; \tau-t')F(|\boldsymbol{k}-\frac{\boldsymbol{q}}{2}|,t')
    \label{eq:bare_source_Xq0+}
    \end{equation}
and 
    \begin{equation}
        \chi_{\boldsymbol{q}}^{(0-)}(\boldsymbol{k}; t ; t') = \int_0^{t'}\mathrm{d}\tau F(|\boldsymbol{k}-\frac{\boldsymbol{q}}{2}|, t-t')\tilde{v}_{21}^{\mathrm{reg}}(\boldsymbol{k}-\frac{\boldsymbol{q}}{2}, \boldsymbol{q} ; \boldsymbol{k}+\frac{\boldsymbol{q}}{2} ; t'-\tau)F(|\boldsymbol{k}+\frac{\boldsymbol{q}}{2}|, \tau).
    \label{eq:bare_source_Xq0-}
    \end{equation}
Note that we have defined $\tilde{v}_{12}^{\mathrm{reg}}(\boldsymbol{k} ; \boldsymbol{q}, \boldsymbol{q}' ; t) = k^{-1}v_{12}^{\mathrm{reg}}(\boldsymbol{k} ; \boldsymbol{q}, \boldsymbol{q}' ; t)$ and $\tilde{v}_{21}^{\mathrm{reg}}(\boldsymbol{q}, \boldsymbol{q}'; \boldsymbol{k} ; t) = k^{-1}v_{21}^{\mathrm{reg}}(\boldsymbol{q}, \boldsymbol{q}' ; \boldsymbol{k} ;  t)$ following the convention outlined in the main text. Since $\chi_{\boldsymbol{q}}^\pm(\boldsymbol{k}; t; t')$ depends explicitly on both the observation time $t$ and the perturbation time $t'$, it is convenient to first perform a one-sided Fourier transform with respect to $t'$:
    \begin{equation}
        \chi_{\boldsymbol{q}}^{\pm}(\boldsymbol{k}; t ; \omega) = \int_{0}^t \mathrm{d}t' e^{i\omega t'} \chi_{\boldsymbol{q}}^{\pm}(\boldsymbol{k}; t ; t'),
    \end{equation}
so that the limit $\omega \to 0$ corresponds to the time–integrated response. For the observation time $t$, we then apply a Laplace transform, in analogy with the treatment of the propagator in Sec.~\ref{app:asymptotics_MCT}. Expanding around the critical point with the ansatz of Eq.~\eqref{eq:MCT_asymptotic_expansion1}, we arrive at the following compact linear integral equation to leading order in $\varepsilon$:
    \begin{equation}
    \begin{split}
        \int \frac{\mathrm{d}\boldsymbol{p}}{(2\pi)^d} &\left[ (2\pi)^d\delta(\boldsymbol{k}-\boldsymbol{p}) - \mathcal{M}_{\boldsymbol{q}}(\boldsymbol{k}, \boldsymbol{p}) \right]\chi^{\pm}_{\boldsymbol{q}}(\boldsymbol{p};z ; \omega) = \chi_{\boldsymbol{q}}^{(0\pm)}(\boldsymbol{k}; \hat{z} ; \omega)        
    \end{split}
    \label{eq:dynamic_eigenvalue_susceptibility_app}
    \end{equation}
where the mass operator $\mathcal{M}_{\boldsymbol{q}}(\boldsymbol{k}, \boldsymbol{p})$ is given by 
    \begin{equation}
        \mathcal{M}_{\boldsymbol{q}}(\boldsymbol{k}, \boldsymbol{p}) = n S(|\boldsymbol{k}+\frac{\boldsymbol{q}}{2}|)\left(1-f(|\boldsymbol{k}+\frac{\boldsymbol{q}}{2}|)\right) \tilde{v}_{\boldsymbol{k}+\boldsymbol{q}/2}(\boldsymbol{k}-\boldsymbol{p}, \boldsymbol{p}+\frac{\boldsymbol{q}}{2}) F(|\boldsymbol{k}-\boldsymbol{p}|) \tilde{v}_{\boldsymbol{k}-\boldsymbol{q}/2}(\boldsymbol{k}-\boldsymbol{p}, \boldsymbol{p}-\frac{\boldsymbol{q}}{2}) \left(1-f(|\boldsymbol{k}-\frac{\boldsymbol{q}}{2}|)\right)S(|\boldsymbol{k}-\frac{\boldsymbol{q}}{2}|)
    \label{eq:mass_operator_stability}
    \end{equation}    
The dynamical equation for the integrated response function [\textit{e.g.}~ $\lim_{\omega\to 0}\chi^{\pm}_{\boldsymbol{q}}(\boldsymbol{p};z ; \omega)$] is analogous to that derived by \citet{biroli2006inhomogeneous} in the context of IMCT, except for differences in the source terms\footnote{The discrepancy arises because we consider the response to finite-time perturbations, whereas \citet{biroli2006inhomogeneous} assume a perturbation applied infinitely in the past, leading to non-commuting limits that prevent an exact correspondence.} $\chi_{\boldsymbol{q}}^{(0\pm)}(\boldsymbol{k};\hat{t} ; t')$ \cite{szamel2008divergent}. As indicated by the $\hat{t}$ notation, the susceptibility also inherits the diverging timescale $t_*(\varepsilon)$ associated with the critical $\beta$-regime. This can be seen through the bare source terms $\chi_{\boldsymbol{q}}^{(0\pm)}(\boldsymbol{k};\hat{t} ; t')$ which dependent explicitly on the temporal deviations $g^{(1)}_\pm(\hat{t})$ of the intermediate scattering function upon the expansion around the critical point.

\paragraph*{Static treatment.}
In order to investigate the long-time dynamics in the non-ergodic phase, we take the joint limit $\lim_{z,\omega\to0} \chi_{\boldsymbol{q}}^\pm(\boldsymbol{k};z ; \omega) \equiv \chi_{\boldsymbol{q}}^{\pm}(\boldsymbol{k})$ which gives the integral equation derived by one of us in Ref.~\cite{szamel2013breakdown} and presented in the companion paper:
    \begin{equation}
    \begin{split}
        \chi_{\boldsymbol{q}}^{\pm}(\boldsymbol{k}) =&\ \chi_{\boldsymbol{q}}^{(0\pm)}(\boldsymbol{k}) + \int \frac{\mathrm{d}\boldsymbol{p}}{(2\pi)^d}\ \mathcal{M}_{\boldsymbol{q}}(\boldsymbol{k}, \boldsymbol{p})\chi_{\boldsymbol{q}}^\pm(\boldsymbol{p}).
    \end{split}
    \label{eq:static_eigenvalue_susceptibility}
    \end{equation}
In this limit, the left-handed source $\lim_{\hat{z},\omega\to0} \chi_{\boldsymbol{q}}^{(0-)}(\boldsymbol{k}; \hat{z} ; \omega) \equiv \chi_{\boldsymbol{q}}^{(0-)}(\boldsymbol{k})$ reads (see Fig.~\ref{fig:rainbow_insertions} for momentum label convention)
    \begin{equation}
    \begin{split}
        \chi_{\boldsymbol{q}}^{(0-)}(\boldsymbol{k}) =&\ F(k_-) \frac{n\tilde{v}_{\boldsymbol{k}_+}(\boldsymbol{k}_-, \boldsymbol{q})}{m_{c\mathrm{MCT}}^{\mathrm{irr}}(k_+)} F(k_+).
    \end{split}
    \end{equation}
The $m_{c\mathrm{MCT}}^{\mathrm{irr}}(k)^{-1}$ contribution comes from the long-time limit of the regularized vertex $\lim_{t\to\infty}\mathcal{v}_{21}^{\mathrm{reg}}(\boldsymbol{k}-\boldsymbol{q}/2, \boldsymbol{q} ; \boldsymbol{k}+\boldsymbol{q}/2 ; t)$. Similarly, the right-handed source is given by $\lim_{\hat{z},\omega\to0} \chi_{\boldsymbol{q}}^{(0+)}(\boldsymbol{k}; \hat{z} ; \omega) \equiv \chi_{\boldsymbol{q}}^{(0+)}(\boldsymbol{k}) = \chi_{\boldsymbol{q}}^{(0-)}(\boldsymbol{k}) / n$. 

We recognize in the mass operator Eq.~\eqref{eq:mass_operator_stability} the form of the stability operator Eq.~\eqref{eq:MCT_stab_matrix_3D}. To this end, we introduce a generalized stability operator
    \begin{equation}
        C^{(1)}_{\boldsymbol{q}}(\boldsymbol{k}, \boldsymbol{p}) = S(|\boldsymbol{k}-\frac{\boldsymbol{q}}{2}|)\left(1-f(|\boldsymbol{k}-\frac{\boldsymbol{q}}{2}|)\right) \tilde{v}_{\boldsymbol{k}-\boldsymbol{q}/2}(\boldsymbol{k}-\boldsymbol{p}, \boldsymbol{p}-\frac{\boldsymbol{q}}{2}) F(|\boldsymbol{k}-\boldsymbol{p}|)\tilde{v}_{\boldsymbol{k}+\boldsymbol{q}/2}(\boldsymbol{k}-\boldsymbol{p}, \boldsymbol{p}+\frac{\boldsymbol{q}}{2})\left(1-f(|\boldsymbol{k}+\frac{\boldsymbol{q}}{2}|)\right) S(p)
    \label{eq:Generalized_stability_operator}
    \end{equation}    
which reduces to the standard stability operator $\lim_{\boldsymbol{q}\to\boldsymbol{0}}C^{(1)}_{\boldsymbol{q}}(\boldsymbol{k}, \boldsymbol{p}) = C^{(1)}(\boldsymbol{k}, \boldsymbol{p})$ [Eq.~\eqref{eq:MCT_stab_matrix_3D}] in the long wavelength limit.

From general theoretical considerations, it is natural to expect long-wavelength critical fluctuations to accompany the emergence of the non-ergodic state as the dynamical transition essentially corresponds to a spinodal instability \cite{berthier2020finite}. These fluctuations, if they exist, are encoded in the small-$q$ contributions to the spectrum of $C^{(1)}_{\boldsymbol{q}}(\boldsymbol{k}, \boldsymbol{p})$. For the rest of this section, we consider the analysis of the critical fluctuations around the mean-field, meaning that we should understand the propagators and other related functions to be taken in the mode-coupling approximation.

We analyze the spectrum of the generalized stability operator perturbatively in $\boldsymbol{q}$ around $C^{(1)}(\boldsymbol{k}, \boldsymbol{p})$. To leading order in $\varepsilon$ and $q$, this yields
    \begin{equation}
    \begin{split}
        \int \frac{\mathrm{d}\boldsymbol{p}}{(2\pi)^d} C^{(1)}_{\boldsymbol{q}}(\boldsymbol{k}, \boldsymbol{p}) h_0^{\mathrm{R}}(p) =\left[E_0(\varepsilon) - \Gamma q^2\right] h_0^{\mathrm{R}}(k),    
    \end{split}
    \label{eq:Generalized_stability_eigen}
    \end{equation}
where we remind the reader that $E_0(\varepsilon)$ is given by Eq.~\eqref{eq:critical_eigval_MCT} and $\Gamma$ is a coefficient that quantifies the leading correction due to finite $q$. The absence of linear terms in $q$ follows from rotational symmetry, and terms of order $\mathcal{O}(q^4)$ are neglected since we are ultimately interested in the long wavelength ($q\ll1$) regime. A microscopic expression for $\Gamma$ in terms of the structure and form factors is derived in Appendix \ref{app:Gamma_derivation}.

We can express the mass-operator $\mathcal{M}_{\boldsymbol{q}}(\boldsymbol{k}, \boldsymbol{p})$ in terms of $C^{(1)}_{\boldsymbol{q}}(\boldsymbol{k}, \boldsymbol{p})$ and subsequently invert equation Eq.~\eqref{eq:static_eigenvalue_susceptibility} for the susceptibilities $\chi_{\boldsymbol{q}}^\pm(\boldsymbol{k})$. In an isotropic approximation, where we can factorize microscopic and long-wavelength contributions to the three-point functions, the solution can be written as 
    \begin{equation}
        \chi_{\boldsymbol{q}}^{\pm}(\boldsymbol{k}) = \frac{1}{\varepsilon^{1/2}}\frac{b_{\boldsymbol{q}}^{\pm}h_0^{\mathrm{R}}(k) S(k)}{1 + (\varepsilon^{-1/4}\xi_0 q)^2},
    \label{eq:susceptibility_scaling_longtime}
    \end{equation}
in which the bare correlation length $\xi_0$ is given by
    \begin{equation}
        \xi_0 = \sqrt{\frac{\Gamma}{2g(1 - \lambda)}}.
    \end{equation}
The prefactor is given by the contraction of the source term with the left eigenfunction of the stability matrix
    \begin{equation}
        b^{\pm}_{\boldsymbol{q}} = \frac{1}{2g(1 - \lambda)} \int \frac{\mathrm{d}\boldsymbol{p}}{s_d} \left(\frac{h_0^{\mathrm{L}}(p)}{p^{d-1}}\right) \frac{\chi^{(0_{\pm})}_{\boldsymbol{q}}(\boldsymbol{p})}{S(p)},
    \end{equation} 
and can be systematically studied perturbatively in $q$.
These results are in agreement with the results of Refs.~\cite{biroli2006inhomogeneous, szamel2013breakdown}. Specifically, we identify the diverging correlation length $\xi_{\mathrm{d}} \propto \xi_0\varepsilon^{-1/4}$ when approaching the critical point from the non-ergodic phase. 

\paragraph*{Dynamic treatment.}
We next focus on the behavior of the susceptibilities in the $\beta$-regime, where we can approximate the propagator $F(k;t) \approx F_c(k) + \mathcal{O}(\sqrt{\varepsilon})$ as determined above. Following the spirit of the static calculation, we introduce a generalized Ansatz of the form
    \begin{equation}
        \chi^{\pm}_{\boldsymbol{q}}(\boldsymbol{k};z ; \omega) = \frac{1}{\varepsilon^{1/2}}\frac{\mathcal{B}_{\boldsymbol{q}}^{\pm}( \hat{z} ; \omega) S(k)h_0^{\mathrm{R}}(k)}{1 + (\varepsilon^{-1/4}\xi_0q)^2}. 
    \label{eq:susceptibility_scaling_app}
    \end{equation}
Substitution into the integral equation Eq.~\eqref{eq:dynamic_eigenvalue_susceptibility} and projecting onto the corresponding left eigenfunction $h_0^{\mathrm{L}}(k)/k^2$, the scaling function $\mathcal{B}^{\pm}_{\boldsymbol{q}}(\hat{z};\omega)$ is obtained as
    \begin{equation}
    \mathcal{B}^{\pm}_{\boldsymbol{q}}(\hat{z} ; \omega) = \frac{1}{2g(1-\lambda)} \int \frac{\mathrm{d}\boldsymbol{k}}{s_d} \left( \frac{h_0^{\mathrm{L}}(k)}{k^{d-1}} \right) \frac{\chi^{(0\pm)}_{\boldsymbol{q}}(\boldsymbol{k}; \hat{z} ; \omega)}{S(k)}.
    \label{eq:scaling_function_susceptibility}
    \end{equation}
Since the source terms $\chi^{(0\pm)}_{\boldsymbol{q}}(\boldsymbol{k}; \hat{z} ; \omega)$ are completely regular in all arguments, so are the scaling function $\mathcal{B}_{\boldsymbol{q}}^\pm(\hat{z} ; \omega)$. This new scaling function can be studied by a systematic expansion in $q$.

In summary, we find that the critical behavior of the dynamical susceptibility $\chi^{\pm}_{\boldsymbol{q}}(\boldsymbol{k}; t ; \omega) $ at finite times is unchanged in the $\beta$-regime and matches exactly that of the static treatment. 

\section{Computation of the constant $\Gamma$} \label{app:Gamma_derivation}

To determine $\Gamma$, we analyze the small-$\boldsymbol{q}$ behavior of the generalized stability operator $\mathcal{M}_{\boldsymbol{q}}(\boldsymbol{k}, \boldsymbol{p})$ Eq.~\eqref{eq:mass_operator_stability}. Since $\Gamma$ is the coefficient of the $q^2$ term in the expansion of the critical eigenvalue of $\mathcal{M}_{\boldsymbol{q}}(\boldsymbol{k}, \boldsymbol{p})$, it suffices to expand all $\boldsymbol{q}$-dependent factors to second order about $\boldsymbol{q}=\boldsymbol{0}$. The generalized stability operator can be written in the small-$q$ limit as
    \begin{equation}
        \mathcal{M}_{\boldsymbol{q}}(\boldsymbol{k}, \boldsymbol{p}) = \mathcal{M}_{\boldsymbol{0}}(\boldsymbol{k}, \boldsymbol{p}) + \sum_{\alpha,\beta} q_{\alpha}q_{\beta} \mathcal{Q}_{\alpha\beta}(\boldsymbol{k}, \boldsymbol{p}) + \mathcal{O}(q^4).
    \end{equation}
Rotational invariance ensures that only the scalar combination $q^2$ contributes. The prefactor of this term defines the constant $\Gamma$ of Eq.~\eqref{eq:mass_operator_stability}
    \begin{equation}
        \Gamma = \int \frac{\mathrm{d}\boldsymbol{k}}{s_d} \left(\frac{h_0^{\mathrm{L}}(k)}{k^2} \right) \frac{1}{3}\operatorname{Tr}\left[\underline{\mathcal{Q}}(k)\right].
    \label{eq:Gamma_def_Qtensor}
    \end{equation}
We begin with the radial Taylor expansion, valid for any sufficiently smooth scalar function $a(k)$,
    \begin{equation}
        a(\boldsymbol{k}\pm\frac{\boldsymbol{q}}{2}) = a(k) \pm \frac{\hat{\boldsymbol{k}}\cdot\boldsymbol{q}}{2}\left(\frac{\partial a(k)}{\partial k} \right) + \frac{1}{8}\left(q^2 - (\hat{\boldsymbol{k}}\cdot\boldsymbol{q})^2 \right)\left(\frac{1}{k}\frac{\partial a(k)}{\partial k}\right) + \frac{(\hat{\boldsymbol{k}}\cdot\boldsymbol{q})^2}{8}\frac{\partial^2 a(k)}{\partial k^2}
    \end{equation}
which we will use repeatedly below. We can then write, up to $\mathcal \mathcal{O}(q^2)$,
    \begin{equation}
    \begin{split}
        S(|\boldsymbol{k}-\frac{\boldsymbol{q}}{2}|) (1-f(|\boldsymbol{k}-\frac{\boldsymbol{q}}{2}|)) S(|\boldsymbol{k}+\frac{\boldsymbol{q}}{2}|) (1-f(|\boldsymbol{k}+\frac{\boldsymbol{q}}{2}|))
        =&\ S(k)^2(1-f(k))^2 - \frac{(\hat{\boldsymbol{k}}\cdot\boldsymbol{q})^2}{4} \left(\frac{\partial}{\partial k} S(k)(1-f(k)) \right)^2 \\
        &\ + S(k)(1-f(k))\left(\frac{q^2 - (\hat{\boldsymbol{k}}\cdot\boldsymbol{q})^2}{4}\right) \left(\frac{1}{k}\frac{\partial}{\partial k} \left(S(k)(1-f(k))\right)\right) \\
        &\ + S(k)(1-f(k)) \frac{(\hat{\boldsymbol{k}}\cdot\boldsymbol{q})^2}{4} \left(\frac{\partial^2}{\partial k^2}\left(S(k)(1-f(k)) \right)\right) + \mathcal{O}(q^3) 
    \end{split}
    \label{eq:q_expansion_symmetric_contrib_Mq}
    \end{equation}
As anticipated from arguments based on rotational symmetry, all linear terms in $\boldsymbol{q}$ cancel. Next we expand the vertex functions entering $\mathcal{M}_{\boldsymbol{q}}(\boldsymbol{k}, \boldsymbol{p})$. A second-order expansion of $\tilde v_{\boldsymbol k+\boldsymbol q/2}$ around $\boldsymbol{q}=\boldsymbol{0}$ reads
    \begin{equation}
    \begin{split}
        \tilde{v}_{\boldsymbol{k}+\boldsymbol{q}/2}(\boldsymbol{k}-\boldsymbol{p}, \boldsymbol{p}+\frac{\boldsymbol{q}}{2}) =&\ \tilde{v}_{\boldsymbol{k}}(\boldsymbol{p}, \boldsymbol{k}-\boldsymbol{p}) + \frac{1}{2}\sum_{\alpha}q_{\alpha}w_{\alpha}(\boldsymbol{k}, \boldsymbol{p}) + \frac{1}{4}\sum_{\alpha\beta} q_{\alpha}q_{\beta} W_{\alpha\beta}(\boldsymbol{k}, \boldsymbol{p})
    \end{split}
    \end{equation}
with the linear and quadratic coefficients
    \begin{equation}
    \begin{split}
        w_{\alpha}(\boldsymbol{k}, \boldsymbol{p}) =&\ \frac{(\boldsymbol{k}-\boldsymbol{p})_{\alpha}}{2k^2} c(|\boldsymbol{k}-\boldsymbol{p}|) + \frac{(\boldsymbol{k}+\boldsymbol{p})_{\alpha}}{2k^2} c(p) + \hat{p}_{\alpha}\frac{(\boldsymbol{k}\cdot\boldsymbol{p})}{2k^2} \frac{\mathrm{d}c(p)}{\mathrm{d}p} - k_{\alpha}\frac{(\boldsymbol{k}\cdot(\boldsymbol{k}-\boldsymbol{p}))}{k^4}c(|\boldsymbol{k}-\boldsymbol{p}|) - k_{\alpha}\frac{(\boldsymbol{k}\cdot\boldsymbol{p})}{k^4} c(p) \\
        =&\ \frac{(\boldsymbol{k}-\boldsymbol{p})_{\alpha}}{2k^2} c(|\boldsymbol{k}-\boldsymbol{p}|) + \frac{(\boldsymbol{k}+\boldsymbol{p})_{\alpha}}{2k^2} c(p) + \hat{p}_{\alpha}\frac{(\boldsymbol{k}\cdot\boldsymbol{p})}{2k^2} \frac{\mathrm{d}c(p)}{\mathrm{d}p} - \frac{k_{\alpha}}{k^3}\tilde{v}_{\boldsymbol{k}}(\boldsymbol{p}, \boldsymbol{k}-\boldsymbol{p})
    \end{split}
    \end{equation}
and
    \begin{equation}
    \begin{split}
        W_{\alpha\beta}(\boldsymbol{k}, \boldsymbol{p}) =&\ -\frac{2}{k^4} \boldsymbol{v}(\boldsymbol{p}, \boldsymbol{k}-\boldsymbol{p})_{\alpha}k_{\beta} - \frac{2c(p)}{k^4}k_{\alpha} k_{\beta} - \frac{2(\boldsymbol{k}\cdot\boldsymbol{p})}{k^4}\frac{\mathrm{d}c(p)}{\mathrm{d}p} \frac{p_{\alpha}k_{\beta}}{p} + \frac{1}{k^2p}\frac{\mathrm{d}c(p)}{\mathrm{d}p} (\boldsymbol{k}+\boldsymbol{p})_{\alpha} p_{\beta} + \frac{c(p)}{k^2}\delta_{\alpha\beta} \\
        &\ + \frac{\tilde{v}_{\boldsymbol{k}}(\boldsymbol{p}, \boldsymbol{k}-\boldsymbol{p})}{k^5} \left( 4k_{\alpha}k_{\beta}-k^2\delta_{\alpha\beta}\right) + \frac{(\boldsymbol{k}\cdot\boldsymbol{p})}{2k^2} \left[ \frac{(\delta_{\alpha\beta}-\hat{p}_{\alpha}\hat{p}_{\beta})}{p} \frac{\mathrm{d}c(p)}{\mathrm{d}p} +\hat{p}_{\alpha}\hat{p}_{\beta} \frac{\mathrm{d}^2c(p)}{\mathrm{d}p^2} \right].
    \end{split}
    \end{equation}
Hence, to quadratic order in $q$ we have
    \begin{equation}
        \tilde{v}_{\boldsymbol{k}-\boldsymbol{q}/2}(\boldsymbol{k}-\boldsymbol{p}, \boldsymbol{p}-\boldsymbol{q}/2)\tilde{v}_{\boldsymbol{k}+\boldsymbol{q}/2}(\boldsymbol{k}-\boldsymbol{p}, \boldsymbol{p}+\boldsymbol{q}/2) = \tilde{v}_{\boldsymbol{k}}(\boldsymbol{p}, \boldsymbol{k}-\boldsymbol{p})^2 - \frac{1}{4}\left(\boldsymbol{q}\cdot\boldsymbol{w}(\boldsymbol{k}, \boldsymbol{p}) \right)^2 + \frac{\tilde{v}_{\boldsymbol{k}}(\boldsymbol{p}, \boldsymbol{k}-\boldsymbol{p})}{2} \sum_{\alpha\beta}q_{\alpha}q_{\beta}W_{\alpha\beta}(\boldsymbol{k}, \boldsymbol{p}).
        \label{eq:vertex_squared_q_expansion}
    \end{equation}
Note that, once more, there are no linear terms in $q$ in Eq.~\eqref{eq:vertex_squared_q_expansion} as they cancel out. 

To obtain the tensor $\underline{\mathcal{Q}}(k)$, we multiply Eq.~\eqref{eq:q_expansion_symmetric_contrib_Mq} with Eq.~\eqref{eq:vertex_squared_q_expansion}. This yields the contraction at order $q^2$:
    \begin{equation}
    \begin{split}
        \sum_{\alpha\beta}q_{\alpha}q_{\beta}\mathcal{Q}^{\alpha\beta}(k) =&\ \frac{1}{S(k)}\int \frac{\mathrm{d}\boldsymbol{p}}{(2\pi)^d} \left[S(k)^2(1-f(k))^2 \sum_{\alpha\beta} q_{\alpha}q_{\beta}\left( - \frac{1}{4} w_{\alpha}(\boldsymbol{k}, \boldsymbol{p})w_{\beta}(\boldsymbol{k}, \boldsymbol{p}) + \frac{\tilde{v}_{\boldsymbol{k}}(\boldsymbol{p}, \boldsymbol{k}-\boldsymbol{p})}{2} W_{\alpha\beta}(\boldsymbol{k}, \boldsymbol{p}) \right)\right] \\
        &\ \hspace{2cm} \times S(|\boldsymbol{k}-\boldsymbol{p}|)f(|\boldsymbol{k}-\boldsymbol{p}|) S(p)h_0^{\mathrm{R}}(p) \\
        &\ + \frac{1}{S(k)}\int \frac{\mathrm{d}\boldsymbol{p}}{(2\pi)^d} S(k)(1-f(k))\sum_{\alpha\beta}q_{\alpha}q_{\beta}\left[\frac{\delta_{\alpha\beta} - \hat{k}_{\alpha}\hat{k}_{\beta}}{4} \frac{1}{k}\frac{\partial}{\partial k} \left( S(k)(1-f(k)) \right) + \frac{\hat{k}_{\alpha}\hat{k}_{\beta}}{4} \frac{\partial^2}{\partial k^2} \left( S(k)(1-f(k))\right)\right] \\
        &\ \hspace{2cm} \times\tilde{v}_{\boldsymbol{k}}(\boldsymbol{p}, \boldsymbol{k}-\boldsymbol{p})^2 S(|\boldsymbol{k}-\boldsymbol{p}|)f(|\boldsymbol{k}-\boldsymbol{p}|)S(p)h_0^{\mathrm{R}}(p)
    \end{split}
    \end{equation}
from which we read the matrix elements 
    \begin{equation}
    \begin{split}
        \mathcal{Q}^{\alpha\beta}(k) =&\ \frac{1}{S(k)}\int\frac{\mathrm{d}\boldsymbol{p}}{(2\pi)^d} \left[S(k)^2(1-f(k))^2 \left( - \frac{1}{4} w_{\alpha}(\boldsymbol{k}, \boldsymbol{p})w_{\beta}(\boldsymbol{k}, \boldsymbol{p}) + \frac{\tilde{v}_{\boldsymbol{k}}(\boldsymbol{p}, \boldsymbol{k}-\boldsymbol{p})}{2} W_{\alpha\beta}(\boldsymbol{k}, \boldsymbol{p}) \right)\right] S(|\boldsymbol{k}-\boldsymbol{p}|)f(|\boldsymbol{k}-\boldsymbol{p}|)S(p)h_0^{\mathrm{R}}(p)\\
        &\ + \frac{1}{S(k)}\int \frac{\mathrm{d}\boldsymbol{p}}{(2\pi)^d} S(k)(1-f(k))\left[\frac{\delta_{\alpha\beta} - \hat{k}_{\alpha}\hat{k}_{\beta}}{4} \frac{1}{k}\frac{\partial}{\partial k} \left( S(k)(1-f(k)) \right) + \frac{\hat{k}_{\alpha}\hat{k}_{\beta}}{4} \frac{\partial^2}{\partial k^2} \left( S(k)(1-f(k))\right)\right] \\
        &\ \hspace{3cm} \times\tilde{v}_{\boldsymbol{k}}(\boldsymbol{p}, \boldsymbol{k}-\boldsymbol{p})^2S(|\boldsymbol{k}-\boldsymbol{p}|)f(|\boldsymbol{k}-\boldsymbol{p}|) S(p)h_0^{\mathrm{R}}(p).
    \end{split}
    \end{equation}
The trace is therefore given by 
    \begin{equation}
    \begin{split}
        \frac{1}{3}\sum_{\alpha}\mathcal{Q}^{\alpha\alpha}(k) =&\ \frac{S(k)(1-f(k))^2}{3}\int \frac{\mathrm{d}\boldsymbol{p}}{(2\pi)^d}\left( - \frac{1}{4} w_{\alpha}(\boldsymbol{k}, \boldsymbol{p})w_{\alpha}(\boldsymbol{k}, \boldsymbol{p}) + \frac{\tilde{v}_{\boldsymbol{k}}(\boldsymbol{p}, \boldsymbol{k}-\boldsymbol{p})}{2} W_{\alpha\alpha}(\boldsymbol{k}, \boldsymbol{p}) \right)\\
        &\ \hspace{3cm} \times S(|\boldsymbol{k}-\boldsymbol{p}|)f(|\boldsymbol{k}-\boldsymbol{p}|)S(p)h_0^{\mathrm{R}}(p) \\
        &\ + \frac{1}{12}\frac{1}{S(k)(1-f(k))} \left[\frac{\partial^2}{\partial k^2} \left( S(k)(1-f(k))\right) \right]
    \end{split}
    \end{equation}
where 
    \begin{equation}
        W_{\alpha\alpha}(\boldsymbol{k}, \boldsymbol{p}) = \frac{1}{k}\left(-\tilde{v}_{\boldsymbol{k}}(\boldsymbol{p}, \boldsymbol{k}-\boldsymbol{p}) + c(p) - \left(2 (\hat{\boldsymbol{k}}\cdot\boldsymbol{p}) - 2(\boldsymbol{k}\cdot\boldsymbol{p}) - p^2 \right)\frac{\mathrm{d}c(p)}{\mathrm{d}p} + \frac{(\boldsymbol{k}\cdot\boldsymbol{p})}{2} \frac{\mathrm{d}^2c(p)}{\mathrm{d}p^2} \right).
    \end{equation}
This provides the explicit form of $\underline{\mathcal{Q}}(k)$, from which $\Gamma$ can be computed entirely in terms of standard MCT observables from Eq.~\eqref{eq:Gamma_def_Qtensor}. For the Percus–Yevick hard-sphere system, \citet{biroli2006inhomogeneous} reported $\Gamma=0.072\sigma^2$, while \citet{szamel2013breakdown} obtained $\Gamma=0.0708\sigma^2$.

\section{Perturbative Treatment of the Stochastic Process: Technical Details} \label{app:stochastic_perturbative_details}
The following subsections provide technical details that pertain to Sec.~\ref{sec:mapping_stochastic_process}. 
\subsection{Derivation of Eq.~\eqref{eq:stochastic_expansion_integral_form}} \label{app:derivation_stochastic_integral_eq}
The starting point is Eq.~\eqref{eq:postulate_microscopic_stochastic_expansion} which we can re-write as 
    \begin{equation}
    \begin{split}
        \frac{\partial }{\partial t}F_u^{(n)}(\boldsymbol{k}_+, \boldsymbol{k}_-; t, 0) &+ \frac{D_0k_+^2}{S(k_+)} \int_0^t \mathrm{d}\tau R_u^{(0)}(k_+ ; t-\tau) F_u^{(n)}(\boldsymbol{k}_+, \boldsymbol{k}_-  ; \tau, 0)\\
        =&\ -D_0 k_+ \int \frac{\mathrm{d}\boldsymbol{p}}{(2\pi)^d} \int_0^t \mathrm{d}\tau\sum_{m=1}^n \begin{pmatrix}
                n \\ m
            \end{pmatrix} R_u^{(m)}(\boldsymbol{k}_+, \boldsymbol{p} ; t, \tau) \frac{p}{S(p)} F_u^{(n-m)}(\boldsymbol{p}, \boldsymbol{k}_-  ; \tau, 0)\\
            &\ + \mathcal{S}^{+,(n)}_u(\boldsymbol{k}_+,\boldsymbol{k}_- ; t, 0) + \mathcal{S}^{-,(n)}_u(\boldsymbol{k}_+,\boldsymbol{k}_- ; t, 0)
    \end{split}
    \label{eq:postulated_stochastic_appendix}
    \end{equation}
by splitting off the $m=0$ contribution. Let us first consider the left-hand side of Eq.~\eqref{eq:postulated_stochastic_appendix} and convolve with $G_{\mathrm{MCT}}(k_+, t)$ from the left. We also note that $R_u^{(0)}(k_+ ; t-\tau) = R(k_+,t-\tau)$ in a mode-coupling approximation [see Eq.~\eqref{eq:resolvent}]. We can therefore write 
    \begin{equation}
    \begin{split}
        &\int_0^T \mathrm{d}t\ G_{\mathrm{MCT}}(k_+, T-t)\left[\frac{\partial}{\partial t}F_u^{(n)}(\boldsymbol{k}_+, \boldsymbol{k}_-; t, 0) + \frac{D_0k_+^2}{S(k_+)} \int_0^t \mathrm{d}\tau R_u^{(0)}(k_+ ; t-\tau) F_u^{(n)}(\boldsymbol{k}_+, \boldsymbol{k}_-  ; \tau, 0)\right] \\
        & \Leftrightarrow \mathrm{use\ Dyson\ Eq.:\ } D_0k^2/S(k) \int_0^t\mathrm{d}\tau R_0(k, t-\tau)G_{\mathrm{MCT}}(k,\tau) = -\partial_t G_{\mathrm{MCT}}(k,t)\\
        &\int_0^T \mathrm{d}t\ G_{\mathrm{MCT}}(k_+, T-t)\frac{\partial}{\partial t}F_u^{(n)}(\boldsymbol{k}_+, \boldsymbol{k}_-; t, 0) - \int_0^T\mathrm{d}t\frac{\partial G_{\mathrm{MCT}}(k_+, T-t)}{\partial (T-\tau) } F_u^{(n)}(\boldsymbol{k}_+, \boldsymbol{k}_- ; t,0) \\
        & \Leftrightarrow \mathrm{IBP\ the\ first\ term} \\
        & F_u^{(n)}(\boldsymbol{k}_+, \boldsymbol{k}_- ; T, 0) - G_{\mathrm{MCT}}(k_+, T)F_u^{(n)}(\boldsymbol{k}_+, \boldsymbol{k}_- ; 0, 0) - \int_0^T\mathrm{d}t \frac{\partial G_{\mathrm{MCT}}(k_+, T-t)}{\partial t} F_u^{(n)}(\boldsymbol{k}_+, \boldsymbol{k}_- ; t, 0) \\
        &\ \hspace{7.8cm} - \int_0^T\mathrm{d}t\frac{\partial G_{\mathrm{MCT}}(k_+, T-t)}{\partial (T-\tau) } F_u^{(n)}(\boldsymbol{k}_+, \boldsymbol{k}_- ; t,0) \\
        &\ \Leftrightarrow \mathrm{The\ last\ two\ terms\ cancel,\ and\ from\ Eq.~\eqref{eq:generalized_susceptibility_init}\ we\ have\ } F_u^{(n)}(\boldsymbol{k}_+, \boldsymbol{k}_- ; 0, 0) = 0 \\
        & \\
        & F_u^{(n)}(\boldsymbol{k}_+, \boldsymbol{k}_- ; T, 0)
    \end{split}
    \end{equation}
Hence, we get for the convolution of Eq.~\eqref{eq:postulated_stochastic_appendix} with with $G_{\mathrm{MCT}}(k_+, t)$
    \begin{equation}
    \begin{split}
        F_u^{(n)}(\boldsymbol{k}_+, \boldsymbol{k}_- ; t, 0) =&\ \int_0^t\mathrm{d}\tau \frac{F_{\mathrm{MCT}}(k_+, t-\tau)}{S(k_+)} \left[-D_0 k_+ \int \frac{\mathrm{d}\boldsymbol{p}}{(2\pi)^d} \int_0^{\tau} \mathrm{d}\tau'\sum_{m=1}^n \begin{pmatrix}
                n \\ m
            \end{pmatrix} R_u^{(m)}(\boldsymbol{k}_+, \boldsymbol{p} ; \tau, \tau') \frac{p}{S(p)} F_u^{(n-m)}(\boldsymbol{p}, \boldsymbol{k}_-  ; \tau', 0) \right]\\
            &\ + \int_0^t\mathrm{d}\tau \frac{F_{\mathrm{MCT}}(k_+, t-\tau)}{S(k_+)} \left[\mathcal{S}^{+,(n)}_u(\boldsymbol{k}_+,\boldsymbol{k}_- ; t, 0) + \mathcal{S}^{-,(n)}_u(\boldsymbol{k}_+,\boldsymbol{k}_- ; t, 0)\right]
    \end{split}
    \end{equation}
which is Eq.~\eqref{eq:stochastic_expansion_integral_form} presented in the main text. 

\subsection{Derivation of Eq.~\eqref{eq:overarching_rainbows_general_stochastic}}\label{app:derivation_overarching_rainbows_general_stochastic_eq}

We start from Eq.~\eqref{eq:stochastic_expansion_integral_form}. The goal is to isolate the term that generates the overarching propagator. To this end, we isolate the $m=n$ contribution on the first term of the right-hand side such that 
    \begin{equation}
    \begin{split}
        &\int_0^t \mathrm{d}\tau\int_0^{\tau} \mathrm{d}\tau' \int \frac{\mathrm{d}\boldsymbol{p}}{(2\pi)^d} \frac{F(k_+, t-\tau)}{S(k_+)} D_0k_+ \sum_{m=1}^n \begin{pmatrix}
                n \\ m
            \end{pmatrix} R_u^{(m)}(\boldsymbol{k}_+, \boldsymbol{p} ; \tau, \tau') \frac{p}{S(p)} F_u^{(n-m)}(\boldsymbol{p}, \boldsymbol{k}_-  ; \tau', 0) \\
        &\ \Leftrightarrow \\
        &\int_0^t \mathrm{d}\tau\int_0^{\tau} \mathrm{d}\tau'  \frac{F(k_+, t-\tau)}{S(k_+)} D_0k_+ R_u^{(n)}(\boldsymbol{k}_+, \boldsymbol{k}_- ; \tau, \tau') \frac{k_-}{S(k_-)} F_{\mathrm{MCT}}(k_-, \tau') + ...
    \end{split}
    \end{equation}
We further focus on the first contribution, which from Eq.~\eqref{eq:relation_M_Mirr_stochastic} contains 
    \begin{equation}
    \begin{split}
        &\int_0^t \mathrm{d}\tau\int_0^{\tau} \mathrm{d}\tau'  \frac{F(k_+, t-\tau)}{S(k_+)} D_0k_+ R_u^{(n)}(\boldsymbol{k}_+, \boldsymbol{k}_- ; \tau, \tau') \frac{k_-}{S(k_-)} F_{\mathrm{MCT}}(k_-, \tau') \\
        &\Leftrightarrow \\
        &\int_0^t \mathrm{d}\tau\int_0^{\tau} \mathrm{d}\tau'  \frac{F(k_+, t-\tau)}{S(k_+)} D_0k_+ \left( \int_{\tau'}^{\tau} \mathrm{d}t_1 \int_{\tau'}^{t_1}\mathrm{d}t_2 R(k_+, \tau-t_1) M_u^{\mathrm{irr},(n)}(\boldsymbol{k}_+, \boldsymbol{k}_- ; t_1,t_2) R(k_-, t_2-\tau') \right) \frac{k_-}{S(k_-)} F_{\mathrm{MCT}}(k_-, \tau') + ... \\
        & \Leftrightarrow \mathrm{Isolate\ term\ } F^{(n)}*F \mathrm{\ in\ } M_u^{\mathrm{irr},(n)}\\
        & \int_0^t \mathrm{d}\tau\int_0^{\tau} \mathrm{d}\tau'  \frac{F(k_+, t-\tau)}{S(k_+)} D_0k_+ R_0(k_+, \tau-t_1)\\
        & \hspace{2cm} \times \left(nD_0\int_{\tau'}^{\tau} \mathrm{d}t_1 \int_{\tau'}^{t_1}\mathrm{d}t_2\int \frac{\mathrm{d}\boldsymbol{p}}{(2\pi)^d} v_{\boldsymbol{k}_+}(\boldsymbol{p}, \boldsymbol{k}_+-\boldsymbol{p}) F_{\mathrm{MCT}}(p; t_1-t_2) F_u^{(n)}(\boldsymbol{k}_+-\boldsymbol{p}, \boldsymbol{k}_--\boldsymbol{p} ; t_1, t_2) v_{\boldsymbol{k}_-}(\boldsymbol{p}, \boldsymbol{k}_--\boldsymbol{p})\right) \\
        &\hspace{2cm} \times R_0(k_+, t_2-\tau') \frac{k_-}{S(k_-)} F_{\mathrm{MCT}}(k_-, \tau') + ... \\
        &\ \Leftrightarrow \mathrm{Change\ order\ of\ time\ integration\ \& \ relabel \ +\ Identify\ regularized\ vertices\ + \ Shift\ momentum\ integral} \\      &\int_0^t\mathrm{d}\tau_4...\int_0^{\tau_2}\mathrm{d}\tau_1 \int \frac{\mathrm{d}\boldsymbol{p}}{(2\pi)^d} F(k_+, t-\tau_4) \mathcal{v}_{12}^{\mathrm{reg}}(\boldsymbol{k}_+ ; \boldsymbol{k}-\boldsymbol{p}, \boldsymbol{p}+\tfrac{\boldsymbol{q}}{2} ; \tau_4-\tau_3) \times \\
        &\hspace{4cm} \times F_{\mathrm{MCT}}(|\boldsymbol{k}-\boldsymbol{p}|, \tau_3-\tau_1) F_u^{(n)}(\boldsymbol{p} + \tfrac{\boldsymbol{q}}{2}, \boldsymbol{p} - \tfrac{\boldsymbol{q}}{2} ; \tau_3, \tau_2)\mathcal{v}_{21}^{\mathrm{reg}}(\boldsymbol{k}-\boldsymbol{p}, \boldsymbol{p}-\tfrac{\boldsymbol{q}}{2} ; \tau_2-\tau_1)F_{\mathrm{MCT}}(k_- , \tau_1) + ...
    \end{split}
    \end{equation}
which we recognize as the term generating the overarching rainbows in Eq.~\eqref{eq:stochastic_expansion_integral_form}. Bundling up all the neglected terms (\textit{i.e.}\ those in the ``$...$") into a term denoted $A_u^{(n)}(\boldsymbol{k}_+, \boldsymbol{k}_- ; t, 0)$ gives Eq.~\eqref{eq:overarching_rainbows_general_stochastic} presented in the main text.

\section{Contributions to $I_u(\boldsymbol{k}_+, \boldsymbol{k}_- ; \hat{z})$~: Technical Details}
\label{app:derivation_stochastic_inhomogeneities}

By direct substitution of the generalized ansatz into the stochastic process, it is possible to show that the inhomogeneities are given by 
    \begin{equation}
    \begin{split}
        I_u(\boldsymbol{k}_+, \boldsymbol{k}_- ; \hat{z}) =&\ - \frac{1}{f(|\boldsymbol{k}+\boldsymbol{q}/2|)}\int \frac{\mathrm{d}\boldsymbol{p}}{(2\pi)^d}\frac{1}{f(p)} S(\frac{|\boldsymbol{k}-\boldsymbol{q}/2+\boldsymbol{p}|}{2}) h_0^{\mathrm{R}}(\frac{|\boldsymbol{k}+\boldsymbol{q}/2+\boldsymbol{p}|}{2})h_0^{\mathrm{R}}(\frac{|\boldsymbol{k}-\boldsymbol{q}/2+\boldsymbol{p}|}{2})\\
        &\ \hspace{3.5cm} \times \hat{z}^2g_u(\boldsymbol{k}+\frac{\boldsymbol{q}}{2}-\boldsymbol{p}, \hat{z})g_u(\boldsymbol{k}-\frac{\boldsymbol{q}}{2}-\boldsymbol{p}, \hat{z}) \\
        &\ + \frac{S(|\boldsymbol{k}-\boldsymbol{q}/2|)}{f(|\boldsymbol{k}+\boldsymbol{q}/2|)} \int \frac{\mathrm{d}\boldsymbol{p}}{(2\pi)^d} \frac{1}{f(p)(1-f(p))} h_0^{\mathrm{R}}(\frac{|\boldsymbol{k}+\boldsymbol{q}/2+\boldsymbol{p}|}{2})h_0^{\mathrm{R}}(\frac{|\boldsymbol{k}-\boldsymbol{q}/2+\boldsymbol{p}|}{2})\\
        &\ \hspace{3.5cm} \times\hat{z}^2g_u(\boldsymbol{k}+\frac{\boldsymbol{q}}{2}-\boldsymbol{p}, \hat{z})g_u(\boldsymbol{k}-\frac{\boldsymbol{q}}{2}-\boldsymbol{p}, \hat{z}) \\
        &\ - \frac{S(|\boldsymbol{k}-\boldsymbol{q}/2|)}{f(|\boldsymbol{k}+\boldsymbol{q}/2|)} \int \frac{\mathrm{d}\boldsymbol{p}}{(2\pi)^d}\frac{\mathrm{d}\boldsymbol{p}'}{(2\pi)^d}C_{\boldsymbol{q}}^{(2)}(\boldsymbol{k}, \boldsymbol{p}, \boldsymbol{p}')h_0^{\mathrm{R}}(\frac{|\boldsymbol{p}+\boldsymbol{p}'|}{2}) h_0^{\mathrm{R}}(\frac{|2\boldsymbol{k}-\boldsymbol{p}-\boldsymbol{p}'|}{2})\\
        &\  \hspace{3.5cm} \times \hat{z}\mathcal{L}\{g_u(\boldsymbol{p}-\boldsymbol{p}', \hat{t})g_u(\boldsymbol{q}-\boldsymbol{p}+\boldsymbol{p}', \hat{t})\}(\hat{z})
    \end{split}
    \label{eq:I_u_direct_sub}
    \end{equation}
in which we have defined the integral kernel
    \begin{equation}
    \begin{split}
        C^{(2)}_{\boldsymbol{k}-\boldsymbol{k}'}(\frac{\boldsymbol{k}+\boldsymbol{k}'}{2}, \boldsymbol{p}, \boldsymbol{p}') =&\ \frac{n_c}{2}S(k)(1-f(k)) \tilde{v}_{\boldsymbol{k}}(\frac{\boldsymbol{k}+\boldsymbol{k}'}{2}-\boldsymbol{p}, \boldsymbol{p}+\frac{\boldsymbol{k}-\boldsymbol{k}'}{2}) \tilde{v}_{\boldsymbol{k}'}(\frac{\boldsymbol{k}'+\boldsymbol{k}}{2}-\boldsymbol{p}, \boldsymbol{p}+\frac{\boldsymbol{k}'-\boldsymbol{k}}{2})\\
        &\ \times S(\frac{|\boldsymbol{k}-\boldsymbol{p}|}{2}) S(\frac{|\boldsymbol{p}+\boldsymbol{p}'|}{2})(1-f(k')).        
    \end{split}
    \end{equation}
Writing it in the off-diagonal basis $(\boldsymbol{k}+\boldsymbol{q}/2, \boldsymbol{k}-\boldsymbol{q}/2)$ we recover the mean-field definition Eq.~\eqref{eq:def_C2_MF} via $C^{(2)}(\boldsymbol{k}, \boldsymbol{p}) = C^{(2)}_{\boldsymbol{q}}(\boldsymbol{k}, \boldsymbol{p}, \boldsymbol{p}')\delta(\boldsymbol{q})\delta(\boldsymbol{p}-\boldsymbol{p}')$.
We next focus on each of these three-terms separately, focusing on the long-wavelength contributions.
\subsection{Derivation of Eq.~\eqref{eq:Iu_1}}
The first term of Eq.~\eqref{eq:I_u_direct_sub} can be rearranged to 
    \begin{equation}
    \begin{split}
        & \frac{1}{f(|\boldsymbol{k}+\boldsymbol{q}/2|)}\int \frac{\mathrm{d}\boldsymbol{p}}{(2\pi)^d}\frac{1}{f(p)} S(\frac{|\boldsymbol{k}-\boldsymbol{q}/2+\boldsymbol{p}|}{2}) h_0^{\mathrm{R}}(\frac{|\boldsymbol{k}+\boldsymbol{q}/2+\boldsymbol{p}|}{2})h_0^{\mathrm{R}}(\frac{|\boldsymbol{k}-\boldsymbol{q}/2+\boldsymbol{p}|}{2})\hat{z}^2g_u(\boldsymbol{k}+\frac{\boldsymbol{q}}{2}-\boldsymbol{p}, \hat{z})g_u(\boldsymbol{k}-\frac{\boldsymbol{q}}{2}-\boldsymbol{p}, \hat{z}) \\
        &\Leftrightarrow \mathrm{change\ of\ variable\ } \boldsymbol{p}\rightarrow \boldsymbol{p}' = \boldsymbol{k}+\boldsymbol{q}/2-\boldsymbol{p} \\
        & \frac{1}{f(|\boldsymbol{k}+\boldsymbol{q}/2|)} \int \frac{\mathrm{d}\boldsymbol{p}'}{(2\pi)^d} \frac{1}{f(|\boldsymbol{k}+\boldsymbol{q}/2-\boldsymbol{p}'|)}S(\frac{|2\boldsymbol{k}-\boldsymbol{p}'|}{2}) h_0^{\mathrm{R}}(\frac{|2\boldsymbol{k}+\boldsymbol{q}-\boldsymbol{p}'|}{2})h_0^{\mathrm{R}}(\frac{|2\boldsymbol{k}-\boldsymbol{p}'|}{2})\hat{z}^2g_u(\boldsymbol{p}', \hat{z})g_u(\boldsymbol{q}-\boldsymbol{p}', \hat{z}) \\
        &\Leftrightarrow \mathrm{Consider\ integrand\ in \ } |\boldsymbol{q}|,|\boldsymbol{p}'|\ll1 \mathrm{\ limit\ and\ relabel\ \ integration\ variable\ to\ } \boldsymbol{p} \\
        & \approx \frac{S(k)}{f(k)}\frac{h_0^{\mathrm{R}}(k)^2}{f(k)} \hat{z}^2 \int \frac{\mathrm{d}\boldsymbol{p}}{(2\pi)^d} g_u(\boldsymbol{p};\hat{z})g_u(\boldsymbol{q}-\boldsymbol{p};\hat{z})
    \end{split}
    \end{equation}
which corresponds to Eq.~\eqref{eq:Iu_1} of the main text.
\subsection{Derivation of Eq.~\eqref{eq:Iu_2}}
The second term of Eq.~\eqref{eq:I_u_direct_sub} can be rearranged to
    \begin{equation}
    \begin{split}
        &\frac{S(|\boldsymbol{k}-\boldsymbol{q}/2|)}{f(|\boldsymbol{k}+\boldsymbol{q}/2|)} \int \frac{\mathrm{d}\boldsymbol{p}}{(2\pi)^d} \frac{1}{f(p)(1-f(p))} h_0^{\mathrm{R}}(\frac{|\boldsymbol{k}+\boldsymbol{q}/2+\boldsymbol{p}|}{2})h_0^{\mathrm{R}}(\frac{|\boldsymbol{k}-\boldsymbol{q}/2+\boldsymbol{p}|}{2}) \hat{z}^2g_u(\boldsymbol{k}+\frac{\boldsymbol{q}}{2}-\boldsymbol{p}; \hat{z})g_u(\boldsymbol{k}-\frac{\boldsymbol{q}}{2}-\boldsymbol{p}; \hat{z}) \\
        &\ \approx \frac{S(k)}{f(k)} \frac{h_0^{\mathrm{R}}(k)^2}{f(k)(1-f(k))} \hat{z}^2\int \frac{\mathrm{d}\boldsymbol{p}}{(2\pi)^d} g_u(\boldsymbol{p};\hat{z})g_u(\boldsymbol{q}-\boldsymbol{p};\hat{z})
    \end{split}
    \end{equation}
using the same integration change and small integrand limit as above. This corresponds to Eq.~\eqref{eq:Iu_2} of the main text.

\subsection{Derivation of Eq.~\eqref{eq:Iu_3}}
Starting from the third term of Eq.~\eqref{eq:I_u_direct_sub}, we can write 
    \begin{equation}
    \begin{split}
        & \frac{S(|\boldsymbol{k}-\boldsymbol{q}/2|)}{f(|\boldsymbol{k}+\boldsymbol{q}/2|)} \int \frac{\mathrm{d}\boldsymbol{p}}{(2\pi)^d}\frac{\mathrm{d}\boldsymbol{p}'}{(2\pi)^d} C_{\boldsymbol{q}}^{(2)}(\boldsymbol{k}, \boldsymbol{p}, \boldsymbol{p}')h_0^{\mathrm{R}}(\frac{|\boldsymbol{p}+\boldsymbol{p}'|}{2}) h_0^{\mathrm{R}}(\frac{|2\boldsymbol{k}-\boldsymbol{p}-\boldsymbol{p}'|}{2})\hat{z}\mathcal{L}\{g_u(\boldsymbol{p}-\boldsymbol{p}'; \hat{t})g_u(\boldsymbol{q}-\boldsymbol{p}+\boldsymbol{p}'; \hat{t})\}(\hat{z}) \\
        &\Leftrightarrow \mathrm{let\ } \boldsymbol{p}' \to \boldsymbol{p}'' = \boldsymbol{p}-\boldsymbol{p}' \\
        &\ \frac{S(|\boldsymbol{k}-\boldsymbol{q}/2|)}{f(|\boldsymbol{k}+\boldsymbol{q}/2|)} \int \frac{\mathrm{d}\boldsymbol{p}}{(2\pi)^d} \frac{\mathrm{d}\boldsymbol{p}''}{(2\pi)^d} C^{(2)}_{\boldsymbol{q}}(\boldsymbol{k}, \boldsymbol{p}, \boldsymbol{p}-\boldsymbol{p}'') h_0^{\mathrm{R}}(\frac{|2\boldsymbol{p}-\boldsymbol{p}''|}{2})h_0^{\mathrm{R}}(\frac{|2\boldsymbol{k}-2\boldsymbol{p}+\boldsymbol{p}''|}{2}) \hat{z}\mathcal{L}\{g_u(\boldsymbol{p}''; \hat{t})g_u(\boldsymbol{q}-\boldsymbol{p}''; \hat{t})\}(\hat{z}) \\
        &\Leftrightarrow \mathrm{Consider\ integrand\ in \ } |\boldsymbol{q}|,|\boldsymbol{p}''|\ll1 \mathrm{\ limit\ and\ relabel\ \ integration\ variable\ } \boldsymbol{p}'' \mathrm{\ to \ } \boldsymbol{p}' \\
        &\approx \frac{S(k)}{f(k)} \int \frac{\mathrm{d}\boldsymbol{p}}{(2\pi)^d}C_{\boldsymbol{q}\to\boldsymbol{0}}^{(2)}(\boldsymbol{k}, \boldsymbol{p}, \boldsymbol{p}) h_0^{\mathrm{R}}(p)h_0^{\mathrm{R}}(|\boldsymbol{k}-\boldsymbol{p}|) \hat{z}\int \frac{\mathrm{d}\boldsymbol{p}'}{(2\pi)^d} \mathcal{L}\{ g_u(\boldsymbol{p}'; \hat{t})g_u(\boldsymbol{q}-\boldsymbol{p}'; \hat{t})\}(\hat{z}) \\
        & \Leftrightarrow \mathrm{\ use\ } C_{\boldsymbol{q}\to\boldsymbol{0}}^{(2)}(\boldsymbol{k}, \boldsymbol{p}, \boldsymbol{p}) = C^{(2)}(\boldsymbol{k}, \boldsymbol{p}) \\
        & \frac{S(k)}{f_c(k)} \int \frac{\mathrm{d}\boldsymbol{p}}{(2\pi)^d} C^{(2)}(\boldsymbol{k}, \boldsymbol{p})h_0^{\mathrm{R}}(p)h_0^{\mathrm{R}}(|\boldsymbol{k}-\boldsymbol{p}|) \hat{z}\mathcal{L}\left\{ \int\frac{\mathrm{d}\boldsymbol{p}'}{(2\pi)^d}g_u(\boldsymbol{p}'; \hat{t})g_u(\boldsymbol{q}-\boldsymbol{p}'; \hat{t})\right\}(\hat{z}).
    \end{split}
    \end{equation}
This corresponds to Eq.~\eqref{eq:Iu_3} of the main text.
\end{widetext}

\end{document}